\documentclass[conference]{IEEEtran}

\usepackage{amsfonts}
\usepackage{tabu}
\usepackage{booktabs}
\usepackage{gensymb}
\usepackage{textcomp}
\usepackage{amsmath}
\usepackage{algorithm}
\usepackage{algpseudocode}
\usepackage{pifont}
\usepackage{verbatim}
\usepackage{graphicx}
\PassOptionsToPackage{hyphens}{url}
\usepackage{hyperref}
\usepackage{seqsplit}
\usepackage{caption}
\usepackage{array}
\usepackage{subcaption}
\usepackage{booktabs}
\usepackage{breqn}
\usepackage[capitalise,noabbrev]{cleveref} \crefname{appsec}{Appendix}{Appendices} %
\usepackage{amssymb}
\usepackage{xcolor}
\usepackage{enumerate}
\usepackage{stfloats}

\begin{document}

\title{\emph{Zoom on the Keystrokes:} Exploiting \\Video Calls for Keystroke Inference Attacks \thanks{Research reported in this publication was supported by the Division of Computer and Network Systems (CNS) of the National Science Foundation (NSF) under award number 1943351.}}

\author{\IEEEauthorblockN{Mohd Sabra}
\IEEEauthorblockA{University of Texas at San Antonio\\
\href{mailto:mohd.sabra@utsa.edu}{\small\texttt{mohd.sabra@utsa.edu}}}
\and
\IEEEauthorblockN{Anindya Maiti}
\IEEEauthorblockA{University of Oklahoma\\
\href{mailto:a.maiti@ieee.org}{\small\texttt{a.maiti@ieee.org}}}
\and
\IEEEauthorblockN{Murtuza Jadliwala}
\IEEEauthorblockA{University of Texas at San Antonio\\
\href{mailto:murtuza.jadliwala@utsa.edu}{\small\texttt{murtuza.jadliwala@utsa.edu}}}
}

\IEEEoverridecommandlockouts
\makeatletter\def\@IEEEpubidpullup{6.5\baselineskip}\makeatother
\IEEEpubid{\parbox{\columnwidth}{
    Network and Distributed Systems Security (NDSS) Symposium 2021\\
    21-24 February 2021, San Diego, CA, USA\\
    ISBN 1-891562-61-4\\
    https://dx.doi.org/10.14722/ndss.2021.23xxx\\
    www.ndss-symposium.org
}
\hspace{\columnsep}\makebox[\columnwidth]{}}

\maketitle

\begin{abstract}
Due to recent world events, video calls have become the new norm for both personal and professional remote communication. However, if a participant in a video call is not careful, he/she can reveal his/her private information to others in the call. In this paper, we design and evaluate an attack framework to infer one type of such private information from the video stream of a call -- keystrokes, i.e., text typed during the call. 
{ We evaluate our video-based keystroke inference framework using different experimental settings and parameters, including different webcams, video resolutions, keyboards, clothing, and backgrounds. Our relatively high keystroke inference accuracies under commonly occurring and realistic settings highlight the need for awareness and countermeasures against such attacks.}
Consequently, we also propose and evaluate effective mitigation techniques that can automatically protect users when they type during a video call.

\end{abstract}

 \vspace{-0.1in}

\section{Introduction}
\label{sec:introduction}

Catalyzed by the ubiquity of the Internet, audio-video calling has become a mainstream method of remote communication \cite{talkative-video}. The trend has recently seen a further boost due to the COVID-19 pandemic \cite{covid19}, whereby audio-video calls became the default medium for professionals to confer remotely and for students to attend lectures from home. Nonetheless, audio-video calls (often referred to as just video calls) should be considered privacy sensitive as participants may have spoken or displayed private information during the call. To protect video calls from eavesdropping threats, secure video calling protocols are usually end-to-end encrypted. However, even if we disregard widespread system weaknesses \cite{zoom-hack-1,zoom-hack-2}, end-to-end encryption may be ineffective when an adversary is present at one end of the video call. 

Can an adversary, who is at one end of a video call, infer some potentially sensitive information about the participant at the other end which is not trivially visible/audible from the call? Modern video calling softwares such as Skype  \cite{skype}, Hangouts \cite{googlehangout} and Zoom \cite{zoom} already provide features such as background-blurring to enable users to potentially blur/hide everything in the users' background, except their body. In this work, we want to investigate what sensitive information can be inferred by just observing a target users' body and physiological features in an audio/video call. More specifically, we would like to investigate the feasibility of inferring keystrokes of a target user on a traditional QWERTY keyboard by just observing their video feed on a video calling application such as Skype, Hangouts and Zoom.

Prior efforts in the literature have shown that the sound (audio) of keystrokes typed during a video call can be exploited to infer the text typed \cite{compagno2017don,anand2018keyboard}. But, audio-based keystroke inference is not very practical primarily because of naturally occurring (audio) noises in an audio-video call signal, such as background sound and participants talking \cite{anand2018keyboard}. Moreover, such audio-based attacks may not work for the relatively \emph{quieter} membrane and dome-switch keyboards, because (i) the low amplitude keystroke audio emanations are not effectively captured by many entry-level microphones \cite{van2011ability}, and (ii) non-vocal low amplitude audio frequencies are often filtered out by many video calling applications \cite{gmeet-noise-cancel,zoom-noise-cancel}. 

We believe that the video signal in such calls is less prone to naturally occurring noises and can be exploited for effective keystroke inference attacks. It is also a relatively unexplored modality for keystroke inference. 
{
 Our contributions in this paper includes modeling commonly observed typing behaviors during a video call, and utilizing them to construct a novel video-based keystroke and typing detection framework. We then create a text inference framework that uses the keystrokes detected from the video to predict words that were most likely typed by the target user (\cref{sec:system}). We then comprehensively evaluate both the keystroke/typing detection and text inference frameworks using data collected from a large number of human subject participants in two different settings: (i) an \emph{In-Lab} setting (\cref{sec:experimental_setup}) where the video call setup, including the device(s) used for the calls, participants' sitting position during the call, and the text typed by the participants is fixed, and (ii) an \emph{At-Home} setting (\cref{sec:experimental_setup_home}) where the video call data is collected in a realistic environment without any constraints or requirements. Evaluation results for the In-Lab setting are outlined in \cref{sec:evaluation,sec:audiovideo_evaluation}, while results for the At-Home setting appear in \cref{sec:home_eva}. 
}
We also propose and evaluate multiple techniques which can help in the mitigation of such keystroke inference attacks from video calls (\cref{sec:mitigations}).

 \vspace{-0.1in}
\section{Related Work}
\label{sec:related}

The research literature is rich with various modalities of side-channel inference threats targeting different types of private information. We limit our literature review discussion in two closely related directions as follows.

\noindent
\textbf{Keystroke Inference Threats.}
This direction deals with the class of inference threats that aim to reconstruct text typed by a target user, using one or more modalities of side-channels. Keystroke inference attacks can have potentially dangerous consequences as the text typed can often be private in nature, and can sometime even contain sensitive information, such as credit card numbers, authentication codes, and addresses. Side-channel keystroke inference threats have utilized electromagnetic emanations from keyboards \cite{Vuagnoux:2009}, optical or visual cues \cite{simon2013pin,sun2016visible,chen2018eyetell}, Wi-Fi channel state information \cite{ali2015keystroke,li2016csi}, audio or acoustic signals from keyboards \cite{AsonovA:2004,BergerWY:2006,ZhuangZT:2009,HaleviS:2012,ZhuMZL:2014,compagno2017don,anand2018keyboard}, typing-related table vibrations captured by a nearby sensor \cite{MarquardtVCT:2012}, smartphone motion sensors (to infer text typed on the smartphone) \cite{CaiC:2011,miluzzo2012tapprints,OwusuHDPZ:2012,xu2012taplogger}, and wearable motion sensors \cite{wang2015mole,liu2015good,wang2016friend,maiti2016smartwatch,maiti2018side}.
Among these prior works, \cite{compagno2017don} and \cite{anand2018keyboard} are the most closely related research efforts to ours. Both works demonstrated the feasibility of accurate keystroke inference threats from keystroke sounds propagated over a video call. However, as mentioned earlier, such sound or audio based threats may not be practical because of naturally occurring interferences (such as participants talking) and background noise-cancellation techniques being used by many video calling applications \cite{gmeet-noise-cancel,zoom-noise-cancel} that also eliminates/reduces the propagation of keystroke sounds.

\noindent
\textbf{Inference Attacks Using Visual Side-Channels.}
With the recent ubiquity of video capturing hardware embedded in many consumer electronics, such as smartphones, tablets, and laptops, the threat of information leakage through visual channel has amplified. Moreover, high-end digital cameras and lenses have also trickled down to the consumer market at competitive prices, making them easily accessible to an adversary. Earlier works leveraged upon optical emanations from monitors \cite{Kuhn:2002} or from eyes \cite{BackesTDLW:2009} to infer information such as content being displayed or watched. More recent works studied information leakage due to outdoor light effusions, such as inference of multimedia consumption \cite{maiti2019light,xu2014watching} and private data exfiltration using smart lights \cite{maiti2019light,ronen2016extended}.
A few research works also leveraged on visual side-channel for keystroke inference threats. Simon and Anderson \cite{simon2013pin} were able to infer keystrokes made on a smartphone based on visual movements, captured from the on-device camera, that occurred as a result of individual keystrokes. Similarly, Sun et al. \cite{sun2016visible} were able to infer keystrokes typed on a tablet just from visual observation of the rear side of the tablet. Chen et al. \cite{chen2018eyetell} leveraged on users' eye movements for touchscreen keystroke inference. To the best of our knowledge, our paper is the first work that proposes and evaluates a keystroke inference framework that solely leverages body movements observable in video calls.

 \vspace{-0.1in}
\section{Background}
\label{sec:background}

In this section, we describe the different factors affecting typing-related body movements and the characteristics of the corresponding data available from a video call. %

\noindent
\textbf{Muscles, Joints and Motor Control.}
Human bodily movements, clinically known as motor functions, are achieved primarily through movement of joints. Joints are formed where two or more bones are connected via ligaments, a flexible fibrous connective tissue which binds together bones and/or cartilages. These joints allow several degrees of freedom in the human body. With the help of muscles that wrap around the bones and connect to the bones via tendons, along with motor control signals from the brain that appropriately engages the required set of muscles, the human body can carry out tasks through coordinated joint movements. Tasks may range from basic balance and stability of the body to more complex actions, such as running or typing on a keyboard. Also, most tasks are actuated through movement of multiple joints simultaneously, rather than just one at a time. %
\cref{fig:arms} (\cref{appendix:background-arms-typing}) shows an overview of the shoulder and arm bones, and their joints, which when engaged in a series of harmonious movements can enable a human to type hundreds of words per minute \cite{ostrach1997typing}.

When a user starts typing a sentence, the initial set of joint movements are significantly dependent on the keyboard positioning and user's typing habits (i.e., the set of motor control signals learned by the brain). A user can exercise an elbow joint, shoulder joint or joints from the Carpus and Metacarpus regions or a combination of all of them, that ultimately positions the user's finger on the initial key. The joint movements associated with keystrokes following the initial keystroke depends primarily on the user's typing style, e.g., \emph{hunt-and-peck}, \emph{touch-typing}, or \emph{hybrid} (more details on typing styles in \cref{appendix:typing-styles}).
Certain typing styles, such as hunt-and-peck, result in significant upper hand movements (not just fingers or wrist) between keystrokes, than others. %
For instance, in hunt-and-peck typing the elbow, shoulder, and Carpus joints are heavily utilized, whereas in touch-typing the Carpus, Metacarpus, and Phalanges joints are heavily utilized.

\noindent
\textbf{Reaction Force of a Keystroke.}
A common phenomenon observed across all typing styles is that whenever the user presses a key, a reaction force is produced in the opposite direction (Newton's third law of motion). This reaction force propagates throughout the arm and shoulder muscles and joints until the force is absorbed by the body. As a result, even if the user uses only the joints in the Phalanges bones to press a key, one can visually observe subtle arm and shoulder movements due to the reaction force exerted by the key on the user's hand. However, the \emph{five fingers} of each hand are connected through different bones in the wrist, that have different joints in the Carpus area as shown in Figure \ref{fig:arms} (\cref{appendix:background-arms-typing}). %
As a result, the reaction force of a keystroke propagates slightly differently through the arm and shoulder muscles and joints, depending on which finger was used to press the key. Visually, this translates to distinguishable types of upper hand movements (which is observable through a webcam during a video call) for key presses with different fingers.

\noindent
\textbf{Characteristics of the Available Call Data.}
During a typical video call, an adversary can leverage two different types of information for inferring keystrokes made during the call.

\noindent
\emph{(1) Video Data or Feed:}
Most modern video calling applications employ a webcam to capture the visual and/or audio signals during the call. %
The webcam's camera sensor captures the visual image of the target subject and his/her surroundings as a series of two-dimensional images, also known as \emph{frames}. In other words, a video is a chronological ordering of two dimensional images or frames, displayed in that order at a very high rate or speed (often measured in frames/second or $fps$ in short). A typical modern webcam can capture 30 or 60 $fps$, and at $1280\times720$ (921,600 $pixels$) or $1920\times1080$ (2,073,600 $pixels$) resolution per frame. %
Based on the traditional position of a webcam during a video call, lateral movements of hand and shoulder can easily be observed in the captured video (\cref{fig:heatmap-l,fig:heatmap-r} in  \cref{appendix:background-arms-typing}).

\noindent
\emph{(2) Audio Data or Feed:}
The sound during a video call is typically captured either using a microphone sensor integrated within the webcam or using an external microphone. The captured sound often contains both the user's voice (or speech) and background/ambient noise, including sound related to the keystrokes made by the user or any other activities performed by the user during the call. Video calling softwares also often implement audio optimizations, such as dampening/filtering of non-vocal frequencies \cite{gmeet-noise-cancel,zoom-noise-cancel} and echo suppression/cancellation.
Most modern microphones can record audio signals at $44,100 Hz$ or more, and such high-fidelity audio can capture fine-grained audio characteristics. However, a microphone may not always be enabled by the user during a video call, as observed in the recent popularity of the push-to-talk feature provided by most video calling software that lets a user mute sound at the push of a button. Also, during multi-participant video conference calls, it is a common courtesy or etiquette for participants to mute their microphone when they are not actively speaking. Nonetheless, while our attack framework employs only the available video data for keystroke inference, we assume the availability of audio data for comparative evaluation with a prior work \cite{compagno2017don}.

 \vspace{-0.1in}
\section{Adversary Model}
\label{sec:adversary}

The goal of an adversary in our setting is to infer keystrokes typed by a target user at the opposite end of a video conference/call by just employing the video feed from the call. To undertake a purely video-based keystroke inference attack, we assume that the adversary first records the video feed of the call where the target user was a participant, and that the target user typed private text on her/his keyboard during the call. 
{
More specifically, we assume a field-of-view of a typical desktop or laptop webcam where the video stream would consist of only visible upper body movements of the user (and not the actual keys being typed as then the attack would be trivial), as shown in \cref{fig:background-removal} (\cref{appendix:preprocessing-details}). Additionally, we assume that both shoulders and upper arms are within the field-of-view of the webcam, which is a practical assumption because desktop and laptop webcams are often positioned centrally with respect to the user. 
In addition to targeting participants in live video calls, an adversary could also potentially target videos obtained from public video sharing/streaming platforms such as YouTube \cite{youtube} and Twitch \cite{twitch}.
Many live streamers interact with their laptop or desktop during a live video exposition, which could include sensitive typed information, and thus, potentially targeted by an adversary.
As outlined earlier, the recorded video is nothing but a series of picture frames sampled at a constant frequency, and the adversary's goal is to utilize the observable upper body movements across all the recorded frames to infer the private text typed by the target. The adversary does not have any other medium of inferring the private text typed, and must rely entirely on the video stream. This makes our attack more practical as it can target not only real-time video calls (both in a public and private setting), as well as, archived videos of live exposition/events -- all an adversary needs for the attack is a video stream.
}

 \vspace{-0.1in}
\section{Attack Framework}
\label{sec:system}

In this section, we outline the technical details of our keystroke inference and word prediction framework.

\vspace{-0.1in}
\subsection{Overview}
\begin{figure}[t]
\centering
\includegraphics[width=\linewidth]{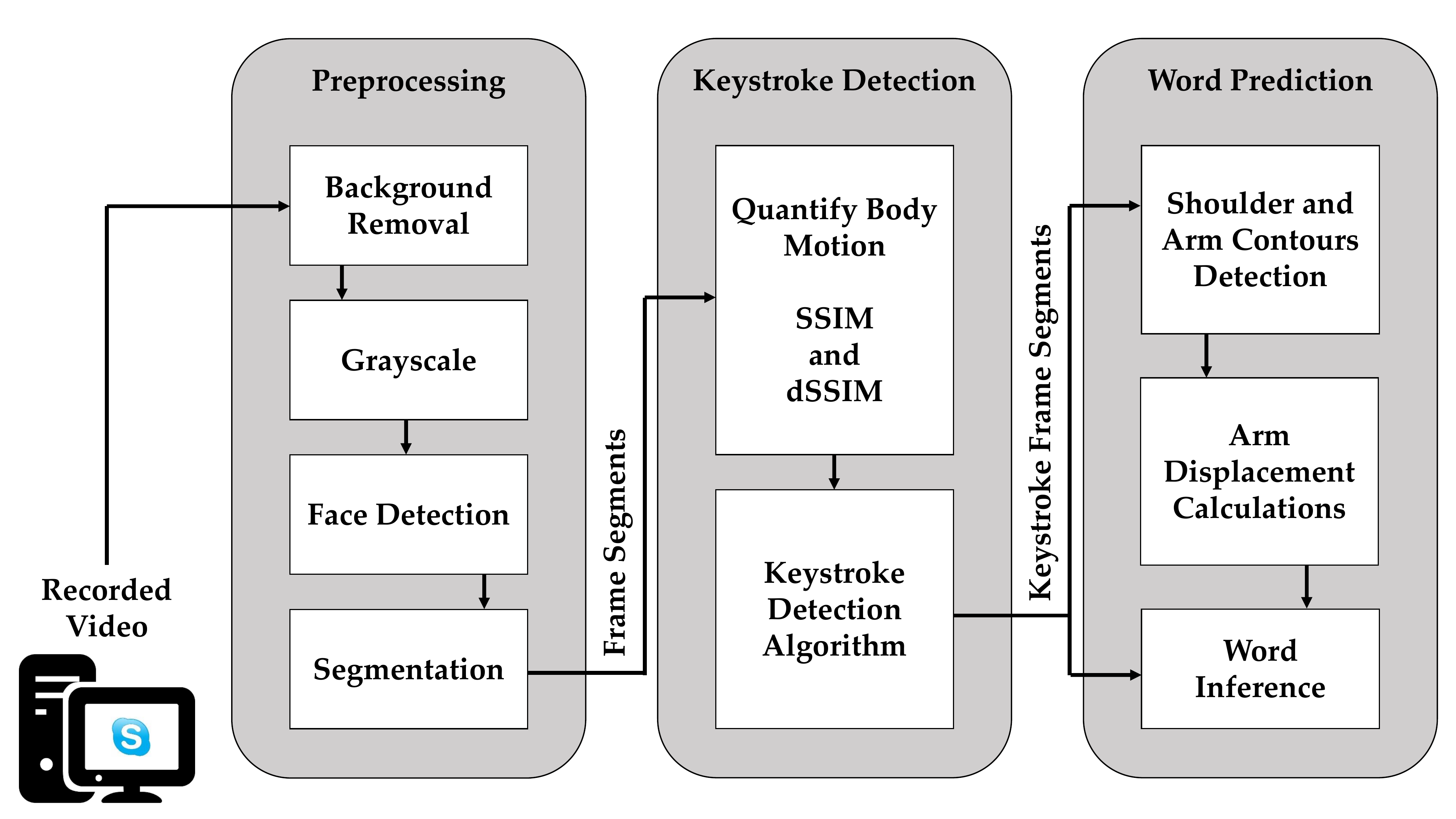}
\caption{Overview of the keystroke inference framework.}
\label{fig:framework-overview}
\vspace{-0.2in}
\end{figure}

To draw a relationship between typing related body movements observable in the video (\cref{fig:upper-body} in \cref{appendix:background-arms-typing}) and the text being typed, the adversary has to formulate two key procedures. 
First, within the video stream the adversary must be able to accurately identify the occurrences of keystrokes based on the upper body movements. Second, by modeling the body movement characteristics immediately before, after, and in-between detected keystrokes, the adversary must be able to accurately predict words and sentences typed by the target user. Let's first intuitively describe how the adversary can accomplish these objectives by giving an overview of the different components of our video-based keystroke inference framework (\cref{fig:framework-overview}). We will later provide details of each of these components.

The recorded video first undergoes multiple pre-processing steps in the following order: (i) \emph{background removal}, (ii) conversion to \emph{grayscale}, (iii) \emph{face detection}, and (iv) \emph{segmentation} of left and right arm regions based on their relative position with respect to the face. After pre-processing, the framework employs a keystroke detection algorithm based on Structural SIMilarity (\emph{SSIM}) index \cite{wang2004image} across all the frames in each of the left and right side video segments. Finally, the framework computes several motion features from the video segments immediately before and after each detected keystroke, and employs them in a dictionary-based prediction algorithm for word inference. Let's now provide details of each component.

\vspace{-0.1in}
\subsection{Pre-processing}
\label{reprocessing}

Given a video $v$ composed of $l_v$ frames recorded at $30Hz$, let us denote the set of frames in the video as $ v = \{f_1, f_2, f_3, \ldots, f_{l_v}\} $.
Assuming that the video resolution is constant, each frame in $v$ is composed of $m$ rows and $n$ columns of pixel values such that each pixel $p_{i,j}$ (where $i\in\{1, 2, 3, \ldots, m\}$ and \linebreak $j\in\{1, 2, 3, \ldots, n\}$) in a frame represents a RGB value. %
The RGB value of a pixel represents the hue (color), saturation, and brightness of that particular pixel in the frame. With this representation of a recorded video, we now describe the four pre-processing steps, in sequential order.

\noindent
\textbf{Background Removal.}
The background removal process is applied to all frames ($f_i \in v$), in order to identify the location of the body in the frame. We utilize the \texttt{DeepLabv3} model \cite{chen2017rethinking} for this task, which employs \emph{Atrous Convolution} with upsampled filters to extract dense feature maps and to capture long range context. Training of the model is done using he \texttt{Microsoft COCO} dataset \cite{lin2014microsoft}, which contains a rich set of human body related training samples. With the background removed, we can focus purely on the body's relative movements vis-\`{a}-vis typing. Example outputs of this background removal process is shown in \cref{fig:background-removal} (\cref{appendix:preprocessing-details}). This background removal step makes our proposed framework agnostic to any moving elements in the background. Let's denote background-removed frames as $^{r}f_i$.

\noindent
\textbf{Grayscale.}
We next convert all background-removed frames ($\{^{r}f_1, ^{r}f_2, ^{r}f_3, \ldots, ^{r}f_{l_v}\}$) to colorimetric grayscale \cite{stokes1996standard}, using the RGB values %
of individual pixels in a frame. This conversion to grayscale simplifies all following steps by making them color-independent. Let's denote such background-removed grayscale frames as $^{rg}f_i$.

\noindent
\textbf{Face Detection.}
Our next objective is to focus specifically on the two arms, where typing related motion is most perceptible. However, as webcam setups may not be predetermined or homologous, we require a webcam and webcam setup agnostic methodology for automatically and accurately identifying arm regions across all background-removed grayscale frames ($\{^{rg}f_1, ^{rg}f_2, ^{rg}f_3, \ldots, ^{rg}f_{l_v}\}$). To do so, we leverage on the consistency in relative position of the target user's arms with respect to their face (\cref{fig:face+segmentation} in \cref{appendix:preprocessing-details}). The intuition is that the left arm will be located around the bottom-right of the face in the frame, and similarly the right arm will be located around the bottom-left of the face in the frame. Face detection is a matured research topic, with several state-of-the-art frameworks and training datasets readily available. We utilize the CPU-friendly \texttt{Faceboxes} model \cite{zhang2017faceboxes}, that employs Rapidly Digested Convolutional Layers (RDCL) and the Multiple Scale Convolutional Layers (MSCL), to detect target user's face in each frame. For training the \texttt{Faceboxes} face detection model, we used the \texttt{WIDER FACE} dataset \cite{widerface}, that consists of 12,880 diverse facial images.

\noindent
\textbf{Segmentation.}
The \emph{facebox} generated by \texttt{Faceboxes} identifies the user's face and draws a rectangular boundary around it (solid green rectangle in \cref{fig:face+segmentation}). The objective of this last part of the pre-processing is to utilize this \emph{facebox} in order to automatically segment the left and right arms in the background-removed grayscale frame. Assume that in a given frame $^{rg}f_i$, the four vertices of the generated \emph{facebox} are located at pixels $p_{j,k}$, $p_{j,(k+a)}$, $p_{(j+b),k}$, and $p_{(j+b),(k+a)}$, where $a$ and $b$ are the width and height of the \emph{facebox} (in pixels), respectively. Using these \emph{facebox} vertices, the left arm segment is calculated as the rectangular area of the frame enclosed within the pixels $p_{(j+b),(k+a)}$, $p_{(j+b),n}$, $p_{m,(k+a)}$, and $p_{m,n}$. Similarly, the right arm segment is calculated as the rectangular area of the frame enclosed within the pixels $p_{(j+b),1}$, $p_{(j+b),k}$, $p_{m,1}$, and $p_{m,k}$. Let's denote the left and right arm segments extracted from a frame $^{rg}f_i$ as $L_i^{s}$ and $R_i^{s}$, respectively.

\vspace{-0.1in}
\subsection{(Potential) Keystroke \& Typing Activity Detection}
\label{subsec:detection}

Using the preprocessed left and right arm segments from all frames of the video ($L_i^{s}$ and $R_i^{s}$, respectively, where $i\in\{1, 2, 3, \ldots, l_v\}$), our next objective is to precisely determine the time-stamps, i.e., the frames, when a keystroke was typed using either hand. This is non-trivial because the adversary does not have a view of the lower arm. In this section, we propose a novel keystroke detection algorithm which accurately detects keystroke events using only upper hand arm movements as observed in $L_i^{s}$ and $R_i^{s}$.
The proposed keystroke detection algorithm is applied independently for each arm, i.e., the left and right arm segments $L_i^{s}$ and $R_i^{s}$ ($i\in\{1, 2, 3, \ldots, l_v\}$) are processed independently for the keystroke detection task.

\noindent
\textbf{Intuition.}
Every time the target user presses a key on her/his keyboard, she/he undertakes some degree of hand movement, the extent of which may vary depending on the typing style (\cref{appendix:typing-styles}) and position of the key on the keyboard. This movement may be from a resting position, or from an earlier keystroke using the same hand. Moreover, every keystroke lasts for a few milliseconds, until the user depresses the key, and during this time there is little to no movement. Finally, after the keystroke is completed, the user's hand moves on to another key or back to a resting position.
Intuitively, we should be able to observe this pattern of body movements in the video ($v$). Accordingly, our keystroke detection algorithm (\cref{alg:keystroke-detection} in \cref{appendix:keystroke-detection}) is designed based on empirically observed characteristics of the (left or right) arm segments immediately before, during, and immediately after a keystroke. The empirically observed characteristics that we leverage upon, as described below, are fairly independent of the typing style. For simplicity, going forward we will describe the keystroke detection process only for the left-hand. The process for the right hand is identical. 

\noindent
\textbf{Quantifying Body Motion.}
$SSIM$ \cite{wang2004image} is a well-known metric for measuring the similarity between two images, and a high $SSIM$ index between consecutive frames would denote insignificant body movements, and vice versa. To quantify left-hand body movements across consecutive frame segments, we compute $SSIM$ index between every $L_{i}^{s}$ and $L_{i+1}^{s}$ ($i\in\{1, 2, 3, \ldots, l_v-1\}$). This results in a series of $SSIM$ indices $SSIM^{L^{s}} = \{ L_{1}^{s} \bowtie L_{2}^{s}, L_{2}^{s} \bowtie L_{3}^{s}, \ldots, L_{l_{v} - 1}^{s} \bowtie L_{l_{v}}^{s}\}$, where $\bowtie$ is the $SSIM$ operator. To understand the rate of change in body movements across consecutive frame segments, we also compute the discrete derivative of $SSIM^{L^{s}}$ as $dSSIM^{L^{s}} = \{(L_{1}^{s} \bowtie L_{2}^{s}) - (L_{2}^{s} \bowtie L_{3}^{s}), (L_{2}^{s} \bowtie L_{3}^{s}) - (L_{3}^{s} \bowtie L_{4}^{s}), \ldots, (L_{l_{v} - 2}^{s} \bowtie L_{l_{v}-1}^{s}) - (L_{l_{v} - 1}^{s} \bowtie L_{l_{v}}^{s})\}$. 
In terms of body motion detected between the frame segments, $SSIM^{L^{s}}$ may be viewed as the `speed' and $dSSIM^{L^{s}}$ as the `acceleration'. %
Similarly, we also independently compute $SSIM^{R^{s}}$ and $dSSIM^{R^{s}}$ for the right hand.

\noindent
\textbf{Observed Characteristics.}
We observed a consistent pattern in the $dSSIM^{L^{s}}$ measurements, which is in line with our intuition outlined earlier. We observed that:
\begin{enumerate}[\hspace{0pt}(1)]
\item
If the video is recorded at 30 $fps$, and a keystroke occurred in frame $t$, there exists a local maxima in $dSSIM^{L^{s}}$ at $(L_{t-2}^{s} \bowtie L_{t-1}^{s}) - (L_{t-1}^{s} \bowtie L_{t}^{s})$. 
This is a result of the increase in body motion immediately before a keystroke followed by the lack of body motion for the duration of the key press.
\item
The above local maxima is followed by a local minima within the next 0.05 $sec$. For a video captured at 30 $fps$, this means that the local minima occurs among the next two elements of $dSSIM^{L^{s}}$, i.e, $(L_{t-1}^{s} \bowtie L_{t}^{s}) - (L_{t}^{s} \bowtie L_{t+1}^{s})$ or $(L_{t}^{s} \bowtie L_{t+1}^{s}) - (L_{t+1}^{s} \bowtie L_{t+2}^{s})$. This is a result of the lack of body motion for the duration of key press followed by the body motion immediately after a keystroke when the user's hand moves on to another key or back to a resting position. If no local minima is detected within this time frame, it would imply that the user's body movements are likely not related to keystrokes.
\end{enumerate}

\begin{figure}[t]
\centering
\includegraphics[width=\linewidth]{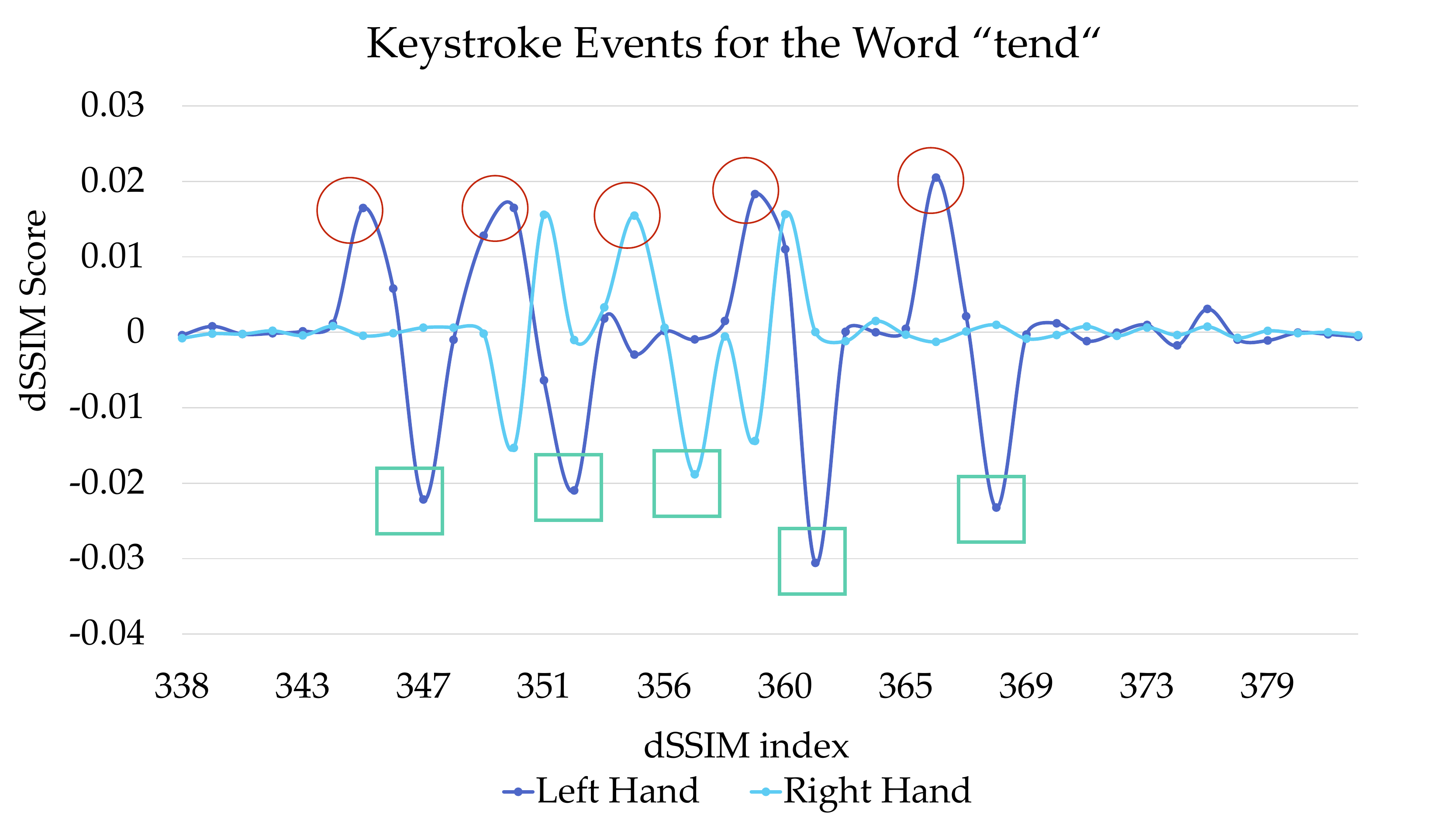}
\caption{An example of $dSSIM^{L^{s}}$ during a typed word. Circles represent the actual keystroke event, whereas squares represent the local minima within two frames of every keystroke.}
\label{fig:keystroke-detection}
\vspace{-0.2in}
\end{figure}

An example of this pattern can be observed in \cref{fig:keystroke-detection}.

\noindent
\textbf{Keystroke Detection Algorithm.}
We utilized the above observed characteristics to design a keystroke detection algorithm (\cref{alg:keystroke-detection}) that automatically labels frames where keystrokes potentially happened. In addition to the above observed characteristics, \cref{alg:keystroke-detection} also employs a \emph{filtering technique} to eliminate body movements that are not related to typing activity, but may still trigger a false positive. \cref{alg:keystroke-detection} filters based on statistical analysis of magnitudes in $dSSIM^{L^{s}}$. According to this filtering technique, a frame $t$ is considered to be a keystroke frame if:
\begin{enumerate}[\hspace{0pt}(1)]
\item
$dSSIM^{L^{s}}$ at $(L_{t-2}^{s} \bowtie L_{t-1}^{s}) - (L_{t-1}^{s} \bowtie L_{t}^{s})$ lies between $\phi_a\sigma_{dSSIM^{L^{s}}}$ and $\phi_b\sigma_{dSSIM^{L^{s}}}$, in addition to being the local maxima. $\sigma_{dSSIM^{L^{s}}}$ is the standard deviation in the distribution of magnitudes in $dSSIM^{L^{s}}$, and optimal values of $\phi_a$ and $\phi_b$ can be determined empirically.
\item
The local minima following $t$ in $dSSIM^{L^{s}}$ ($(L_{t-1}^{s} \bowtie L_{t}^{s}) - (L_{t}^{s} \bowtie L_{t+1}^{s})$, or $(L_{t}^{s} \bowtie L_{t+1}^{s}) - (L_{t+1}^{s} \bowtie L_{t+2}^{s})$) is less than $\phi_c\sigma_{dSSIM^{L^{s}}}$. Again, $\sigma_{dSSIM^{L^{s}}}$ is the standard deviation in the distribution of magnitudes in $dSSIM^{L^{s}}$, and optimal value of $\phi_c$ can be determined empirically.
\end{enumerate}

The effect of different $\phi_a$, $\phi_b$, and $\phi_c$ values on keystroke detection is presented in \cref{sec:eva-detection}. Appropriately chosen values of $\phi_a$, $\phi_b$, and $\phi_c$ would ideally eliminate false positives related to frequently occurring minor body movements that are closer to the mean value ($\mu_{dSSIM^{L^{s}}}$), and can be otherwise regarded as noise. Similarly, appropriately chosen values of $\phi_a$, $\phi_b$, and $\phi_c$ will also eliminate false positives related to infrequently occurring major body movement that are far away from the mean value ($\mu_{dSSIM^{L^{s}}}$), and can be otherwise regarded as outliers.

{

\noindent
\textbf{Typing Activity Detection.}
As the target user may type at specific instance(s) in time during the video call, it is necessary for the adversary to detect the time periods (or windows) where typing activity occurred. Typing activity detection is especially needed to effectively eliminate false positives during keystroke detection, which could otherwise result in 
incorrect word prediction results. We next outline a heuristic-based typing activity detection technique which employs our individual keystroke detection algorithm (\cref{alg:keystroke-detection}).

As outlined earlier, \cref{alg:keystroke-detection} returns a set of frames where potential keystrokes could have occurred, but these detected potential keystrokes could also include non-typing activities (false positives). 
We leverage on a few intuitive heuristics in order to distinguish between the (detected) keystrokes that correspond to a typing activity from those that may correspond to non-typing activities similar to typing. The first heuristic, referred as \emph{maximum speed filter}, filters out false positives from the detected keystrokes by observing the maximum rate at which these (potential) keystrokes are detected by \cref{alg:keystroke-detection}. Studies have shown that most users typically type at a rate of about 4 keystrokes per second, and that it is highly unlikely to come across a typing rate of 10 or more keystrokes per second \cite{livechat}. Thus, the maximum speed filter will filter out (as false positives) from the detected keystroke frames those that correspond to a rate of 10 or more keystrokes per second, per hand. 

The second heuristic, referred as \emph{location filter}, filters out false positives by determining if both hands are on or near the keyboard. Here, the basic idea is to first create a set of reference frames ($K$) where the target user is most likely typing (i.e., hands on/near the keyboard), and then use these reference frames to determine (using optical flow \cite{stavens2007opencv}) if their hand(s) are at a significantly different position in the other remaining frames corresponding to potential keystrokes. Specifically, we create the reference set $K$ by including all potential keystroke frames in each 2-second window, if and only if the window contains at least four detected potential keystrokes (both hands combined) with each hand contributing two or more potential keystrokes. This condition represents a very likely case of typing activity, and thus the user's hands being on or near the keyboard. We then use this reference set to filter out as false positive any potential keystroke frame that is not within an empirically evaluated optical flow distance threshold to any frame (of the corresponding hand) in the reference set $K$.

Similar to the maximum speed filter, the next heuristic we employ is the \emph{minimum speed filter} which filters out
detected potential keystrokes as false positives if they occur at a rate of one (or lower)
keystroke per second, combined for both hands (i.e., representing highly unlikely typing rate).
The fourth and final heuristic, referred as \emph{exclusive hand filter}, attempts to detect one-handed non-typing activities (e.g., mouse clicks) that often get classified by \cref{alg:keystroke-detection} as potential keystrokes.
Specifically, in each 10-second window, the exclusive hand filter filters out 10 or more consecutive potential keystrokes with the same hand as false positives. \cref{fig:typing-flow} outlines the order in which these heuristics are applied for filtering out false positives. 

All potential keystrokes that are not filtered based on the above four heuristics represent typing activity and are used for word predictions. \cref{fig:typing-example,fig:typing-example-false} in \cref{appendix:typing-act-ex} further elucidates the working of our heuristic-based approach by means of two real scenarios that we encountered during our experimentation. We present a comprehensive evaluation of its performance later in \cref{sec:home_eva}.

\begin{figure}[t]
\centering
\includegraphics[width=\linewidth]{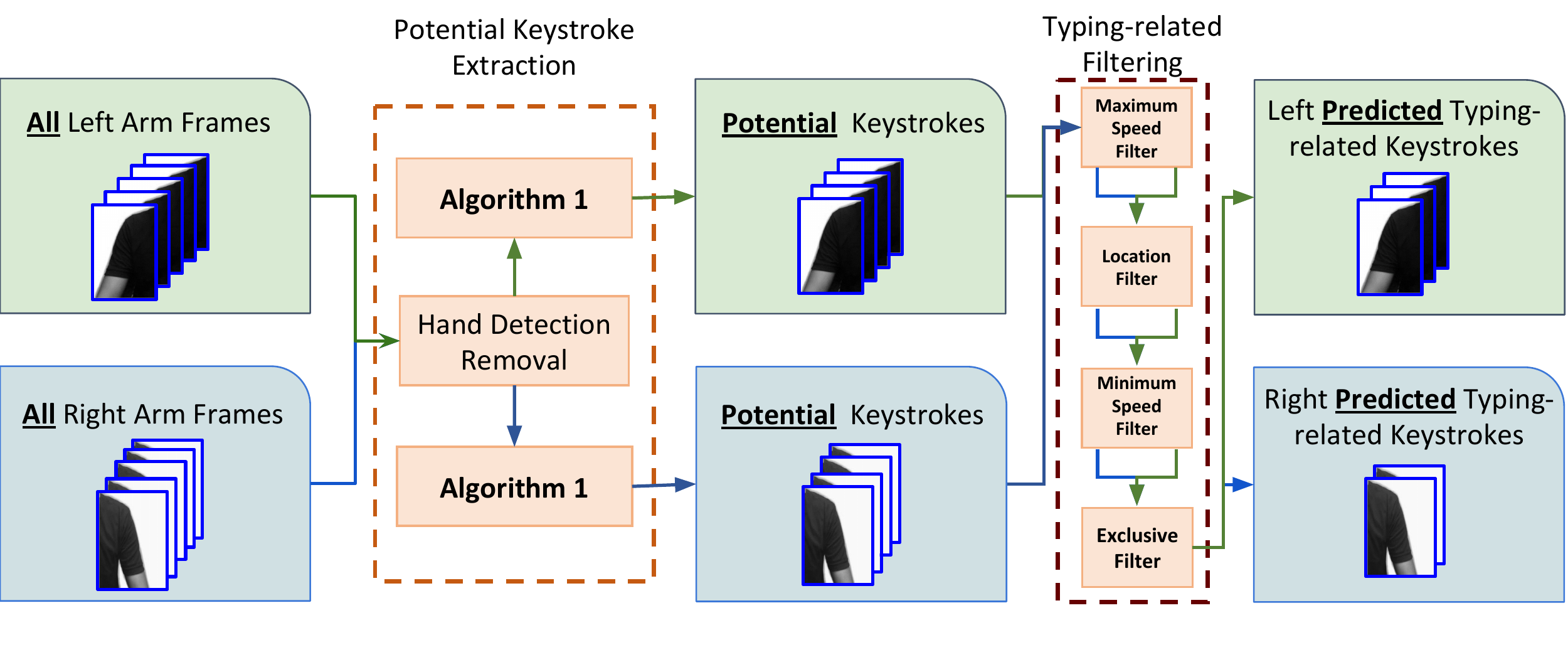}
\caption{Overview of the typing activity detection technique.}%
\label{fig:typing-flow}
\vspace{-0.2in}
\end{figure}

}

\vspace{-0.1in}
\subsection{Word Prediction}
\label{word-prediction}

We now describe how the adversary can infer words that were typed from the detected keystrokes, using two different groups of information. The first group of information is simply the number of keystrokes detected for a word, and the hand (left/right) which was used to conduct individual keystrokes of the word. Let us call this information as \emph{keystroke information}.
The second group of information is the magnitude and direction of body displacement, more specifically the arm displacement, between consecutive keystrokes of the word. Assuming that the target user typed on a standard QWERTY keyboard, mapping the arm displacement between consecutive keystrokes to relative position of the keys can significantly improve the inference accuracy. Let us call this information as \emph{displacement information}.
After executing the typing activity detection (utilizing \cref{alg:keystroke-detection}) the adversary already has knowledge of the \emph{keystroke information} -- all the frame segments in which a keystroke was detected, separately for each hand. However, the \emph{displacement information} is not readily available, and will require us to employ advanced computer vision techniques to effectively measure arm displacement between consecutive keystrokes.

Let $keystrokesFS_{}^{L}$ denote the list of all detected left-hand keystroke frame segments and $keystrokesFS_{}^{R}$ denote the list of the right-hand keystroke frame segments. $keystrokesFS_{}^{L}$ and $keystrokesFS_{}^{R}$ essentially constitute the \emph{keystroke information}. We now describe how to utilize these two lists ($keystrokesFS_{}^{L}$ and $keystrokesFS_{}^{R}$) to derive the \emph{displacement information}. In brief, we (i) identify the outer contour of individual arms in each keystroke frame segment, (ii) calculate the displacement of individual arms by tracking change in position of the outer edge of the arms across consecutive keystroke frame segments, and (iii) interpret calculated arm displacements with respect to the QWERTY keyboard layout. After obtaining the \emph{displacement information}, we describe how the adversary can utilize both the \emph{keystroke} and \emph{displacement information} to carry out word predictions using a dictionary or reference database.

\begin{figure}[t]
\centering
\begin{subfigure}{0.19\linewidth}
\centering
\includegraphics[width=\textwidth]{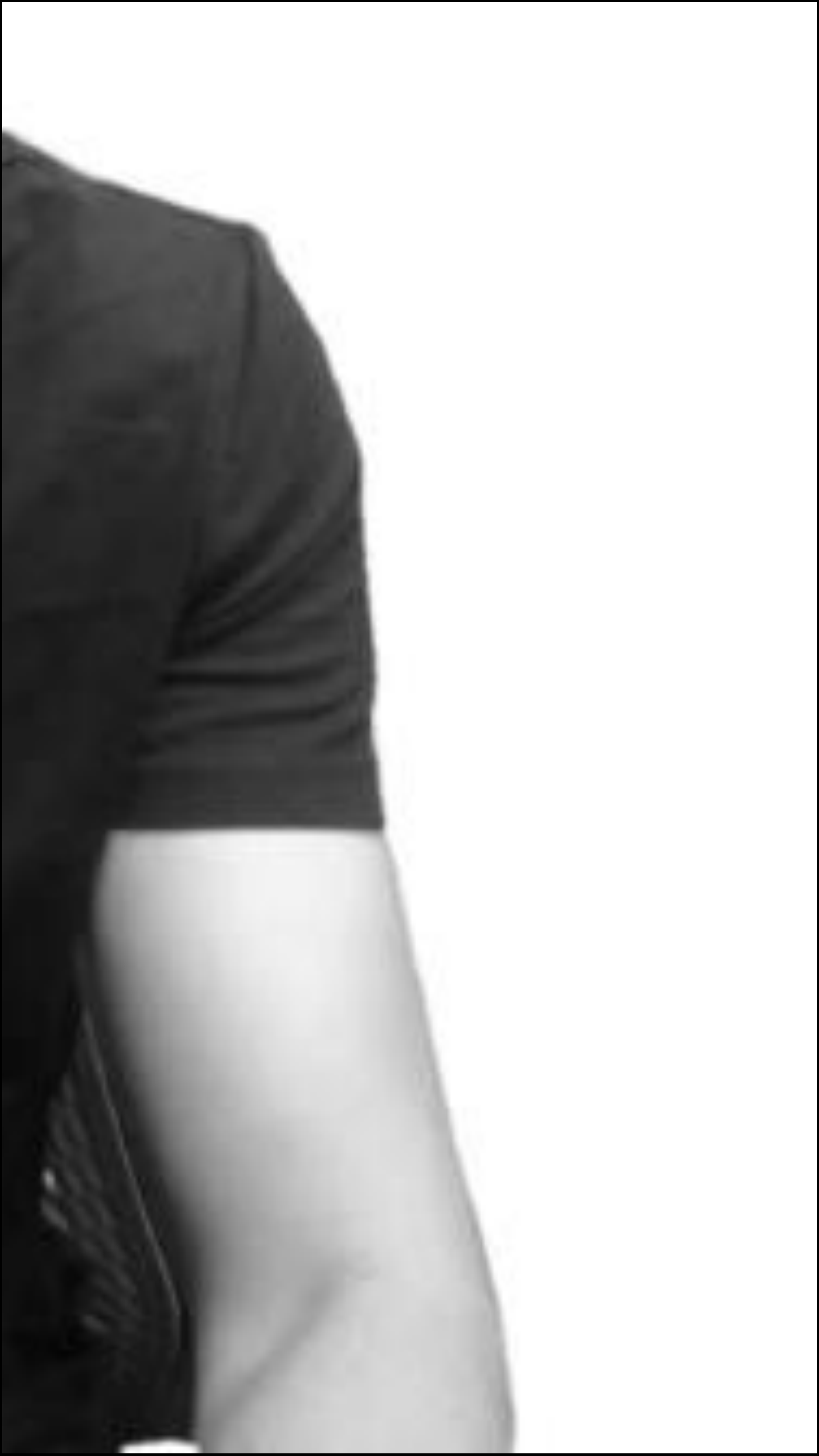}
\caption{}
\label{features-original}
\end{subfigure}\hfill
\begin{subfigure}{0.19\linewidth}
\centering
\includegraphics[width=\textwidth]{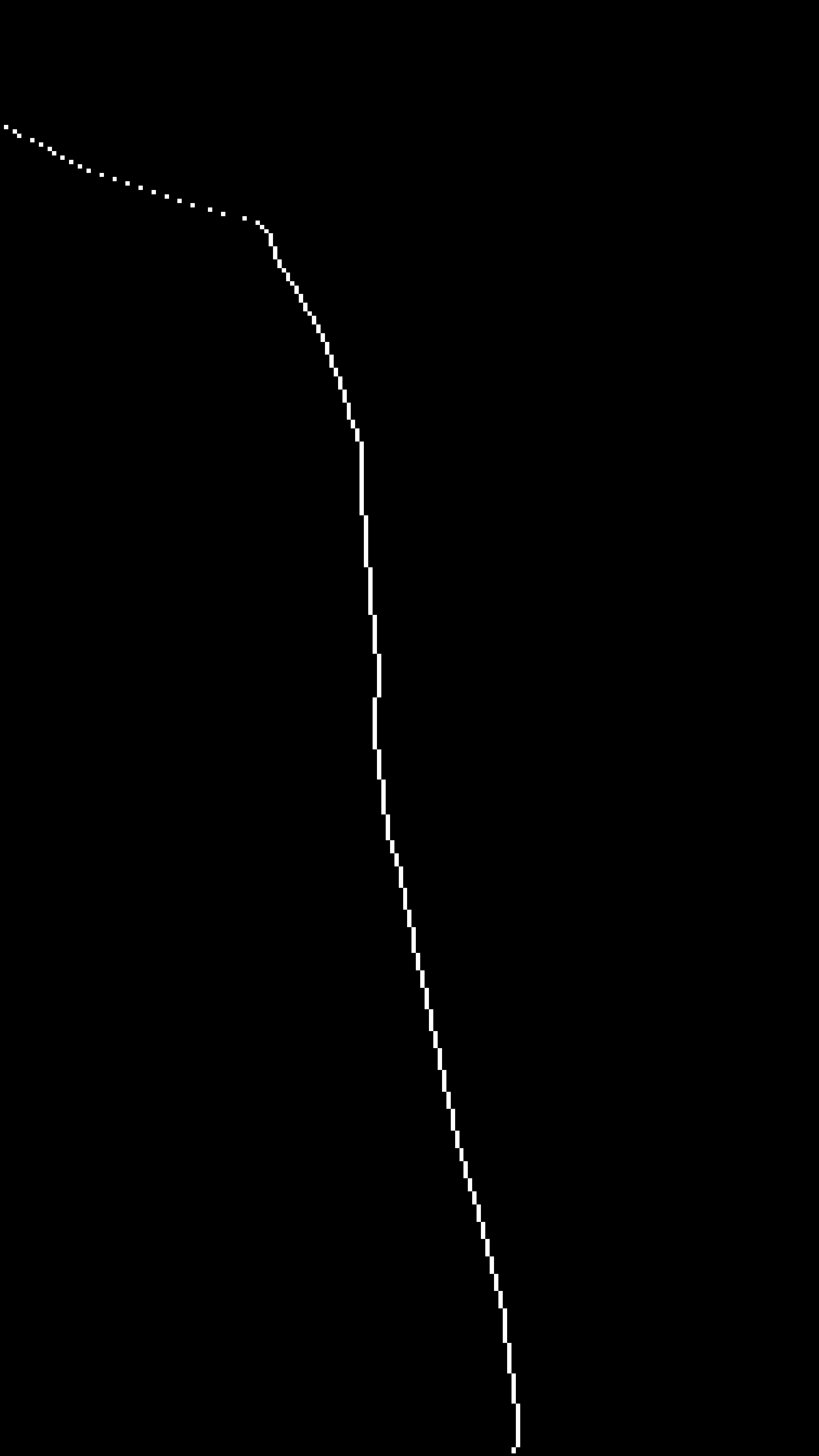}
\caption{}
\label{features-oc}
\end{subfigure}\hfill
\begin{subfigure}{0.19\linewidth}
\centering
\includegraphics[width=\textwidth]{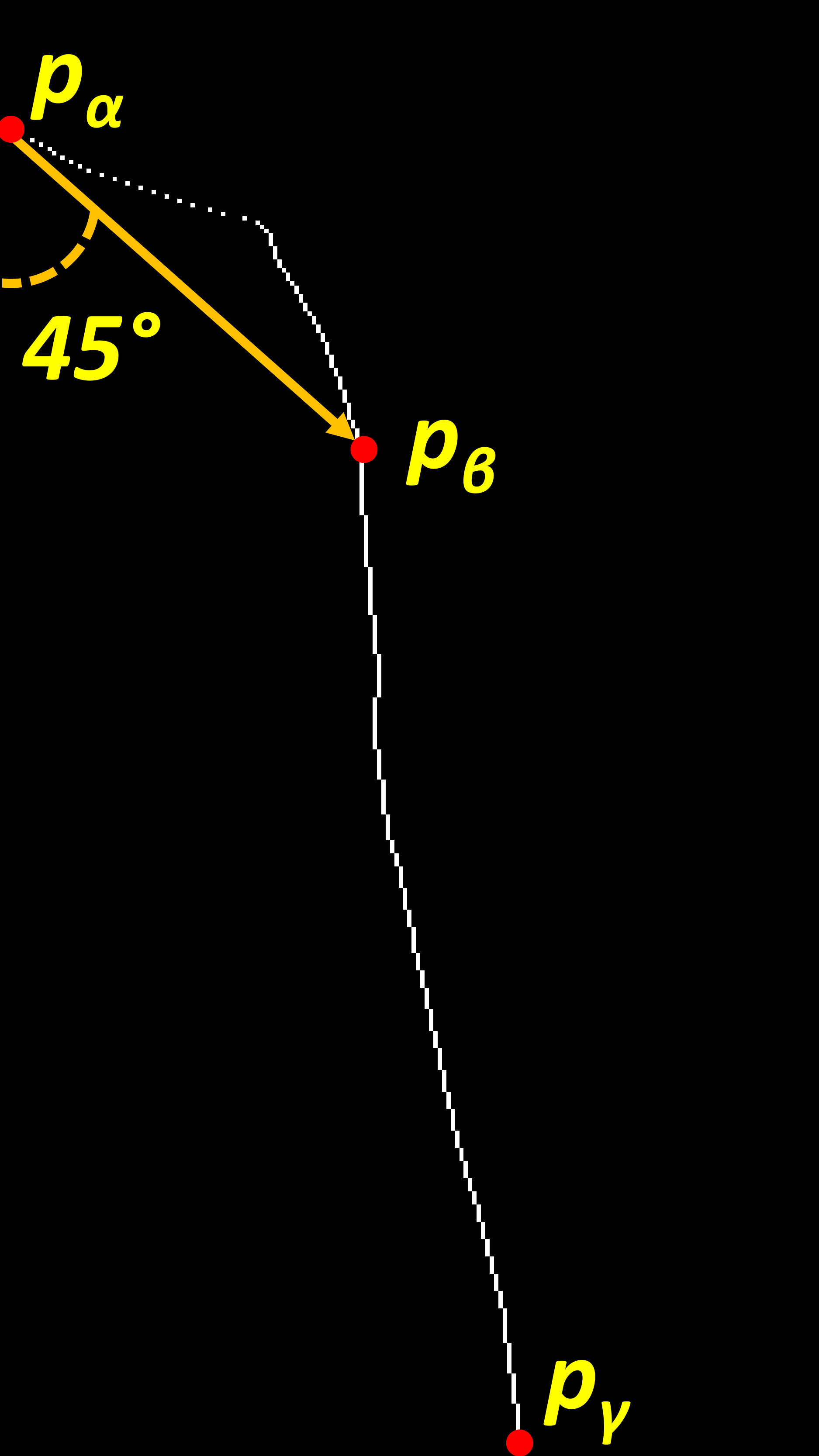}
\caption{}
\label{features-projection}
\end{subfigure}\hfill
\begin{subfigure}{0.19\linewidth}
\centering
\includegraphics[width=\textwidth]{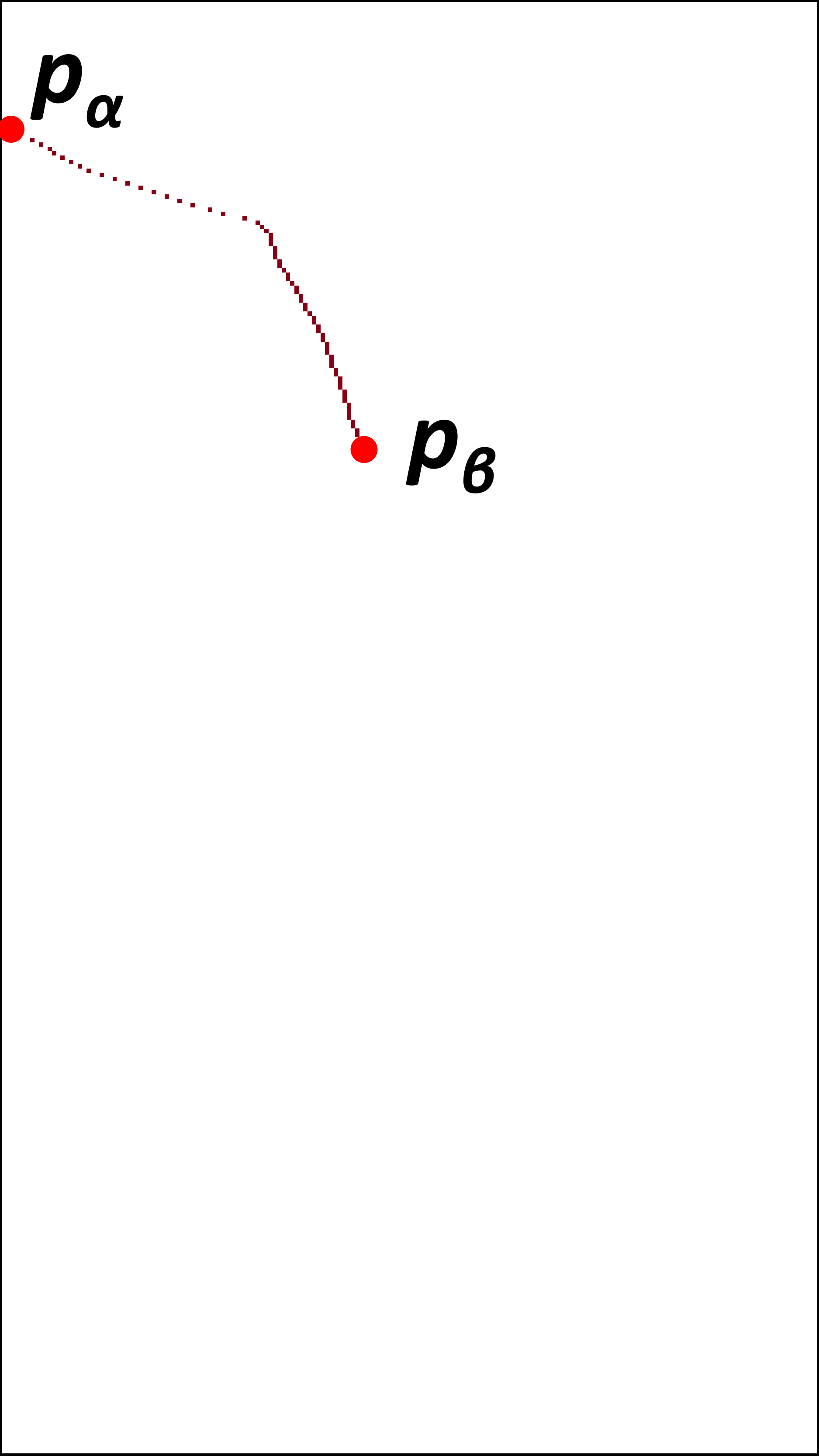}
\caption{}
\label{features-sc}
\end{subfigure}\hfill
\begin{subfigure}{0.19\linewidth}
\centering
\includegraphics[width=\textwidth]{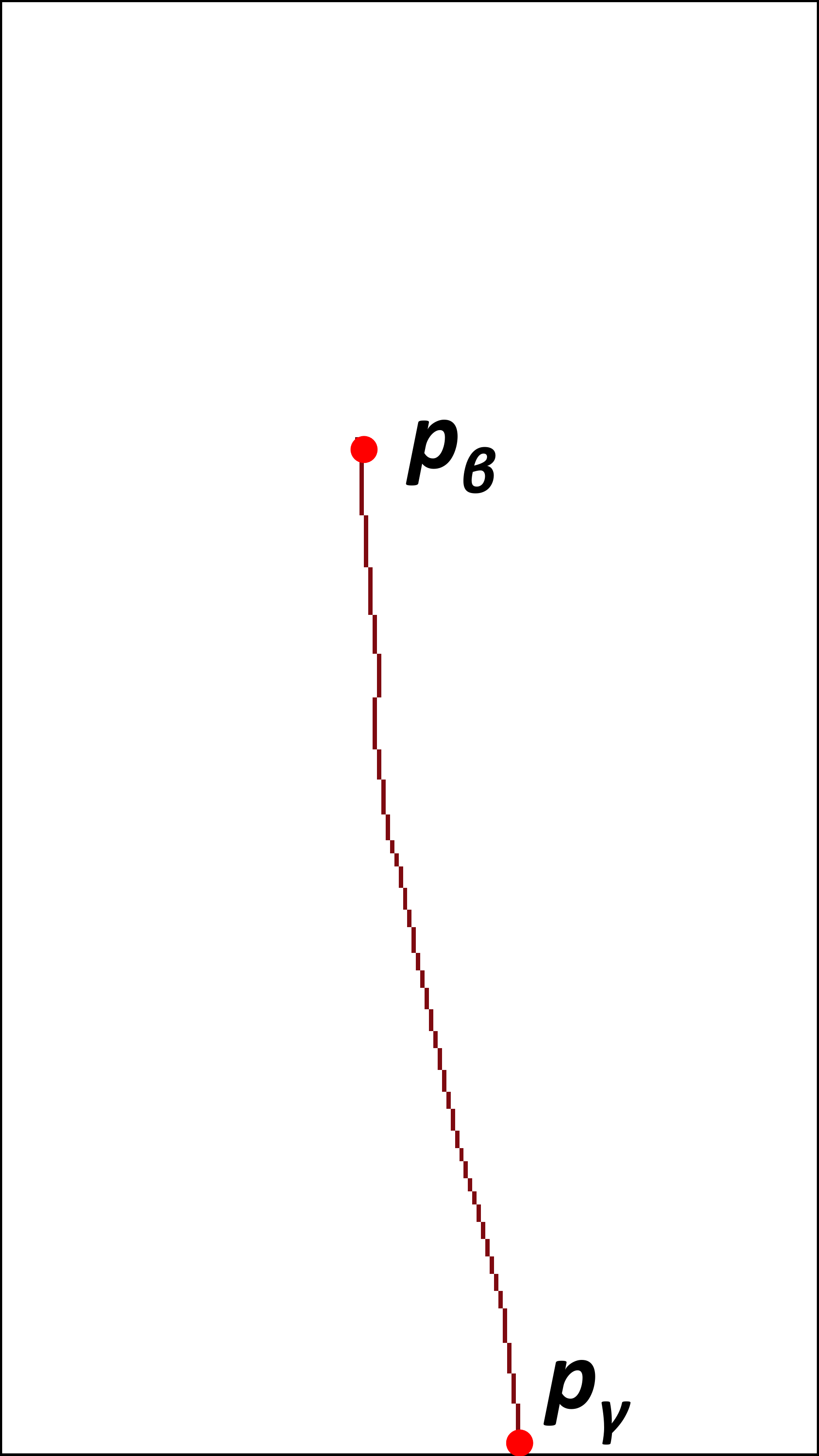}
\caption{}
\label{features-hc}
\end{subfigure}
\caption{\emph{(a)} A keystroke frame segment, \emph{(b)} Outer contour ($OC$), \emph{(c)} $45\degree$ projection from $p_\alpha$ that intersects $OC$ at $p_\beta$, \emph{(d)} Shoulder contour ($SC$), and \emph{(e)} Arm contour ($AC$).}
\label{fig:features}
\vspace{-0.2in}
\end{figure}

\noindent
\textbf{Outer Edge Detection of Arms.}
In order to efficiently measure arm displacement between consecutive keystrokes, we focus on specific regions of the keystroke frame segments. Instead of trying to analyze movement of all pixels between two keystroke frame segments, we focus on pixels covering the outer-edge movements of the arms (\cref{features-original}). The intuitive reasoning behind this design decision is that the characteristics of outer-edge movements are reflective of the movement of the entire upper arm and shoulder. Let us label the subset of pixels in a keystroke frame segment covering the outer contour/edge of the body as the \emph{outer contour}, or $OC$.
To compute the \emph{outer contour} in each of the keystroke frame segments, %
we first detect all \emph{edges} in a keystroke frame segment using Canny edge detection technique \cite{ding2001canny}.
As the background in the frame segment is already removed during the keystroke detection step (\cref{subsec:detection}), outer edges of the arm and shoulder are easily captured by the edge detection process. However, there is a possibility that edges within the arm and shoulder areas, such as creases or patterns on a shirt, could also get detected as an edge.
To overcome this issue, we device a straightforward approach to remove all additional edges (i.e., all edges except the outer edge of the arm and shoulder), as described below.
In case of the left hand, for each row of pixels we keep the rightmost pixel in the edge-detected frame segment that is part of an edge. The intuition is that in the absence of a background, the rightmost pixel in each row has to be part of the \emph{outer contour}. Similarly, in case of the right hand, for each row of pixels we keep the leftmost pixel in the edge-detected frame segment that is part of an edge. An example of \emph{outer contour} can be seen in (\cref{features-oc}).

After the \emph{outer contour} is computed for every keystroke frame segment in $keystrokesFS_{}^{L}$ and $keystrokesFS_{}^{R}$, we next segment the outer contour into \emph{shoulder contour} ($SC$) and \emph{arm contour} ($AC$) based on human physiology (\cref{fig:arms}). This physiology-based division is approximated by drawing a projection from the pixel nearest to the neck ($p_\alpha$) such that the angle between this projection and the vertical boundary of the frame segment is $45\degree$ (\cref{features-projection}). Let $p_\beta$ be the pixel where this projection intersects the \emph{outer contour}, and $p_\gamma$ be the pixel farthest from the neck in the \emph{outer contour}. Pixels in the \emph{outer contour} between $p_\alpha$ and $p_\beta$ becomes the \emph{shoulder contour}, and pixels in the \emph{outer contour} between $p_\beta$ and $p_\gamma$ becomes the \emph{arm contour}. Obviously, this is just an approximate computation of \emph{shoulder} and \emph{arm contours} as the underlying physiological differences between person to person cannot be accurately modeled using available webcam video data. \cref{features-sc} and \cref{features-hc} shows \emph{shoulder} and \emph{arm contours}, respectively, computed for the \emph{outer contour} example mentioned earlier. While the \emph{arm contours} are directly useful in displacement calculations, explained next, \emph{shoulder contours} are also utilized for calibrating the displacement calculations.

\noindent
\textbf{Displacement Calculations.}
We employ sparse optical flow technique \cite{stavens2007opencv} to quantify hand displacements between consecutive keystrokes. Sparse optical flow is a computer vision technique that takes a set of pixels (for example, constituting an object) within an image as input, and outputs a vector set representing the displacement of those pixels (and thus the object) in another image. Sparse optical flow is especially useful to track object movements across chronological frames of a video.
In our framework, we apply sparse optical flow to track the displacement of \emph{shoulder} and \emph{arm contours} across all consecutive keystroke frame segments, individually for each hand. For simplicity, we use the left hand to explain the use of sparse optical flow on two consecutive keystroke frame segments ($\in keystrokesFS_{}^{L}$), say $L_{i}^s$ and $L_{i+1}^s$, with respective \emph{arm contours} $AC_{i}$ and $AC_{i+1}$. By applying sparse optical flow between $AC_{i}$ and $AC_{i+1}$, we obtain a set of displacement vectors representing the direction and magnitude of how each pixel in $AC_{i}$ has shifted in $AC_{i+1}$. We then use this set of displacement vectors to calculate a mean displacement vector for the \emph{arm contour}, and let us call it $\overrightarrow{oa_i}$. Calculated in a similar fashion, let us call the mean displacement vector for the \emph{shoulder contour} as $\overrightarrow{os_i}$. In summary, a $\overrightarrow{os}$ represents the average movement of the shoulder between consecutive keystrokes, whereas a $\overrightarrow{oa}$ represents the average movement of the hand between consecutive keystrokes.
 
Our objective behind computing $\overrightarrow{os}$ and $\overrightarrow{oa}$ is to use these displacement vectors to determine the relative position of the keys corresponding to keystrokes. We carry out two additional operations using the displacement vectors $\overrightarrow{oa}$ and $\overrightarrow{os}$ in order to make our inference framework more generalizable. First, we observed that in certain scenarios the camera itself may move slightly, in addition to the arm. This can be prominently observed in the case of a laptop webcam, where a press on the laptop keyboard can result in a noticeable motion of the webcam which is generally located on top of the display. We solve this by applying sparse optical flow on the background during the pre-processing (\cref{reprocessing}), and negating the mean displacement vector of the background ($\overrightarrow{ob}$) from $\overrightarrow{oa}$ and $\overrightarrow{os}$. Second, we observed that in certain instances the typer changes her/his posture in between consecutive keystrokes, for example due to fatigue. To address this, we utilize the shoulder displacement ($\overrightarrow{os}$) as an approximation of posture changes, and subtract it from $\overrightarrow{oa}$. Combining both of these operations, we obtain $\overrightarrow{om} = \overrightarrow{oa} - \overrightarrow{os} - \overrightarrow{ob}$, where $\overrightarrow{om}$ represents the approximate average arm displacement, free of influence from posture or camera movements, that happened between consecutive keystrokes.

\noindent
\textbf{Interpreting Calculated Arm Displacements.}
From our keystroke detection (\cref{subsec:detection}), we are already aware of which hand was used to type individual letters. While this information alone can be very useful in conducting dictionary-based predictions, we deploy the arm displacement vector ($\overrightarrow{om}$) computed now to further reduce the search space. Reduction in the search space will in turn make our predictions more accurate. 
Between any two consecutive keystrokes using the same hand, we classify the corresponding $\overrightarrow{om_{i}}$ into one of the four intercardinal directions: northeast ($NE$), northwest ($NW$), southeast ($SE$), southwest ($SW$). The classification of a left hand $\overrightarrow{om_{i}}$ is conducted as per conditions listed in \cref{tab:observed-direction-left} (\cref{appendix:keystroke-prediction}). In \cref{tab:observed-direction-left}, $\overrightarrow{om_{i}(x)}$ and $\overrightarrow{om_{i}(y)}$ are the $x$-axis and $y$-axis displacements (i.e., vector components), respectively, measured in pixels. The classification is isomorphic in case of right arm displacements between keystrokes, as listed in \cref{tab:observed-direction-right} (\cref{appendix:keystroke-prediction}).

\noindent
\textbf{Template Inter-keystroke Directions.}
Now, we define template inter-keystroke directions on the standard QWERTY keyboard, which are the \emph{ideal} directions a typer's hand should follow. To define the template inter-keystroke directions, we first divide the QWERTY keyboard into two halves (left and right). The left side of the keyboard contains the letters $\{q, w, e, r, t, a, s, d, f, g, z, x, c, v, b\}$ while the right side of the keyboard contains the letters $\{y, u, i, o, p, h, j, k, l, n, m\}$ as shown in \cref{fig:key-dir} (\cref{appendix:keystroke-prediction}). Similar to prior works that used an analogous modeling \cite{MarquardtVCT:2012,maiti2016smartwatch}, we assume that a typer will predominantly type keys on the left side of the keyboard using her/his left hand, and vice versa. However, every key on the keyboard occupies a rectangular area, and a typer can have some variance in the position within each key where it is pressed. %
Some keys may be pressed in the center, while others could be pressed around the edges. This naturally occurring variance lead us to model the template inter-keystroke directions more flexibly using nine possible scenarios between any two keys $key_{i}$ and $key_{j}$, as detailed in \cref{tab:template-directions} (\cref{appendix:keystroke-prediction}) and exemplified in \cref{fig:key-dir} (\cref{appendix:keystroke-prediction}).

\noindent
\textbf{Word Inference.}
Our word prediction is a dictionary-based search for words based on (i) matching the order and number of left and right handed keystrokes, and (ii) matching the calculated direction of arm displacements with the template inter-keystroke directions. To satisfy the first criterion, a $word_i$ in the dictionary is deemed as a candidate for the typed word if $keystrokesFS_{}^{L}$ and $keystrokesFS_{}^{R}$ contain a combined number of keystroke frame segments equal to the length of $word_i$. The keystroke frame segments in $keystrokesFS_{}^{L}$ and $keystrokesFS_{}^{R}$ should also be chronologically interleaved according to the alphabets in the left and right sides of the keyboard (\cref{fig:key-dir}). To satisfy the second criterion, a $word_i$ in the dictionary is deemed as a candidate if the calculated arm displacements $\overrightarrow{om_j}$ between every letter of the $word_i$ satisfies the template mappings outlined in \cref{tab:template-directions}. We also sort the dictionary based on how frequently its words are used in the English literature (in descending order), so as to improve inference accuracy when there exists multiple candidate words that satisfy the above two criteria. 
In addition to the top prediction (i.e., the candidate word with the most usage in English literature), we also evaluate if the typed word is contained in top-$k$ of such candidate words, as an adversary can run additional semantical and contextual analyses to improve inference of complete sentences. We, however, limit the scope of this work to only word inferences.
{
We next outline details of the different experimental setups and evaluation experiments that we conduct to evaluate our keystroke detection and word prediction framework. Our first set of evaluation experiments are conducted in a slightly constrained (or ``\emph{In-Lab}") setting to analyze the \emph{best-case} performance of our framework. Our second set of experiments are conducted in a fully unrestricted (or ``\emph{At-Home}") setting to analyze the \emph{worst-case} performance of our framework. All our participant recruitment and data collection experiments were approved by our university's Institutional Review Board (IRB). }

 \vspace{-0.1in}
\section{In-Lab Experimental Setup}
\label{sec:experimental_setup}

{
Our first set of evaluation experiments were conducted by fixing the video call setup, including, the device(s) used for the calls and participants' sitting position during the call, and text typed by the participants. For this set of experiments, which we refer as In-Lab setup, we recruited a diverse set of 20 human subject participants and collected video call data while they were performing typing tasks, details of which are outlined below.

\noindent
\textbf{Participant Demographics.}
Out of the 20 participants recruited for this setting, 9 are females and 11 are males. Based on a screening-survey, 4 participants conducted hunt-and-peck typing, 5 conducted touch typing, and the remaining 11 participants conducted hybrid typing. One of the participants identified as being left-handed while the remaining 19 participants identified themselves as right-handed. 
}

\noindent
\textbf{Participant Tasks.}
Each participant completed six different sessions across different experimental parameters, which are listed below. Each session was conducted on a different day. 
Before every session, the experimental parameters were chosen randomly to cover different combinations, and the participant was informed about those parameters beforehand. 
The data collection sessions were conducted in a controlled setting inside a private office, primarily to limit noise in the audio data collected for equitable comparison (with an audio-only framework by  Compagno et al. \cite{compagno2017don}). 
The participants were positioned in front of a computer with a display, keyboard, and a webcam directly facing them. %
Each participant was shown a random word on-screen in large font and the participant was instructed to naturally type the displayed word followed by a blank space. Upon entry of blank space, a new random word replaced the previous word on-screen, and the participant repeated this process for 300 words in each session. The random words were picked from a dictionary of 4000 most frequently used words (of 4 or more letters) in the English literature \cite{wordfrequencydict}. In order to minimize the impact of fatigue while typing, each session was divided in to three sub-sessions, each consisting of 100 words. Participants were encouraged to take a break between each of the sub-sessions. {
It should be noted that, despite the fixed nature of the In-Lab setup, participants were free to change their body posture and position based on their need and comfort-level, both during and in-between the typing sub-sessions.}

\noindent
\textbf{Data Collected.}
On the data collection computer, our custom application recorded the webcam video (at $1920\times1080$ $pixels$), microphone audio (at 44.1 $kHz$) and time-stamped ground-truth of the keys (characters) typed by the participant. The ground-truth information is used to measure the accuracy of our framework. To obtain realistic results, we later transmitted the recorded video over the Internet through different video calling software and captured it remotely on another computer. Skype \cite{skype} was used for majority of the evaluation, but we also compare it with Hangouts \cite{googlehangout} and Zoom \cite{zoom} in \cref{voip-compare}. The video transmission was achieved using \texttt{ManyCam} \cite{manycam}, a virtual webcam driver that can play pre-recorded videos during a video call. The remote capture of the transmitted video was done using \texttt{OBS Studio} \cite{obsproject}.

\noindent
\textbf{Experimental Parameters.}
We evaluate our attack framework across a diverse set of experimental parameters to showcase its generalizability and practical impact. %
Below is a list of the different parameters that were studied:

\begin{enumerate}[\hspace{0pt}(1)]
\small
\item
\textit{Clothing}: Long-sleeves, Short-sleeves, Sleeveless.
\item
\textit{Keyboard}: Logitech K120 (Wired), Anker A7721 (Bluetooth).
\item
\textit{Webcam}: Anivia W8 (1080p), Logitech C920 (1080p).
\item
\textit{Devices}: Lenovo 330-15IGM Laptop, Dell OptiPlex Desktop.
\end{enumerate}

The laptop was evaluated with its built-in keyboard and webcam, whereas on the desktop we collected data using combinations of two different (external) webcams and keyboards. We instructed each participant to wear clothings such that they covered all three types (long-sleeve, short-sleeve, and sleeveless) over the six sessions. %
Overall, the combination of these parameters resulted in fifteen different experimental settings (within the In-Lab setup), three on the laptop and twelve on the desktop. For evaluating keystroke predictions, we used two different English dictionaries. One is a dictionary of 4K words which was the same dictionary used for data collection, and the other is a more comprehensive dictionary of 65K English words. {It should be noted that we do \emph{not} evaluate the typing activity detection technique in the In-Lab experiments, as the participants were not performing any other tasks besides typing.} 

\noindent
\textbf{Different Backgrounds and Removal.}
As we employ \texttt{DeepLabv3} for background removal, which has been extensively evaluated in the literature, we do not evaluate it as an experimental parameter. Nonetheless, in \cref{fig:background-removal} we show that \texttt{DeepLabv3} was able to remove backgrounds in different indoor and outdoor settings. 
Moreover, in the case that \texttt{DeepLabv3} fails to properly identify and remove a particular background in the recorded video, the adversary can easily substitute it for another background removal technique.

\vspace{-0.1in}
\section{Video-only In-Lab Evaluation}
\label{sec:evaluation}

In this section, we evaluate our prediction framework solely using the video data stream collected during In-Lab experiments. We briefly present results on the performance of our keystroke detection algorithm (\cref{alg:keystroke-detection}), before detailing the various prediction results.

\vspace{-0.1in}
\subsection{Keystroke Detection Performance}
\label{sec:eva-detection}
\begin{figure}{}
\centering
\includegraphics[width=0.8\linewidth]{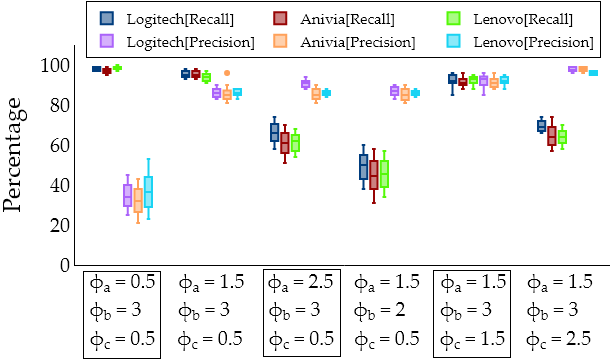}
\caption{Precision and recall of keystroke detection under different $\phi_a$, $\phi_b$, and $\phi_c$.}
\label{fig:eva-precision}
\vspace{-0.2in}
\end{figure}

We evaluate keystroke event detection using the \emph{precision} and \emph{recall} metrics, while also studying the effect of different coefficient values $\phi_a$, $\phi_b$, and $\phi_c$ used in our keystroke detection algorithm (\cref{alg:keystroke-detection}). 
As seen in \cref{fig:eva-precision},
recall increased as $\phi_a$ and $\phi_c$ were decreased, and when $\phi_b$ was increased. 
This is because when $\phi_a$ and $\phi_c$ are small and $\phi_b$ is large, our keystroke detection algorithm will even recognize minute noises as a keystroke event. Corresponding precision values are presented in \cref{fig:eva-precision}.
Balancing between precision and recall is always a trade-off, and based on these empirical results we achieved a good precision-recall balance for the coefficient values $\phi_a=1.5$, $\phi_b=3$, and $\phi_c=1.5$. Using these coefficient values, we obtained an average of 93\% precision with 92\% recall rate. Accordingly, we use keystrokes detected using these coefficient values for the rest of the evaluation, including the false positives and ignoring the false negatives.

\vspace{-0.1in}
\subsection{Keystroke Prediction Performance}

We now present results on word prediction performance for different experimental settings. %

\begin{figure}{}
\centering
\includegraphics[width=0.8\linewidth]{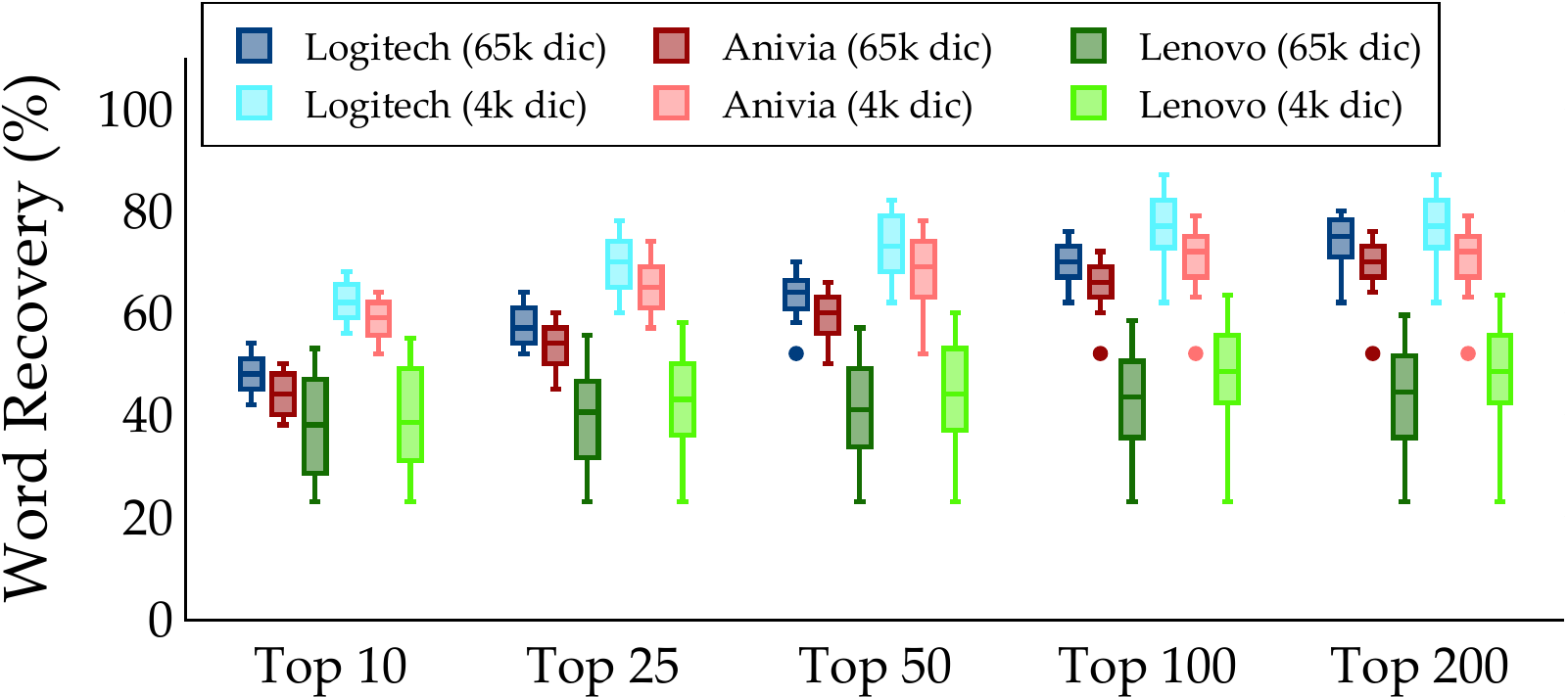}
\caption{Successful word inference within top-$k$ predicted words, for different webcams.}
\label{fig:eva-cam}
\vspace{-0.2in}
\end{figure}

\begin{figure}{}
\centering
\includegraphics[width=0.8\linewidth]{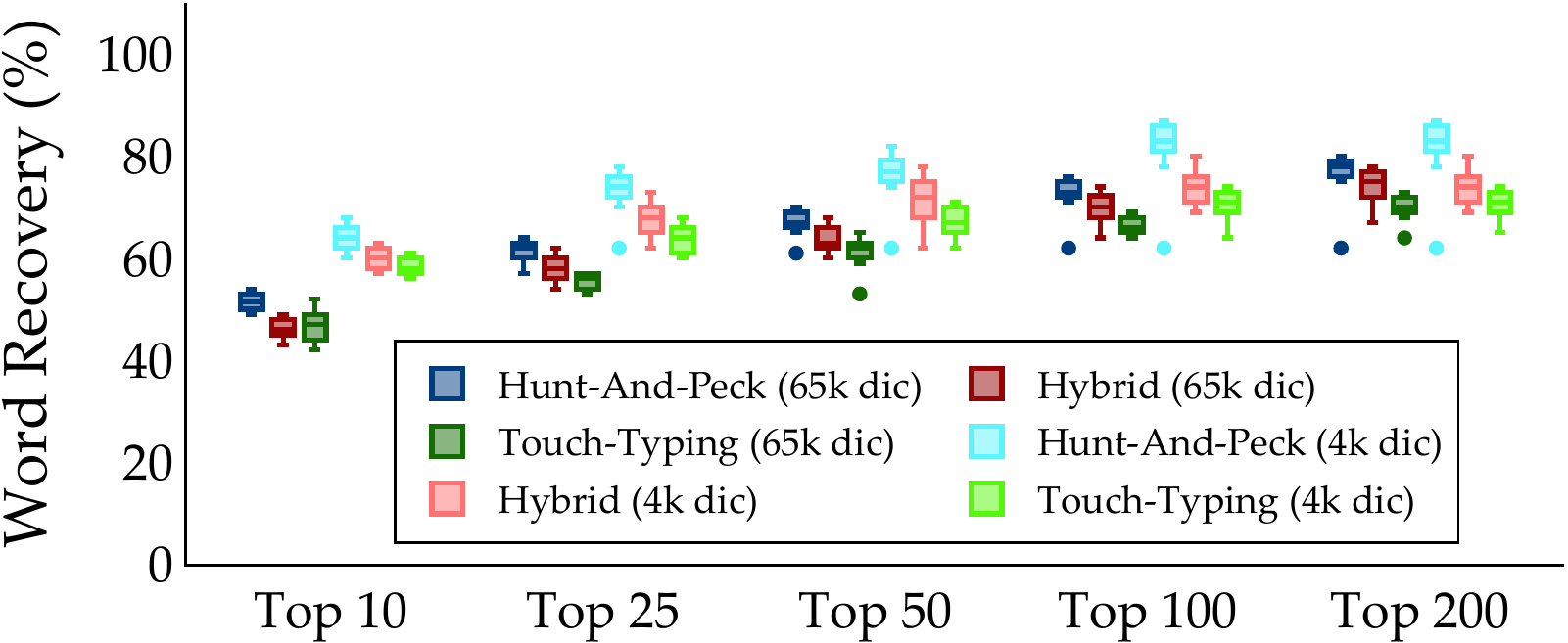}
\caption{Successful word inference within top-$k$ predicted words, for different typing styles (using Logitech webcam).}
\label{fig:eva-type}
\vspace{-0.2in}
\end{figure}

\noindent
\textbf{Different Webcams.}
Quality of the video can intuitively make a significant difference in the prediction accuracy, as low quality video frames are more likely to be erroneously processed by our algorithms. Accordingly, we look at the prediction accuracies obtained for the three experimental webcams (two external webcams, one built-in to the laptop). Both the Anivia and Logitech are able to capture videos at 1080p @ 30 $fps$, but the Anivia webcam features a wide-angle lens when compared to the Logitech webcam. The Lenovo laptop comes with a low-end webcam that can record video only at 720p @ 30 $fps$.
As seen in \cref{fig:eva-cam}, the Lenovo laptop webcam consistently had the worst performance compared to the Anivia and Logitech webcams. For the 65K dictionary, video from the Lenovo laptop webcam resulted in only 44.3\% average word recovery when top-200 words were considered.
The Logitech webcam performed slightly, but consistently, better than the Anivia webcam. Using the 4K dictionary, video from the Logitech webcam resulted in 75\% average word recovery when top-200 words were considered, whereas video from the Anivia webcam resulted in 70\% average word recovery. One of the reasons we speculate why the Anivia webcam did not perform as well as the Logitech webcam is because of its wide-angle lens. A wide-angle view means that the number of pixels capturing the user's body is reduced as more of the background is captured in the fixed video resolution. For many of the following evaluations, we used only the Logitech webcam for better understanding of other parameters.

{
\noindent
\textbf{Different Typing Styles.}
Based on a screening survey, we were able to categorize the typing style of our participants as hunt-and-peck, hybrid, or touch-typing (further explained in \cref{appendix:typing-styles}). Here we analyze if typing styles have any significant impact on the word recovery. \cref{fig:eva-type} shows the word recovery percentage for each of the typing styles. Hunt-and-peck typers were more susceptible with highest mean word recovery of 83\% (top-200, 4K dictionary), followed by hybrid typers at 74\% and touch-typers at 71\%. This is somewhat intuitive as the arm displacements are very subtle for proficient touch-typers, which can lead to a higher number of inaccurate interpretation of the displacement vectors. Nonetheless, we observe that the overall threat is still significant for users with any of the three typing styles.
}

\begin{figure}{}
\centering
\includegraphics[width=0.8\linewidth]{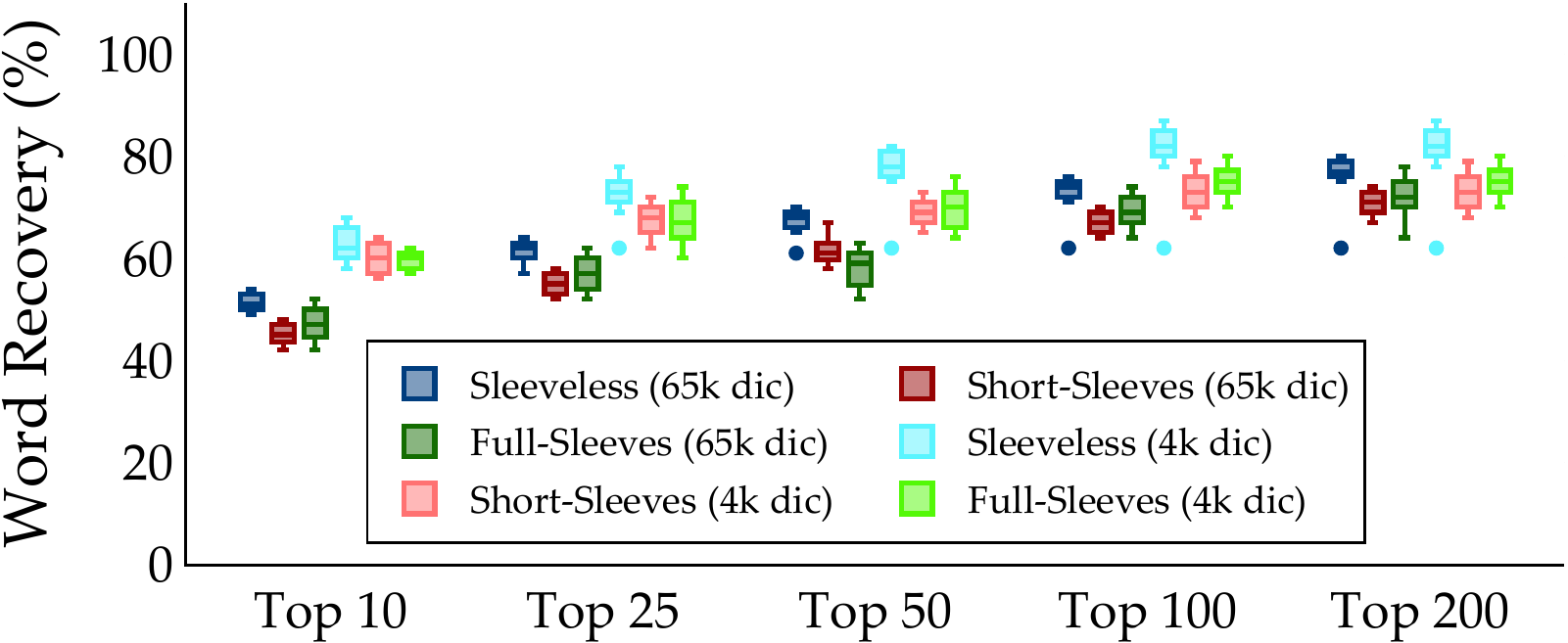}
\caption{Successful word inference within top-$k$ predicted words, for different clothings (using Logitech webcam).}
\label{fig:eva-shirt}
\vspace{-0.2in}
\end{figure}

\begin{figure}{}
\centering
\includegraphics[width=0.8\linewidth]{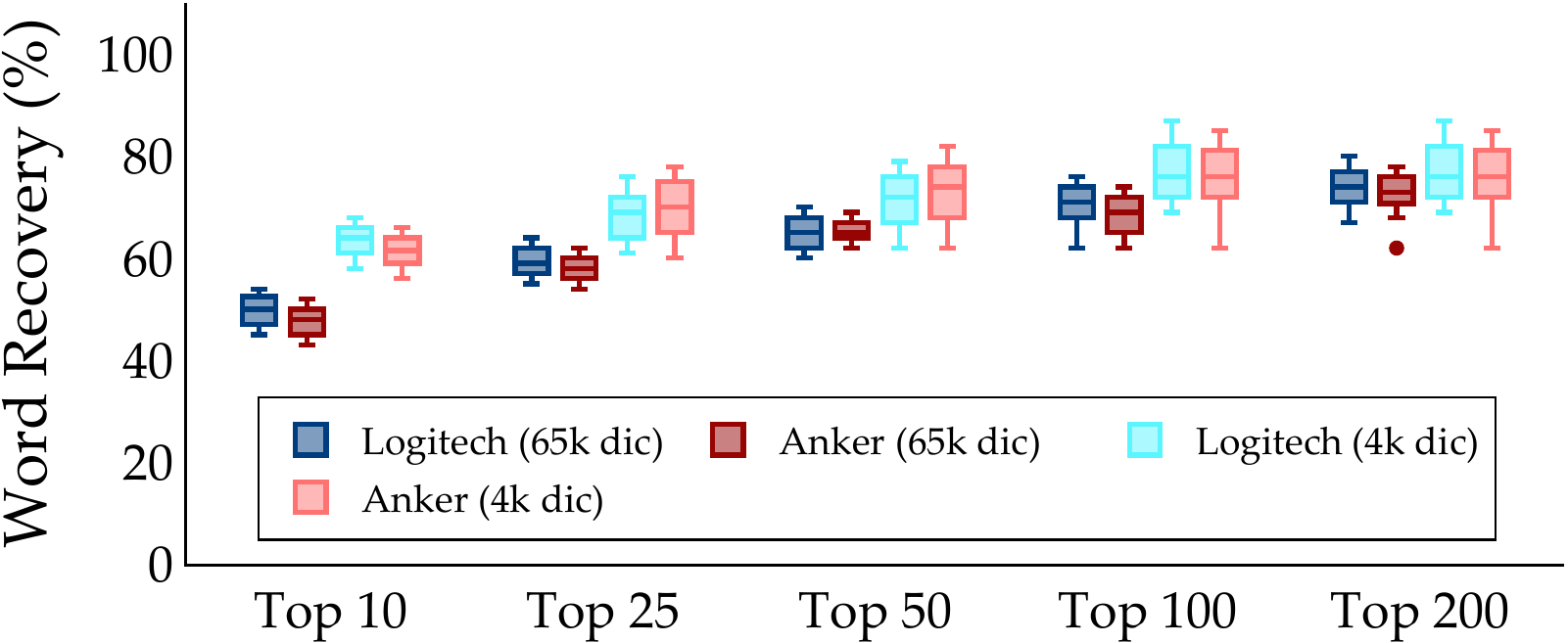}
\caption{Successful word inference within top-$k$ predicted words, for different keyboards (using Logitech webcam).}
\label{fig:eva-keyboard}
\vspace{-0.2in}
\end{figure}

\begin{figure}{}
\centering
\includegraphics[width=0.8\linewidth]{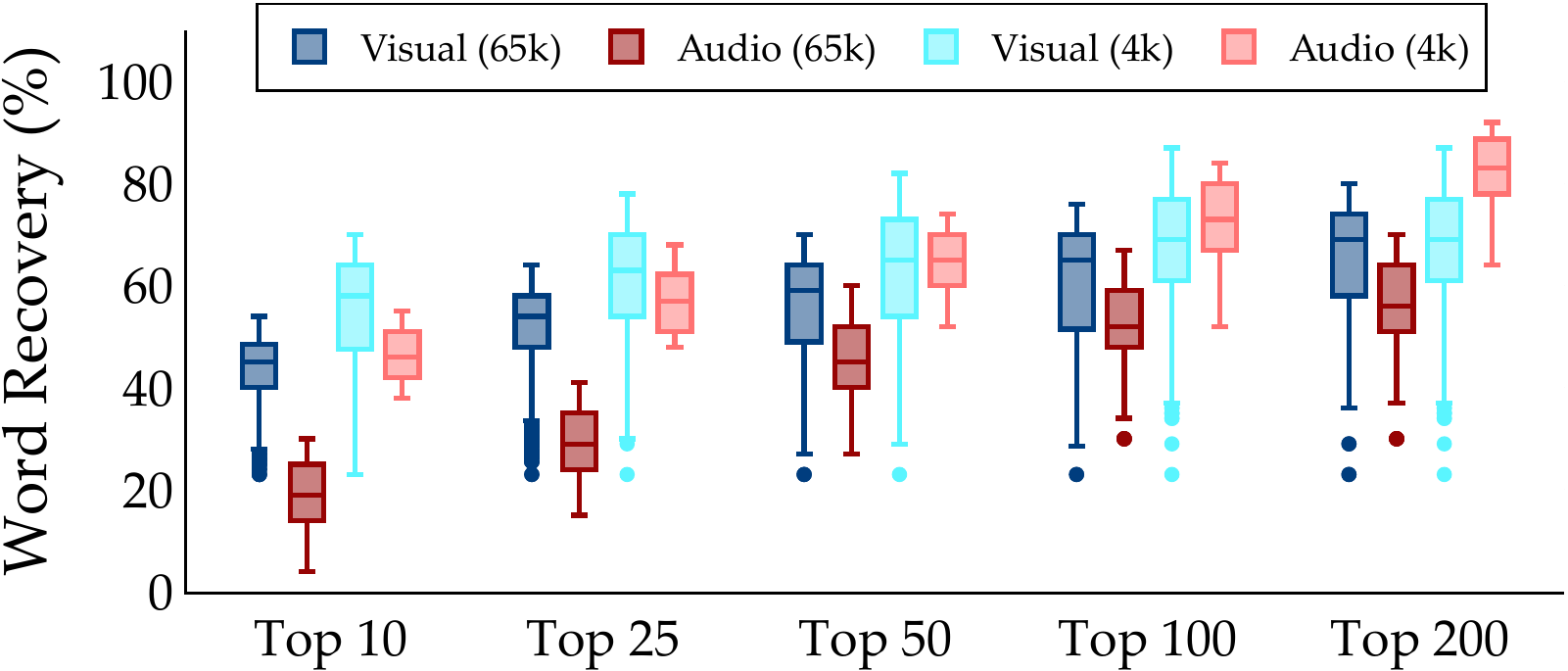}
\caption{Comparison between audio and video inferences.}
\label{fig:eva-all}
\vspace{-0.2in}
\end{figure}

\begin{figure}{}
\centering
\includegraphics[width=0.8\linewidth]{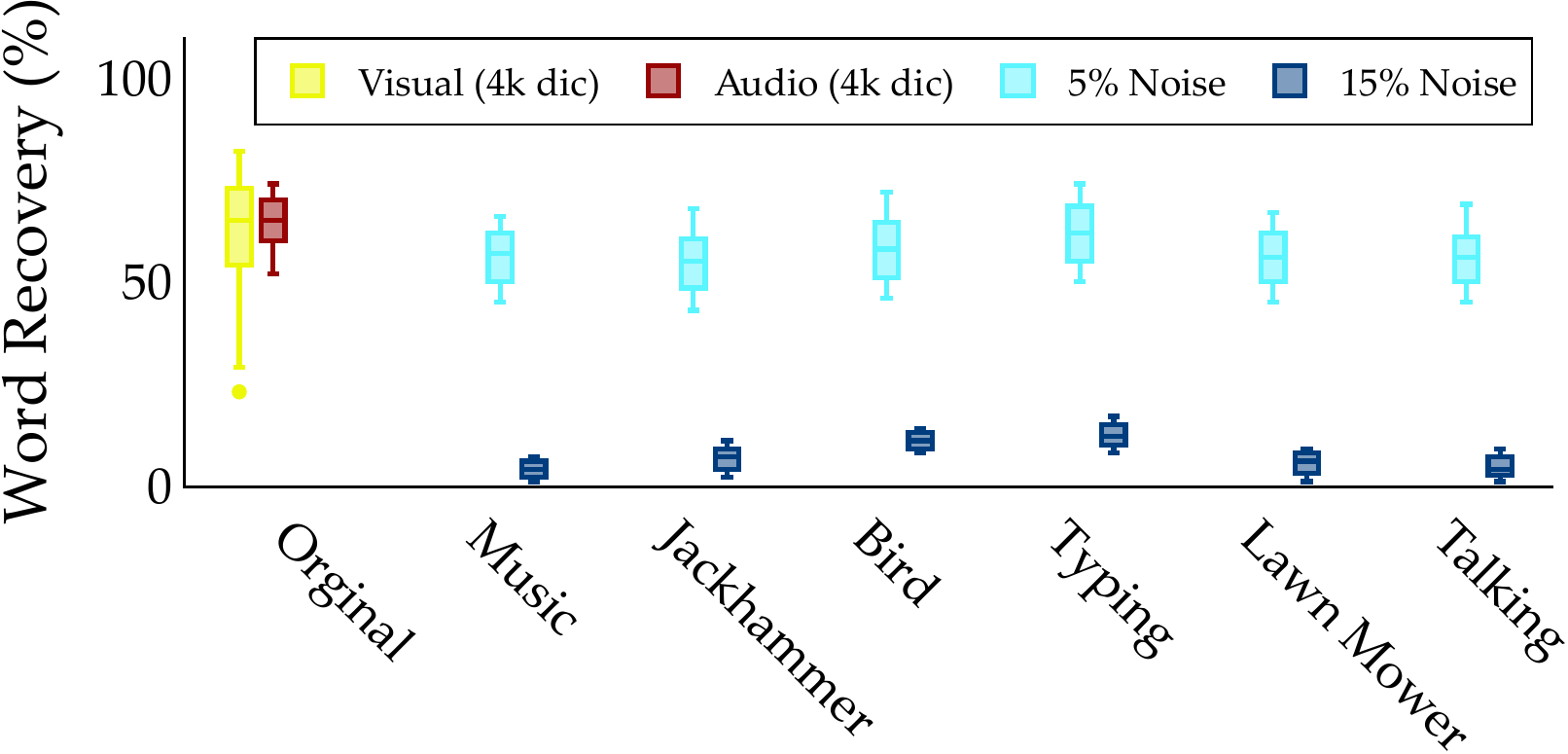}
\caption{Comparison between audio inferences with various acoustic noises (top-50, 4K dictionary).}
\label{fig:eva-noise}
\vspace{-0.2in}
\end{figure}

\noindent
\textbf{Different Clothings.}
We next evaluate if different types of clothing, especially with respect to their sleeve design, can affect our word prediction. In \cref{fig:eva-shirt} we observe that sleeveless typers were more susceptible to our attack, with 81.7\% mean word recovery (top-200, 4K dictionary), compared to typers who wore either full or short sleeved dresses (74.4\% and 73\%, respectively).
We speculate that both short and full sleeves can mask the extent to which the arms are actually displaced, underneath the clothing. As a result, our displacement vector calculations can get affected, resulting in slightly less accurate word predictions.

\noindent
\textbf{Different Keyboards.}
Size of the keyboard, and thus spacing between the keys, can have a significant influence on the arm displacements observed during typing. However, after evaluating the results from the two external keyboards, we did not find any significant difference between them (\cref{fig:eva-keyboard}). Even though the Logitech keyboard is significantly larger than the Anker keyboard, the percentage of successful word predictions were almost identical.

\noindent
\textbf{Different Video Calling Softwares.}
\label{voip-compare}
We tested our attack using three popular video calling softwares: Skype, Hangouts, and Zoom. Analyzing results using the 65K dictionary and top-50 predictions, we found that our evaluation using Skype was marginally better than using Zoom and Hangouts (+3.4\% and +8\% mean word recovery, respectively). The same set of videos were used with all the three video calling softwares, therefore the differences are purely due to factors beyond our control, such as the video compression technique used, network bandwidth utilized, and inconsistent latency in the video call.

\vspace{-0.1in}
\section{Video vs. Audio In-Lab Evaluation}
\label{sec:audiovideo_evaluation}
In this section, we compare our prediction framework (which is based on the video data) with Compagno et al.'s work \cite{compagno2017don}, where they utilized the audio stream of a Skype call for keystroke inference. We utilized Compagno et al.'s implementation of an acoustic-based keystrokes inference framework designed to work over audio/video calling applications, and applied our audio data to train and test the inference model. %

\noindent
\textbf{Audio vs. Visual Performance.}
With the 65K dictionary, our video-based inference was approximately +15\% more successful than Compagno et al.'s audio-based inference (when using top-200 predictions) as shown in \cref{fig:eva-all}. However, with the 4K dictionary, the audio-based inference was +10\% more successful than the video-based inference (using top-200 predictions). The reason why Compagno et al.'s audio-based inference performs poorly for the larger dictionary is because the collision rate significantly increases with the size of the dictionary. Nonetheless, our audio data collection was very controlled, with no one talking and minimum ambient noise levels. A realistic audio/video call will at least have participants talking, which can significantly affect Compagno et al.'s audio-based inference framework. Accordingly, we next evaluate the impact of various types of noise on Compagno et al.'s audio-based inference.

\noindent
\textbf{Noisy Audio vs. Visual Performance.}
We \emph{mixed} six different types of acoustic noises with our audio data: music, typing, lawnmower, bird chirps, jackhammer, and talking. The characteristics of these six acoustic noises are discussed in \cref{appendix:noise-char}. Also, to mimic real-life background noises, we mixed the acoustic noise at only 5\% and 15\% levels after amplitude normalization. 
As speculated earlier, after adding only 5\% noise the average word recovery dropped from 65\% to about 56\% (top-50, 4K dictionary) as shown in \cref{fig:eva-noise}. The variance in word recovery across the six different noise types was not very significant. Interestingly, after adding 15\% noise the word recovery sharply dropped to about 7\%.
These results highlight how even minimal noise levels can significantly affect the audio-based keystroke inference framework. In contrast, our video-based inference framework is not at all affected by acoustic noises which is a common occurrence in audio-video calls.

 \vspace{-0.1in}
{ %
\section{At-Home Experimental Setup}
\label{sec:experimental_setup_home}

To understand our inference framework's effectiveness in the wild, we next evaluate it outside of the lab environment. In this setting, participants were asked to use their own device (a laptop or desktop with a webcam) and setup (sitting position, clothing, background, and positioning of devices) for the video call, including location from where the call is done (e.g., their home). This allowed us to collect typing related video data for a diverse combination of devices and setups, already familiar to the participants and not constrained in any way.
Using such ``At-Home" experiments, we also expand our evaluation beyond just predicting dictated English words. Specifically, we analyze how our framework performs for the inference of user-chosen passwords, websites, and English words.
As in a real setting, users are expected to involve themselves in other activities besides typing (e.g., web surfing, playing online games, etc.). This unconstrained At-Home setup allows us to thoroughly evaluate our typing activity detection technique outlined in \cref{subsec:detection}.

\noindent
\textbf{Participant Demographics.}
We collected data from 10 participants for this in-home evaluation, whose ages ranged between 21 and 29 years. Out of the 10 participants, 3 are females, and 7 are males. Based on a screening-survey, 3 participants conducted hunt-and-peck typing, 5 conducted touch typing, and the remaining 2 participants conducted hybrid typing. 9 participants identified themselves as right-handed and 1 as ambidextrous. The average height of the participants is approximately 170 $cm$, with an average observed typing speed of approximately $3.7$ keystrokes per second and typing accuracy (in relation to typographical errors) of approximately $86.7\%$.

\noindent
\textbf{Participant's Task.}
Participants were invited to join a (maximum) 30 minute Skype video call, using their own device and setup and from their own location of choice, where they had to sporadically (and at their own pace) type 10 email addresses, 10 usernames, 10 passwords, 10 websites, and 10 English words, in no particular order and frequency. The typing was performed in a pre-shared online spreadsheet, which was later used as the ground-truth of the typed text/information. The spreadsheet also automatically recorded edit timestamp for each cell in the spreadsheet, which is useful for evaluating the typing activity detection technique. To ensure participants covered a reasonable amount of time on non-typing activities, we asked participants to take at least three 1-minute breaks doing one of the following three activities: watch a YouTube video, read a Wikipedia article, or play a digital game on their computer that only requires a mouse to play. Participants had the liberty to take additional or longer breaks and/or do any other activity on their computer that does not require keyboard usage. Unlike in the in-lab experiments, participants were allowed to use backspace in case they wanted to rectify a typing error and were allowed to use a larger set of keys/characters on the keyboard for their typing tasks (alphabet keys, number keys directly above the alphabet keys, keys corresponding to ``.", ``-",  and ``@" characters, and the enter and backspace keys).

\noindent
\textbf{Data Collected.}
In addition to the ground-truth text and timestamp information contained in the online spreadsheet where participants typed, participant's Skype video was recorded remotely using \texttt{OBS Studio} \cite{obsproject}.
After each typing experiment was completed, we also collected supplementary information from our participants related to their employed device and setup, as summarized in \cref{tab:demographics}.

\noindent
\textbf{Webcam Hardware and Positioning.}
We observed that three of our 10 participants used an external webcam in a similar fashion as in the in-lab setting, placed approximately at eye-level and focused directly on the participant. 
However, the remaining 7 participants who participated using their laptops, the webcam angle and distance varied noticeably, as shown in \cref{tab:demographics}.
The native webcam resolution across participants also varied between 720p or 1080p.%

}
 \vspace{-0.1in}
{

\section{At-Home Evaluation}
\label{sec:home_eva}
In this section, we evaluate the performance of our proposed typing activity detection and keystroke (or text/word) prediction techniques using video call data collected from the At-Home experiments. During these experiments, we observed that one of the participant's hair completely obscured his/her shoulder area for the entire experiment's duration, thus making the corresponding video frames unusable within our framework. Due to these At-Home experiments' uncontrolled nature, this participant was not asked to re-position his/her hair or change his/her posture.
This points to a limitation of our inference framework. However, this has already been highlighted in our assumed adversary model, where we clearly state that both shoulders and upper arms should be visible (to the adversary) in the recorded video. Thus, our presented evaluation results below are based only on data collected from the remaining 9 participants.

\subsection{Typing Activity Detection Performance}

For evaluation of our typing activity detection technique (\cref{fig:typing-flow}), we employ the same optimal values for parameters $\phi_{a}$, $\phi_{b}$, and $\phi_{c}$ as determined earlier in \cref{sec:eva-detection}. Across all the 9 participants, our typing activity detection technique resulted (\cref{fig:typing-recognition-eva}) in an average of $40.22$ true positives, $12.4$ false positives and $9.78$ false negatives, for an average precision of $77.6\%$ and recall of $80.4\%$. 
This shows that our proposed activity detection technique is fairly accurate enabling 
the adversary to not only detect a majority of the typing activity during a video call, but also successfully differentiate between typing versus non-typing activities.

In addition to the overall results, let's further highlight some interesting special cases. For participants using a laptop, we observed that the location filter (of the typing activity detection technique) was not very effective due to the proximity of the laptop's touchpad to its keyboard. Also, a few participants (at least, two) claimed to have used both their hands for interacting with the laptop touchpad, making the exclusive filter of the typing activity detection technique ineffective at times and resulted in a higher number of false positives. Significant movement and posture changes (between typing and non-typing activities) also resulted in degradation of detection accuracy, as was observed (\cref{tab:demographics}) in the case of at least one participant whose left shoulder was not visible for a significant portion of the video call because of movement/posture changes. This limitation can also be attributed to the fact that while using a laptop, a user's position is a bit constrained (given the webcam's restricted field-of-view) and small movements/posture changes can result in the user's shoulders/upper arms becoming invisible/unavailable to the adversary.

\subsection{Typing Accuracy}
Before presenting our word prediction results, we briefly analyze the rate of typographical errors made by our participants. As our inference framework does not have provisions for handling rectifications made after typographical errors, participants' typing accuracies have a direct correlation to our framework's prediction error. \cref{tab:demographics} lists the typing accuracies of all our participants. As a case in point, participants I and C  had the worst typing accuracy (73.1\% and 49\%, respectively), and our word prediction performance for participants I and C was also the lowest. The following prediction performance results are inclusive of all the typographical errors made by our participants in the At-Home setting.

\subsection{Keystroke or Text Prediction Performance}
Next, we present results for word prediction in the At-Home setting, separated based on the category of typed words. It should be noted that, in contrast to the In-Lab setting, participants in the At-Home setting typed their own words (for each of the five categories), and that four out of the five categories (of typed words) would most likely include words that would not be present in a typical English language dictionary. Thus, rather than using a standard English dictionary for prediction, we first create a ranked reference database of likely words in each category (which could be contextually created based on the target participant) and then employ it for the prediction task. As our framework predicts the possible combinations of typed characters based on the movements, and not individual characters themselves, such a reference database is \emph{required} to complete the prediction task. 

\noindent
\textbf{Websites.} 
For the prediction of websites typed by target users, we created a reference database of 1 million most-visited websites \cite{majestic-million}. In our dataset, all participants typed at least two websites ranked in the top-20 of the reference database, with an overall median rank of $140$ and a mean rank of $36,745.3$ (mean is significantly higher than the median due to a few websites that are not popular and thus have very high ranks in the reference database). Our inference framework was successfully able to infer $ 66.7\%$ of the websites typed by participants, within the top-25 predictions. An adversary may further reduce the search space based on contextual information about the target user.

\noindent
\textbf{Passwords.}
For prediction of passwords typed by target users, we created a reference database of 1 million most commonly used passwords \cite{common-passwords}. Only $18.9\%$ of the passwords were successfully recovered within top-50 predictions, which can be attributed to the fact that $74.4\%$ of the passwords typed by our participants were not found in the reference database used for prediction.
Considering only the passwords that were present in the reference database, $74\%$ of them were successfully recovered within top-50 predictions. %

\noindent
\textbf{English Words.}
For prediction of English words, we use the 65K-words dictionary used earlier for the In-Lab evaluation. 
Similar to passwords, not all (25.6\%) of the words typed by our participants existed in the dictionary used for prediction, and $21.1\%$ of the English words were successfully recovered within top-50 predictions. 
One of the reasons our accuracy is worse than the In-Lab setting is because the reference dictionary's rank sorting is based on word-usage frequency in English language sentences, not based on random words produced by people. 
In other words, the ranking of words within the reference dictionary is not appropriate for prediction of randomly typed words.
If an adversary could produce a more accurate contextual reference dictionary based on the target user, we expect better performance from our framework. 

\noindent
\textbf{Usernames and Email Addresses.}
Usernames and email addresses are commonly used as an identifier for authentication, but they are also often publicly known and not sensitive information by themselves. However, knowing \emph{when} a target user typed their username or email address can be valuable to an adversary, as a password is likely to be typed immediately afterwards during an authentication. Therefore, instead of predicting the usernames and email addresses typed by our participants, we try to predict when their known username and email address was typed by them. 
On average, we were able to correctly predicted when $91.1\%$ of the usernames and $95.6\%$ of the email addresses were typed.

\begin{figure}[]
\centering
\includegraphics[width=0.8\linewidth]{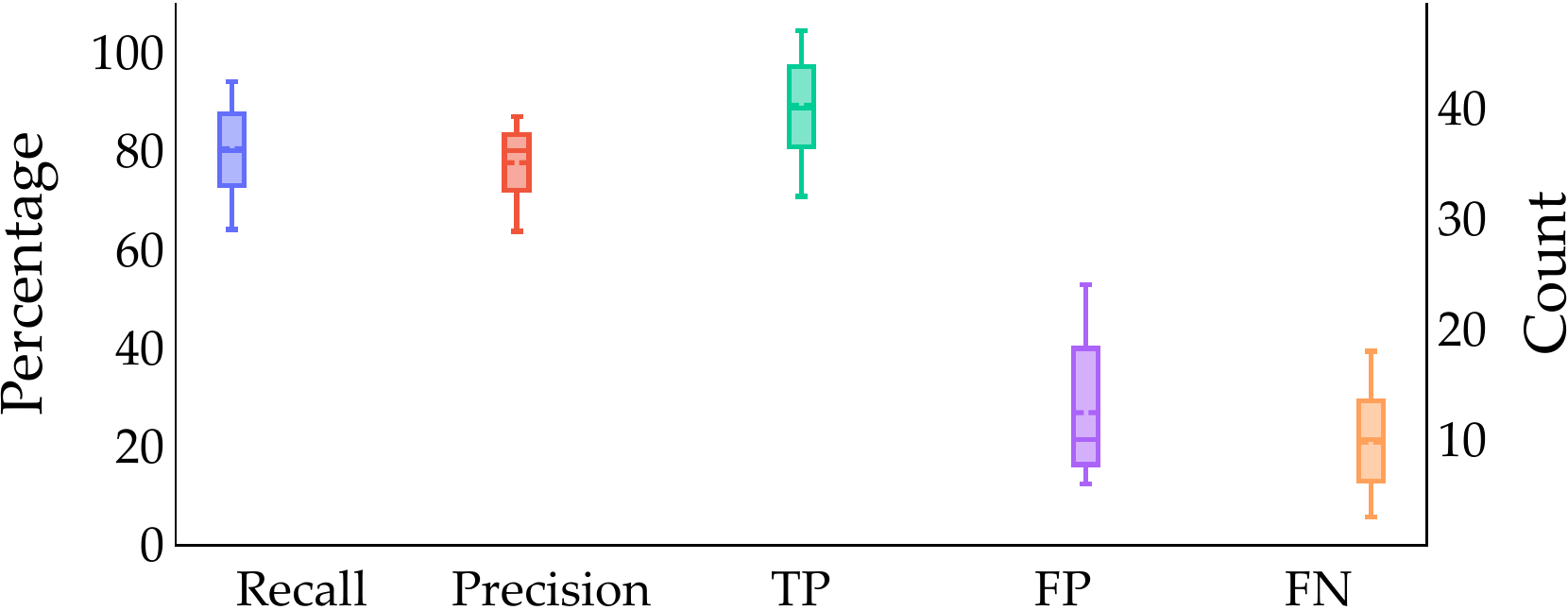}
\caption{Performance of the typing activity detection technique.}
\label{fig:typing-recognition-eva}
\vspace{-0.2in}
\end{figure}

\begin{figure}[]
\centering
\includegraphics[width=0.8\linewidth]{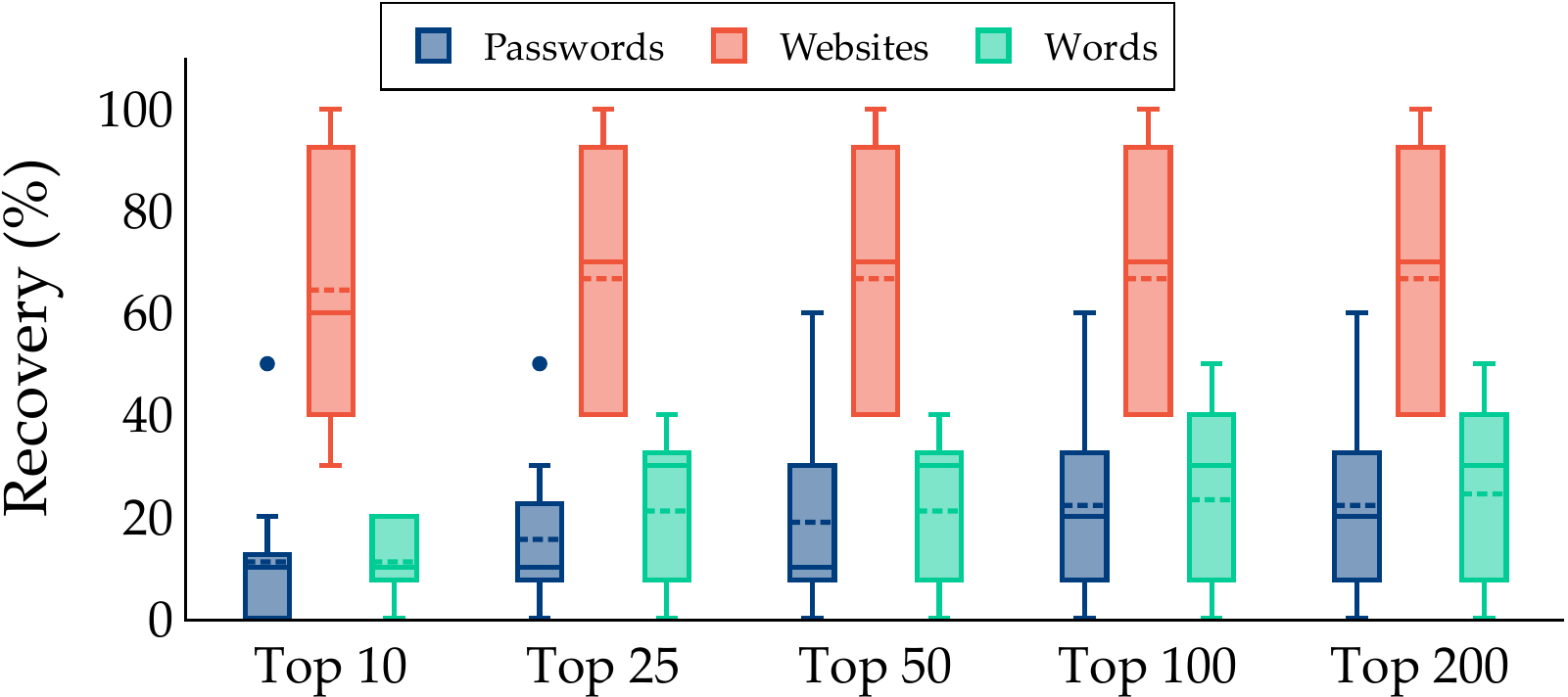}
\caption{Successful inference of different text predictions.}
\label{fig:entry-res}
\vspace{-0.2in}
\end{figure}

}

 \vspace{-0.1in}
\section{Threat Mitigation}
\label{sec:mitigations}

In this section, we outline and evaluate potential mitigation techniques to the video-based keystroke inference threat. {We evaluate these mitigation measures by applying them to the In-Lab video dataset prior to using them in our keystroke inference framework, and then measuring the performance of our framework on these modified video data. We evaluate the mitigation techniques using the In-Lab dataset instead of the At-Home dataset, because with its higher inference success, the In-Lab dataset can better illustrate the effectiveness of the proposed mitigation techniques. %
}
We measure the performance of our framework under the influence of these mitigation techniques using the metrics of (i) \emph{effectiveness}, (ii) \emph{efficiency}, and (iii) \emph{video quality}, which we describe next followed by a description and evaluation of the mitigation techniques.

\begin{enumerate}[\hspace{0pt}(1)]
\item \textbf{Effectiveness} measures the average reduction in word recovery due to the mitigation technique. 
\item \textbf{Efficiency} measures the average time to process each frame. %
\item \textbf{Video Quality}
measures the image quality in the modified (edited) frames using %
\emph{SSIM} index \cite{wang2004image} as a measure of the structural quality of the frames within the video. 
\end{enumerate}

\vspace{-0.1in}
\subsection{Mitigation Techniques}
We now outline three frame manipulation strategies as mitigation techniques against the video-based keystroke inference threat presented earlier, and present performance results for them using the metrics defined above. It must be mentioned that, although these techniques can be applied to all the frames in the entire video call, it makes much more sense to apply them to frames in the vicinity of the target user's actual keystrokes. As keystroke detection for mitigation can be easily accomplished using OS-interrupts on the user side, it should be relatively straightforward to identify frames just before, during and after the keystroke on which the proposed manipulation strategies should be applied. However, to effectively manipulate the frames immediately before a keystroke, we must maintain a buffer of those frames before they are transmitted out. Obviously, a large buffer can introduce significant latency in the video call, which is detrimental to the overall quality. We employ a buffer size of 2 frames (in a 30 $fps$ video) for the first two mitigation techniques, and we use a variable buffer size in the third mitigation technique.

\begin{figure}[]
\centering
\begin{subfigure}{0.13\linewidth}
\centering
\includegraphics[width=\textwidth]{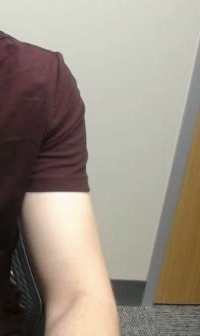}
\caption{}
\label{fig:mitigations-a}
\end{subfigure}
\begin{subfigure}{0.13\linewidth}
\centering
\includegraphics[width=\textwidth]{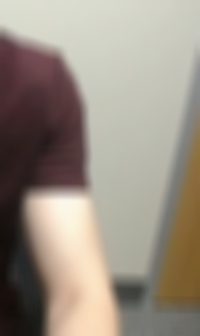}
\caption{}
\label{fig:mitigations-blur5}
\end{subfigure}
\begin{subfigure}{0.13\linewidth}
\centering
\includegraphics[width=\textwidth]{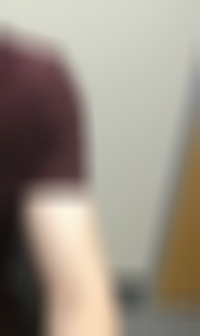}
\caption{}
\label{fig:mitigations-blur10}
\end{subfigure}
\begin{subfigure}{0.13\linewidth}
\centering
\includegraphics[width=\textwidth]{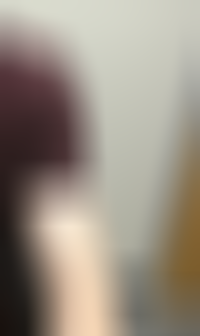}
\caption{}
\label{fig:mitigations-blur20}
\end{subfigure}
\begin{subfigure}{0.13\linewidth}
\centering
\includegraphics[width=\textwidth]{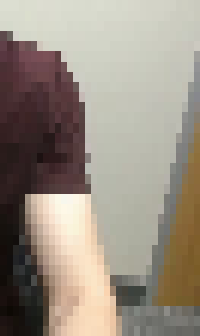}
\caption{}
\label{fig:mitigations-pix3}
\end{subfigure}
\begin{subfigure}{0.13\linewidth}
\centering
\includegraphics[width=\textwidth]{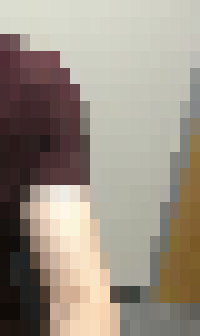}
\caption{}
\label{fig:mitigations-pix5}
\end{subfigure}
\begin{subfigure}{0.13\linewidth}
\centering
\includegraphics[width=\textwidth]{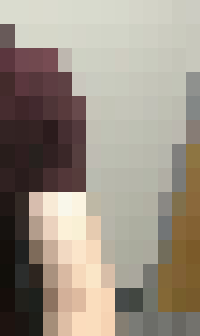}
\caption{}
\label{fig:mitigations-pix7}
\end{subfigure}
\caption{A left hand frame segment from a frame \textit{(a)} unaltered, \textit{(b)} after blurring with $z_b$ = 5\%, \textit{(c)} after blurring with $z_b$ = 10\%, \textit{(d)} after blurring with $z_b$ = 20\%, \textit{(e)} after pixelation with $z_p$ = 3\%, \textit{(f)} after pixelation with $z_p$ = 5\%, and \textit{(g)} after pixelation with $z_p$ = 7\%.}
\label{fig:mitigations}
\vspace{-0.2in}
\end{figure}

\noindent
\textbf{Blurring.}
The first approach is to manipulate (sensitive) frames using a Box blur approach \cite{stavens2007opencv}. This approach produces a \emph{blurring} effect on the original frame by employing an adjustable kernel.
The size of the Box blur kernel is chosen as some proportion ($z_{b}$) of the original frame size and populated with `1's.
Once the kernel is fixed, blurring is done as follows: For each pixel, $p_{i,j}$, of the original image frame, the kernel is centered on that pixel and a new pixel value is computed. This new pixel value is the average of the neighboring pixel values weighted using the kernel. This new pixel value then replaces the original pixel $p_{i,j}$.
This process is repeated for all the pixels of the frame. Some visual examples of the impact of blurring on a sample image frame for different values of the kernel parameter $z_b$ are depicted in \cref{fig:mitigations-blur5,fig:mitigations-blur10,fig:mitigations-blur20}. At the press of a keystroke we blur all the buffered frames and four following frames (total 6 frames) for a total duration of about 200 $ms$, which is the mean duration of keystrokes \cite{card1983psychology}.

\begin{figure}{}
\centering
\includegraphics[width=0.8\linewidth]{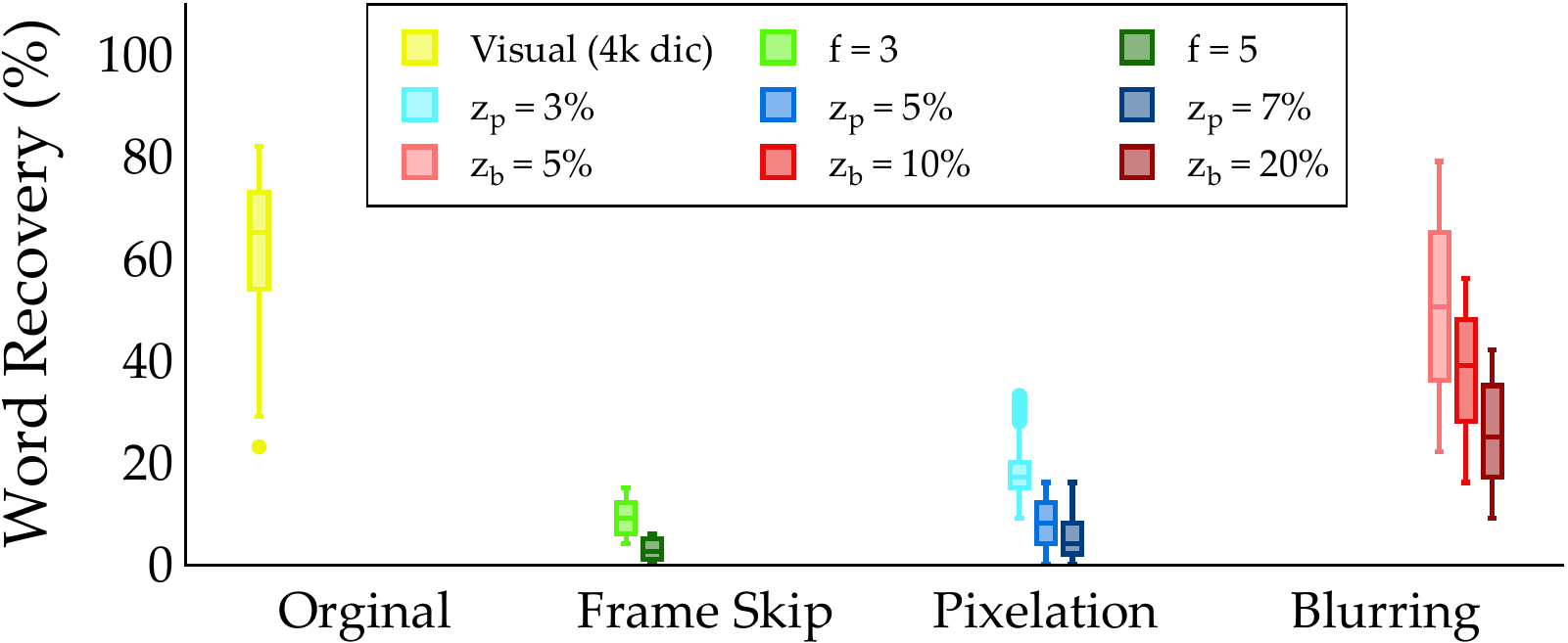}
\caption{Different mitigations techniques and resultant word recovery (top-50, 4K dictionary).}
\label{fig:mig}
\vspace{-0.2in}
\end{figure}

Our experimentation with using blurring within our inference framework shows that we are able to reduce the average word recovery from 65\% to as low as 13\% for $z_b$ = 20\% (\cref{fig:mig}). In other words, we see a mitigation effectiveness of about -52\% in top-50 prediction. We also observed that using higher $z_b$ resulted in less words being recovered, as the frames were more blurry.
For $1920 \times 1080$ sized frames, we observed that blurring takes around 17 $ms$ per frame with a kernel factor ($z_b$) of 5\%, on a laptop with an Intel i7-7700HQ (2.8 $GHz$) processor and 32 GB RAM.
In terms of image quality, we saw an average \emph{SSIM} index of 78.2\% for $z_b$ = 20\%. A high \emph{SSIM} index implies that the manipulated frame is similar to the original frame, and vice versa.
These results show that blurring is an effective mitigation technique, which imposes little efficiency and quality overheads.

\noindent
\textbf{Pixelation.}
The second approach we analyze is pixelation, where the frame is first pixelated (partitioned) into areas defined by a proportion parameter $z_p$. In other words, the frame (of size $m \times n$) is partitioned into $1/ {z_p}^{2}$ areas of size $\frac{m}{z_p} \times \frac{n}{z_p}$. Then, for each such area, the average of all pixel values within that area is computed, and each pixel $p_{i,j}$ within that area is reassigned this new average value. Some visual examples of the impact of pixelation on an image frame for different values of the pixelation proportion parameter $z_p$ is shown in \cref{fig:mitigations-pix3,fig:mitigations-pix5,fig:mitigations-pix7}. Similar to blurring, at the hit of a keystroke we pixelate all the buffered frames and four following frames (total 6 frames) for a total duration of about 200 $ms$.

Our experimentation with using pixelation within our inference framework shows that we are able to reduce the average word recovery from 65\% to as low as 4.3\% for $z_p$ = 7\% (\cref{fig:mig}). In other words, we see a mitigation effectiveness of about -60\% in top-50 prediction. We also observed that using higher $z_p$ resulted in less words being recovered, as the frames were more pixelated.
For $1920 \times 1080$ sized frames, we observed that pixelation takes around 1.41 $ms$ per frame with a $z_p$ of 3\%, which is significantly faster than blurring.
In terms of image quality, we saw an average \emph{SSIM} index of 74\% for $z_p$ = 7\%. 
These results show that pixelation is even more effective than blurring, and it imposes significantly lesser efficiency overhead, with a slight trade-off in quality.

\noindent
\textbf{Frame Skipping.}
The final mitigation approach we analyze is frame skipping, where as the name suggests, not all frames (captured during the video call on the target user side) are sent to the receiver (adversary). 
More specifically, the approach continuously buffers $f$ frames during the call on the target user side. If typing is detected, all the buffered frames, as well as, an additional $f$ frames after the detected key press, are dropped (i.e., not sent).
Our experimentation shows that frame skipping is the most effective method in reducing word recovery rate, with only around 3\% of the words recovered on an average (for $f=5$). In other words, we see a mitigation effectiveness of about -62\% in top-50 prediction.
Frame skipping successfully eliminates all movement relationship between consecutive key strokes, resulting in such a high mitigation effectiveness. Frame skipping does not impact image quality of individual frames as the original frames are never modified.
However, the downside of frame skipping is that user's video will appear to be stuck at a frame just prior to the keystroke, to the other participants in the video call. This can be confusing to the uninformed, but can be remedied with a notice such as \emph{``John Smith is typing''} to other participants in the video call. 

 \vspace{-0.1in}
\section{Discussion \& Limitations}
\label{sec:discussion}

\noindent
\textbf{Generalizability.} 
Let us comment on the generalizability and limitations of our study. First, we believe that our results are very generalizable to real-life scenarios based on the number of participants from which we collected data (more than prior related studies \cite{compagno2017don,anand2018keyboard}) and the different choices of webcams, keyboards, devices, participant clothings, and video calling software used in our experiments, which we believe are well representative samples.
While all our participants were students recruited from a university campus, we observed a huge variety of different typing styles and quirks, which makes us reasonably confident about it being representative of the general population. Moreover, our data collection experiments were designed to reduce all types participant biases, including response bias, and were approved by the university's IRB.

\noindent
\textbf{Limitations.} 
In our framework we only employed video feed to detect keystrokes, but video data can be combined with audio data from the call to further improve keystroke detection. %
The accuracy of our framework also relies significantly on the field of view containing the target user. Obstacles blocking (either completely or partially) the shoulder and arm areas of the target user, such as, microphones, headphone wires, or hair, could adversely affect both keystroke event detection and prediction. Similarly, if a camera's field of view does not fully or partially capture the shoulder and arm areas of the target user, as often observed in laptop webcams as they are generally set at an angle, it could also adversely impact the prediction performance of our framework. Lastly, we have also observed that significant ambient lighting changes (during typing) also disrupts the efficacy of our prediction.
Many target user-specific factors can also disrupt the prediction performance of our framework, for instance if there are significant user movements while typing. This is possible especially if the target user is seated on a movable object, such as a rolling chair.
As seen in our mitigation techniques, video quality is very impactful. If video frames are dropped, or the frames had some quality issues such as blurring or pixelation, then our framework will have poor inference accuracy.

 \vspace{-0.1in}
\section{Conclusion}
\label{sec:conclusion}
We proposed and evaluated a keystroke inference framework which can predict text typed by a user during a video call.
Specifically, we modeled and analyzed hand movements observable in the webcam's field of view, in order to detect keystroke events and then carry out a dictionary-based predictions. We evaluated our framework in a variety of controlled and uncontrolled scenarios, and were able to recover up to 75\% words in some scenarios. We also proposed and evaluated three mitigation techniques which can effectively deter such keystroke inference attack in video calls.

\bibliographystyle{IEEEtranS}
\bibliography{references}

\begin{appendices}

\crefalias{section}{appsec}

\section{Anatomy and Movement of Arms During Typing}
\label{appendix:background-arms-typing}

\begin{figure}[H]
\centering
\begin{subfigure}[b]{0.30\linewidth}
\includegraphics[width=\linewidth]{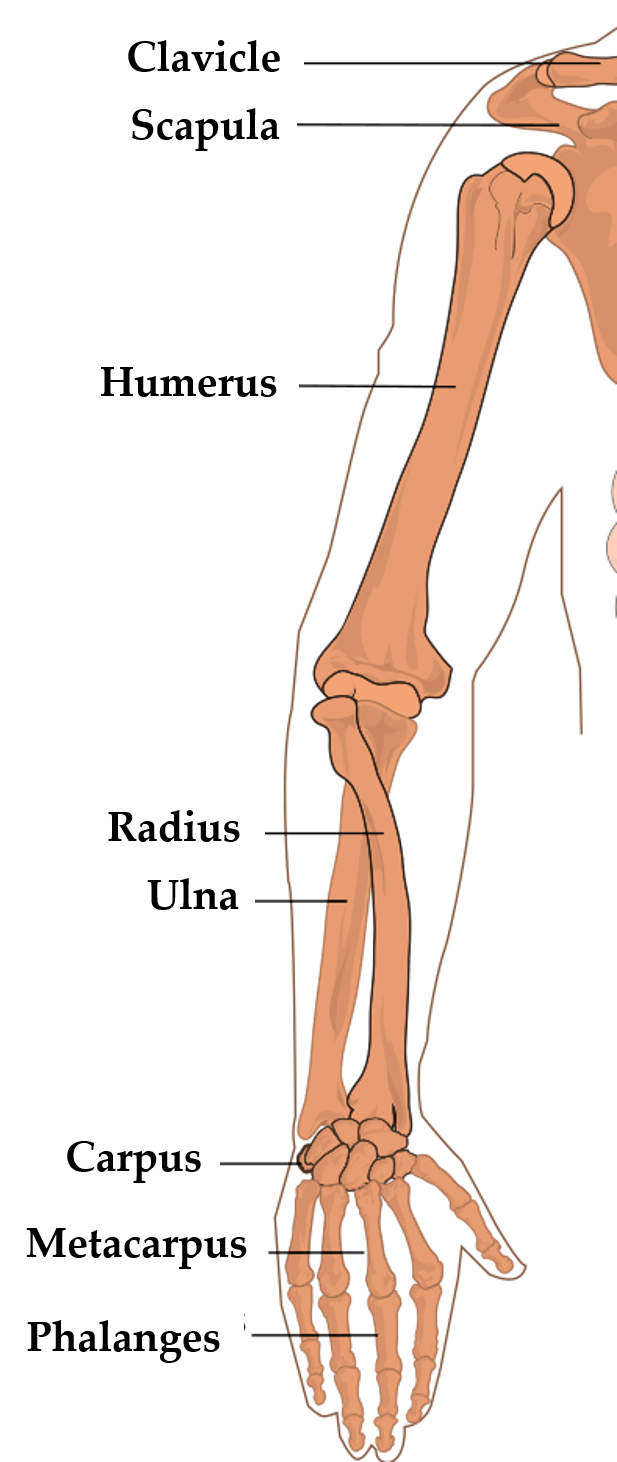}
\caption{}
\label{fig:arms}
\end{subfigure}
\quad\quad\quad
\begin{minipage}[b]{0.42\linewidth}
\begin{subfigure}[b]{\linewidth}
\includegraphics[width=\linewidth]{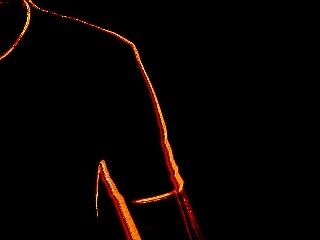}
\caption{}
\label{fig:heatmap-l}
\vspace{-0.15in}
\end{subfigure}\\[\baselineskip]
\begin{subfigure}[b]{\linewidth}
\includegraphics[width=\linewidth]{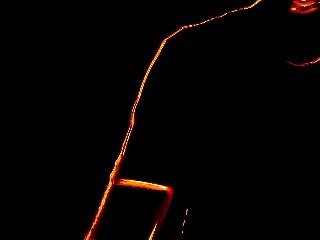}
\caption{}
\label{fig:heatmap-r}
\end{subfigure}
\end{minipage}
\caption{\textit{(a)} Anatomy of the human arm and shoulder bones. \textit{(b)}, \textit{(c)} Pixel-level heatmaps of upper body movements during one minute of typing for left and right sides of the body, respectively.}
\label{fig:upper-body}
\end{figure}

\section{Typing Styles}
\label{appendix:typing-styles}

\noindent
\textbf{Hunt-and-peck typing} is largely regarded as one of the most inefficient typing technique. In hunt-and-peck typing, the typer has sight on the keyboard during typing, as in most cases the typer does not have the keyboard layout memorized. Also, most hunt-and-peck typers heavily use their two index fingers for typing. %
As a result, hunt-and-peck typers' arms undergo significant movement between keystrokes.

\noindent
\textbf{Touch typing} is largely regarded as one of the most efficient typing technique. In touch typing, the typer looks at the screen and types continuously without looking at the keyboard. Touch typers also utilize all ten fingers.

\noindent
\textbf{Hybrid typing}, as the name suggests, is a hybrid of hunt-and-peck and touch typing. Like touch typers, hybrid typers may have memorized the keyboard layout and are able to type while looking at the screen. However, unlike touch typers, hybrid typers utilize fewer fingers, usually between 2 to 6 fingers. %

\section{Preprocessing Figures}
\label{appendix:preprocessing-details}

\begin{figure}[H]
\centering
\includegraphics[width=\linewidth]{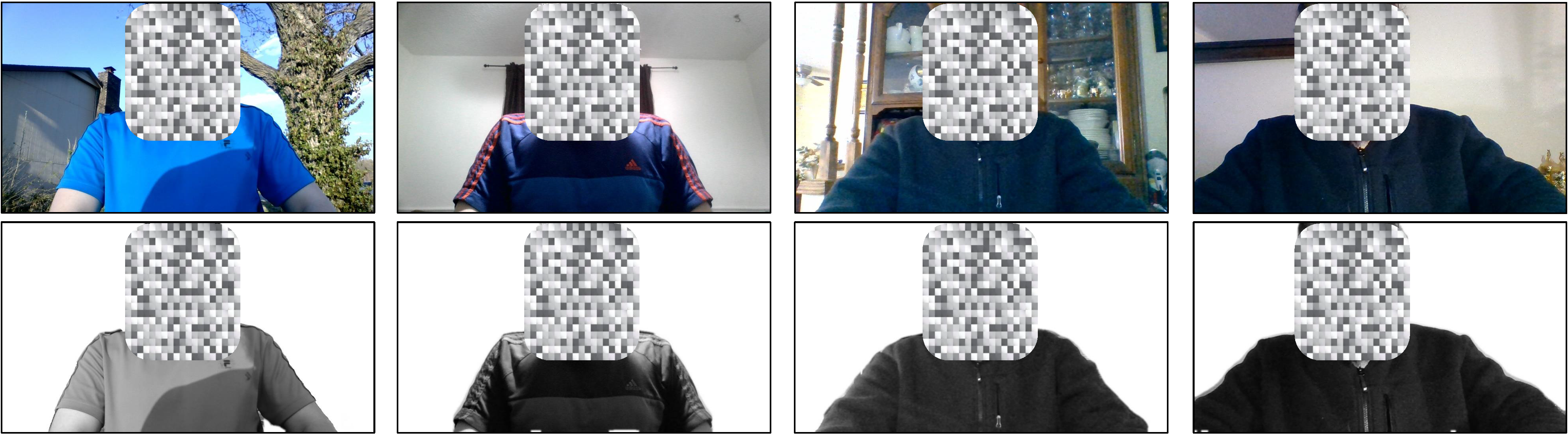}
\caption{Example output of the background removal process, using \texttt{DeepLabv3} and \texttt{Microsoft COCO}, successfully applied in different indoor and outdoor settings. Top four images are the original frames, and bottom four images are corresponding frames after background removal.}
\label{fig:background-removal}
\end{figure}

\begin{figure}[H]
\centering
\includegraphics[width=\linewidth]{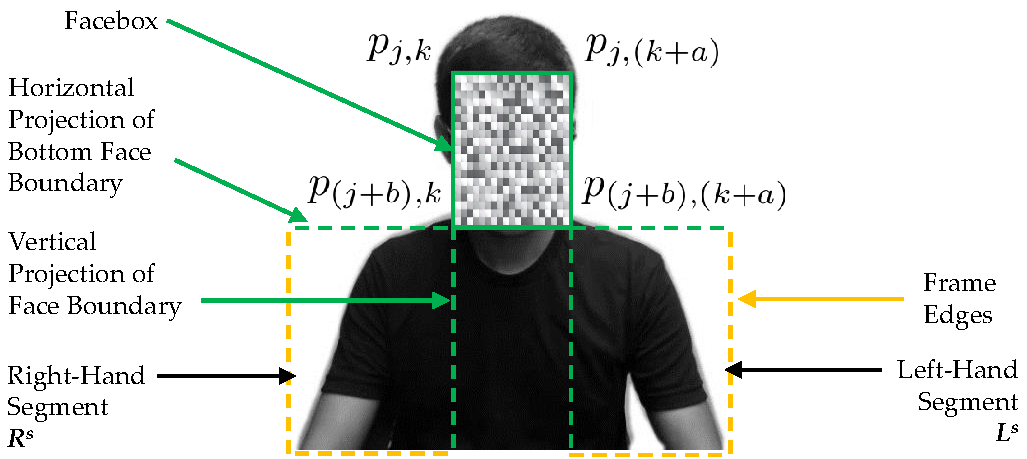}
\caption{Face detection using \texttt{Faceboxes} and segmentation of left and right arms, in a frame from the captured video.}
\label{fig:face+segmentation}
\end{figure}

\section{Keystroke Detection Algorithm}
\label{appendix:keystroke-detection}

\begin{algorithm}[H]
\small
\caption{Keystroke detection algorithm.}
\label{alg:keystroke-detection}
\begin{algorithmic}[1]
\State Input: segmented arm frames stored in $armS[]$ 
\State Output: keystroke frames stored in $keystrokesFS[]$ 
\Procedure{KeystrokeDetect}{}%
	\State $armS[]$ \Comment{$R_{}^{s}$ or $L_{}^{s}$} %
	\State $ssimList[]$ \Comment{Series of SSIM scores}
	\State $ssimDiff[]$ \Comment{$ssimL[i] - ssimL[i+1]$}
	\State $keystrokesFS[]$ \Comment{Store keystroke containing $R_{}^{s}$ or $L_{}^{s}$}
	\For{ $i$ in range($armS.size()- 1$)}
		\State $ssimScore = SSIM(armS[i] - armS[i+1])$
		\State $ssimList.append(ssimScore)$
	\EndFor
	\For{ $i$ in range($ssimList.size()- 1$)}
		\State $ssimDiff.append(ssimList[i] - ssimList[i+1])$
	\EndFor
	\State $mean \leftarrow  mean(ssimDiff)$
	\State $std \leftarrow  standardDeviation(ssimDiff)$	
	\For{ $i$ in range($ssimDiff.size()- 1$)}
		\State $zScore = (ssimDiff[i] - mean)/std$
        	\If{$zScore > \phi_a$ and $zScore < \phi_b$ }
        		\If {$ssimDiff[i]$ is a local max }
        			\If {local min exists between $i \rightarrow i+2$ }
        				\If {$zScore(local min) < \phi_c$}
    					\State $keystrokesFS$.append($armS[i]$)
    					        \EndIf
        				\EndIf
        			\EndIf    		
        		\EndIf		    		
		\EndFor
\EndProcedure
\end{algorithmic}
\end{algorithm}

\section{Typing Activity Detection Example}
\label{appendix:typing-act-ex}

\begin{figure}[H]
\centering
\includegraphics[width=\linewidth]{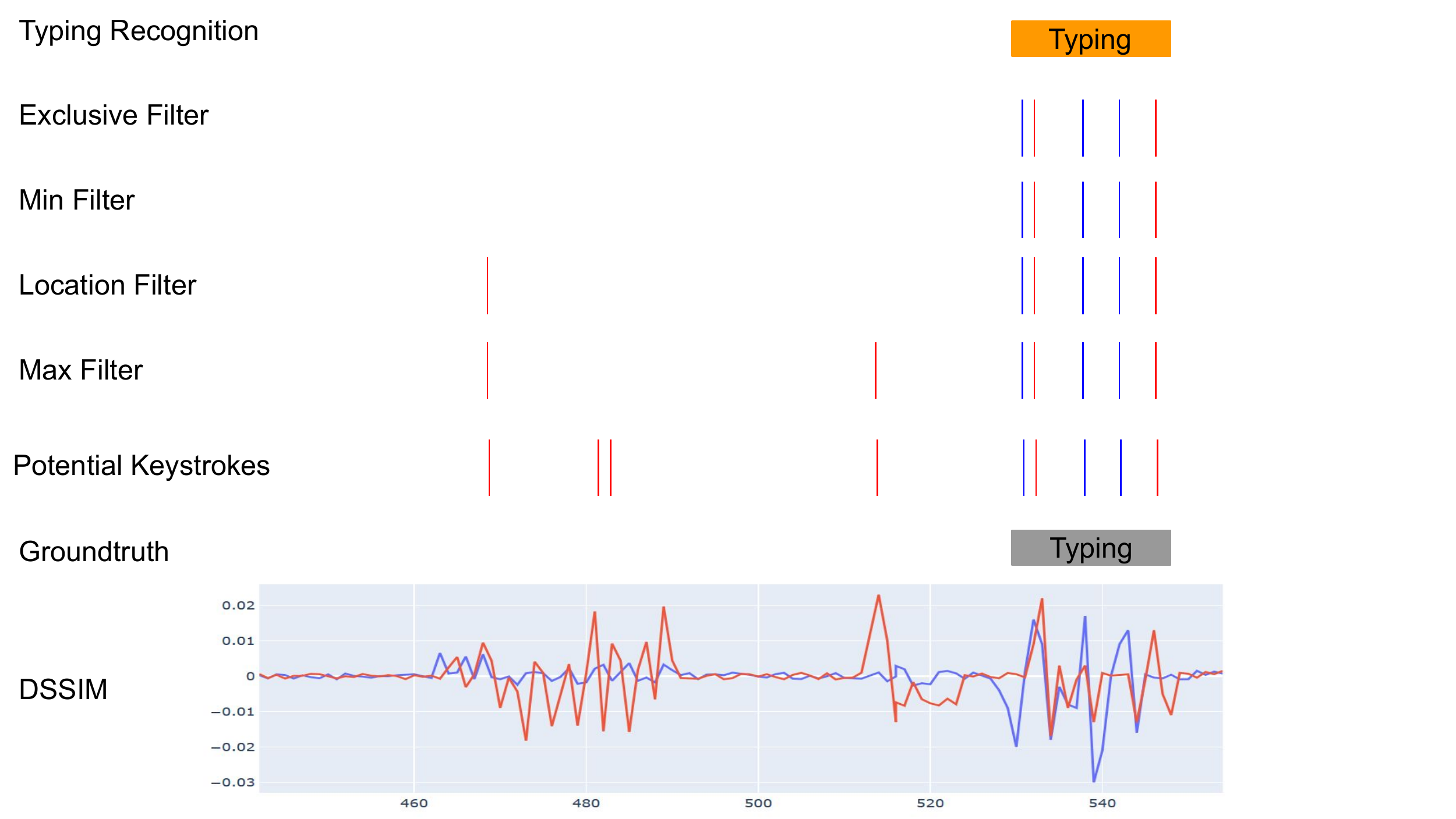}
\caption{An example of our typing activity detection heuristics being applied on potential keystrokes, which resulted in a true positive. Red ticks and lines are for right hand, while blue lines and ticks are for left hand.}%
\label{fig:typing-example}
\end{figure}

\begin{figure}[H]
\centering
\includegraphics[width=\linewidth]{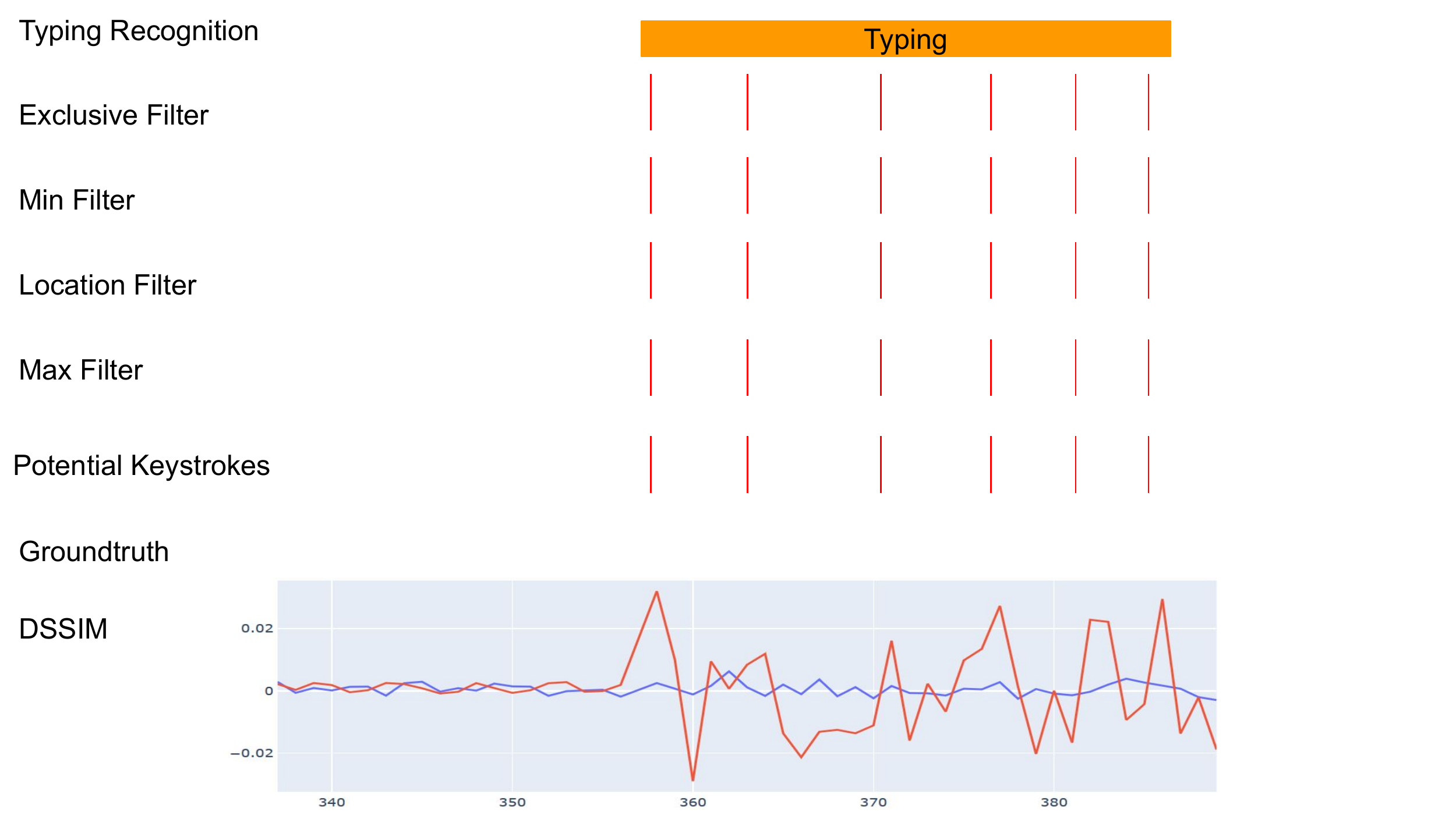}
\caption{An example of our typing activity detection heuristics being applied on potential keystrokes, which resulted in a false positive. Red ticks and lines are for right hand, while blue lines and ticks are for left hand.}%
\label{fig:typing-example-false}
\end{figure}

\cref{fig:typing-example,fig:typing-example-false} elucidates the working of our heuristic-based typing activity detection technique, by means of two real scenarios that we encountered during our experimentation. 

\section{Keystroke Prediction Tables and Figures}
\label{appendix:keystroke-prediction}

{\renewcommand*{\arraystretch}{1.75}
\begin{table}[H]
 \centering
 \caption{Classification of left arm displacements.}
 \label{tab:observed-direction-left}
 \begin{tabular}{|l|c|}
 \hline
 \textbf{Conditions} & \textbf{Direction} \\ \hline
 $\overrightarrow{om_{i}(x)} \geq 0$ and $\overrightarrow{om_{i}(y)} \geq 0$ & $NW$ \\ \hline
 $\overrightarrow{om_{i}(x)} \geq 0$ and $\overrightarrow{om_{i}(y)} \leq 0$ & $SW$ \\ \hline
 $\overrightarrow{om_{i}(x)} \leq 0$ and $\overrightarrow{om_{i}(y)} \geq 0$ & $NE$ \\ \hline
 $\overrightarrow{om_{i}(x)} \leq 0$ and $\overrightarrow{om_{i}(y)} \leq 0$ & $SE$ \\ \hline
 \end{tabular}
\end{table}
}

{\renewcommand*{\arraystretch}{1.75}
\begin{table}[H]
 \centering
 \caption{Classification of right arm displacements.}
 \label{tab:observed-direction-right}
 \begin{tabular}{|l|c|}
 \hline
 \textbf{Conditions} & \textbf{Direction} \\ \hline
 $\overrightarrow{om_{i}(x)} \geq 0$ and $\overrightarrow{om_{i}(y)} \geq 0$ & $NE$ \\ \hline
 $\overrightarrow{om_{i}(x)} \geq 0$ and $\overrightarrow{om_{i}(y)} \leq 0$ & $SE$ \\ \hline
 $\overrightarrow{om_{i}(x)} \leq 0$ and $\overrightarrow{om_{i}(y)} \geq 0$ & $NW$ \\ \hline
 $\overrightarrow{om_{i}(x)} \leq 0$ and $\overrightarrow{om_{i}(y)} \leq 0$ & $SW$ \\ \hline
 \end{tabular}
\end{table}
}
 
\begin{figure}[H]
\centering
\includegraphics[width=\linewidth]{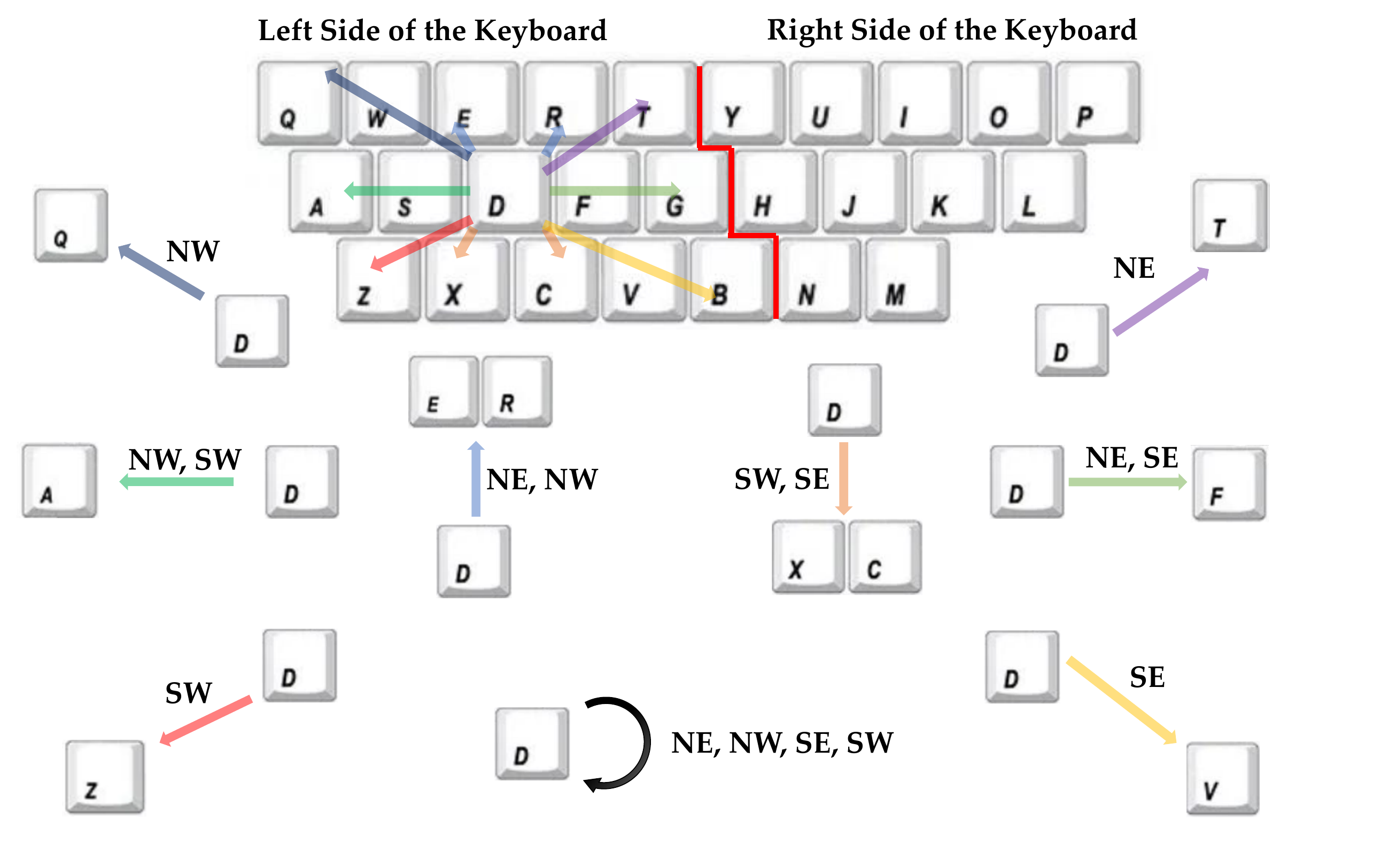}
\caption{Splitting the keyboard into two halves, and the template inter-keystroke directions following the alphabet `D'.}
\label{fig:key-dir}
\end{figure}

{\renewcommand*{\arraystretch}{1.5}
\begin{table}[H]
 \centering
 \caption{Mapping of template inter-keystroke directions.}
 \label{tab:template-directions}
 \begin{tabular}{|p{5.5cm}|c|}
 \hline
 \textbf{Relationship Between $key_{i}$ and $key_{j}$} & \textbf{Template Direction(s)} \\ \hline
 $key_{i}$ is the same key as $key_{j}$ & $NE$, $SE$, $NW$, $SW$ \\ \hline
 $key_{i}$ is in the same row of $key_{j}$ and $key_{j}$ is to the east of $key_{i}$ & $NE$, $SE$ \\ \hline
 $key_{i}$ is in the same row of $key_{j}$ and $key_{j}$ is to the west of $key_{i}$ & $NW$, $SW$ \\ \hline
 $key_{i}$ is in the row above of $key_{j}$ and $key_{j}$ vertically overlaps the key $key_{i}$ & $NE$, $NW$ \\ \hline
 $key_{i}$ is in the row below of $key_{j}$ and $key_{j}$ vertically overlaps the key $key_{i}$ & $SE$, $SW$ \\ \hline
 $key_{i}$ is in the row above of $key_{j}$, $key_{j}$ does not vertically overlap the key $key_{i}$, and $key_{j}$ is to the east of $key_{i}$ & $NE$ \\ \hline
 $key_{i}$ is in the row above of $key_{j}$, $key_{j}$ does not vertically overlap the key $key_{i}$, and $key_{j}$ is to the west of $key_{i}$ & $NW$ \\ \hline
 $key_{i}$ is in the row below of $key_{j}$, $key_{j}$ does not vertically overlap the key $key_{i}$, and $key_{j}$ is to the east of $key_{i}$ & $SE$ \\ \hline
 $key_{i}$ is in the row below of $key_{j}$, $key_{j}$ does not vertically overlap the key $key_{i}$, and $key_{j}$ is to the west of $key_{i}$ & $SW$ \\ \hline
 \end{tabular}
\end{table}
}

\section{Noise Characteristics}
\label{appendix:noise-char}

\begin{figure}[H]
\centering
\begin{subfigure}{\linewidth}
\centering
\includegraphics[width=\textwidth]{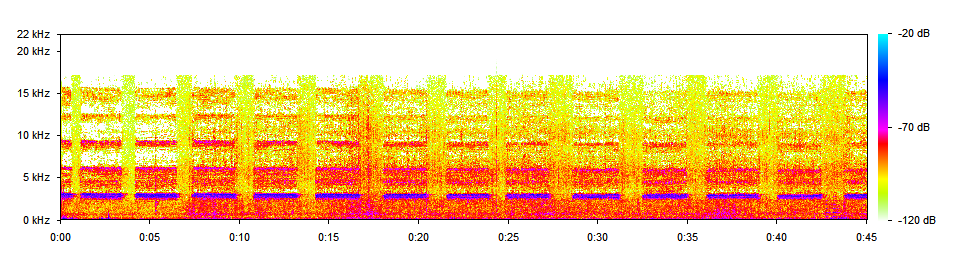}
\caption{Bird Chirps}
\label{fig-birds.png}
\end{subfigure}
\begin{subfigure}{\linewidth}
\centering
\includegraphics[width=\textwidth]{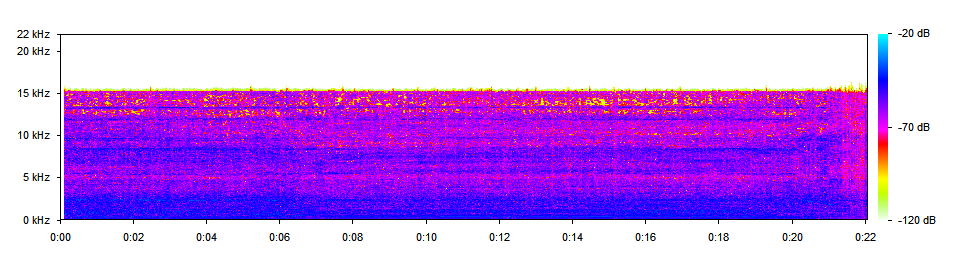}
\caption{Jackhammer}
\label{fig-jackhammer.png}
\end{subfigure}
\begin{subfigure}{\linewidth}
\centering
\includegraphics[width=\textwidth]{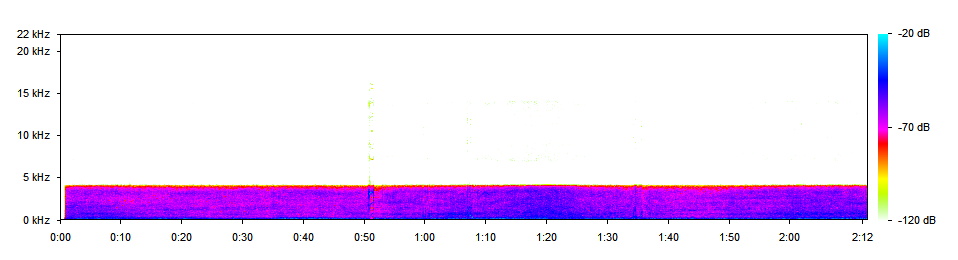}
\caption{Lawn Mower}
\label{fig-lawnmower.png}
\end{subfigure}
\begin{subfigure}{\linewidth}
\centering
\includegraphics[width=\textwidth]{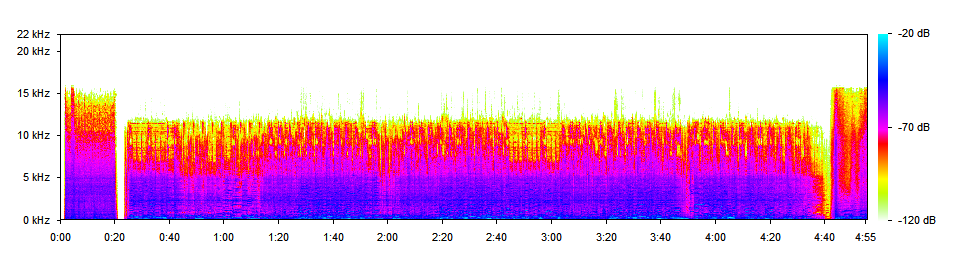}
\caption{Music (Pop/Rock)}
\label{fig-music.png}
\end{subfigure}
\begin{subfigure}{\linewidth}
\centering
\includegraphics[width=\textwidth]{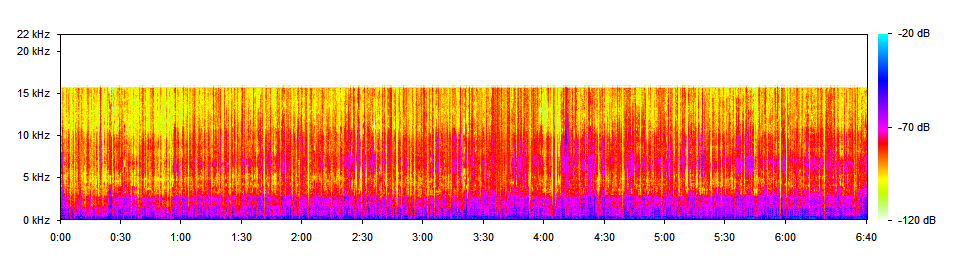}
\caption{Talking (Male)}
\label{fig-talk.png}
\end{subfigure}
\begin{subfigure}{\linewidth}
\centering
\includegraphics[width=\textwidth]{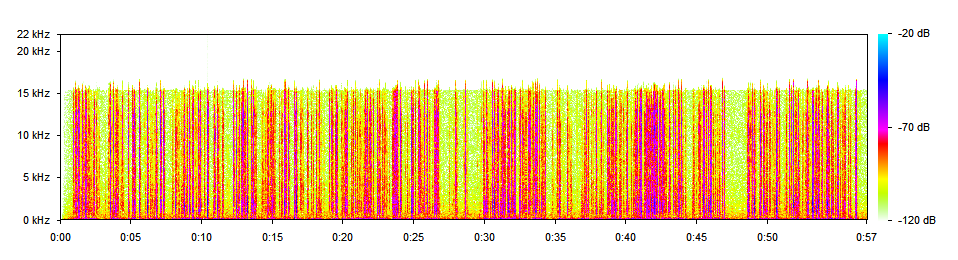}
\caption{Typing}
\label{fig-typing.png}
\end{subfigure}
\caption{Exemplary frequency spectrum plots of different sounds that were evaluated as background noise. These audio files were sourced from the \texttt{AudioSet} dataset \cite{audioset}.}
\label{fig:noise}
\end{figure}

\cref{fig:noise} shows the frequency spectrum plots of sample sounds that were evaluated as background noise. In \cref{fig:noise} we can see that the high amplitude frequencies are within a very narrow band for bird chirps, with periodic patterns. Both the jackhammer and lawn mower sounds have high amplitude frequencies more uniformly spread out across their frequency range and time. Music has a wide range of high and low amplitude frequencies, but some patterns can be observed over time. Talking sound has high amplitudes for lower frequencies, and time-based patterns are not easily identified. Typing sound has high amplitudes for a wide band of frequencies, and sporadic keystrokes can be easily identified.

\section{Changing Position and Posture}
\label{appendix:changing-posture}

\begin{figure}[H]
\centering
\begin{subfigure}[b]{0.31\linewidth}
\includegraphics[width=\linewidth]{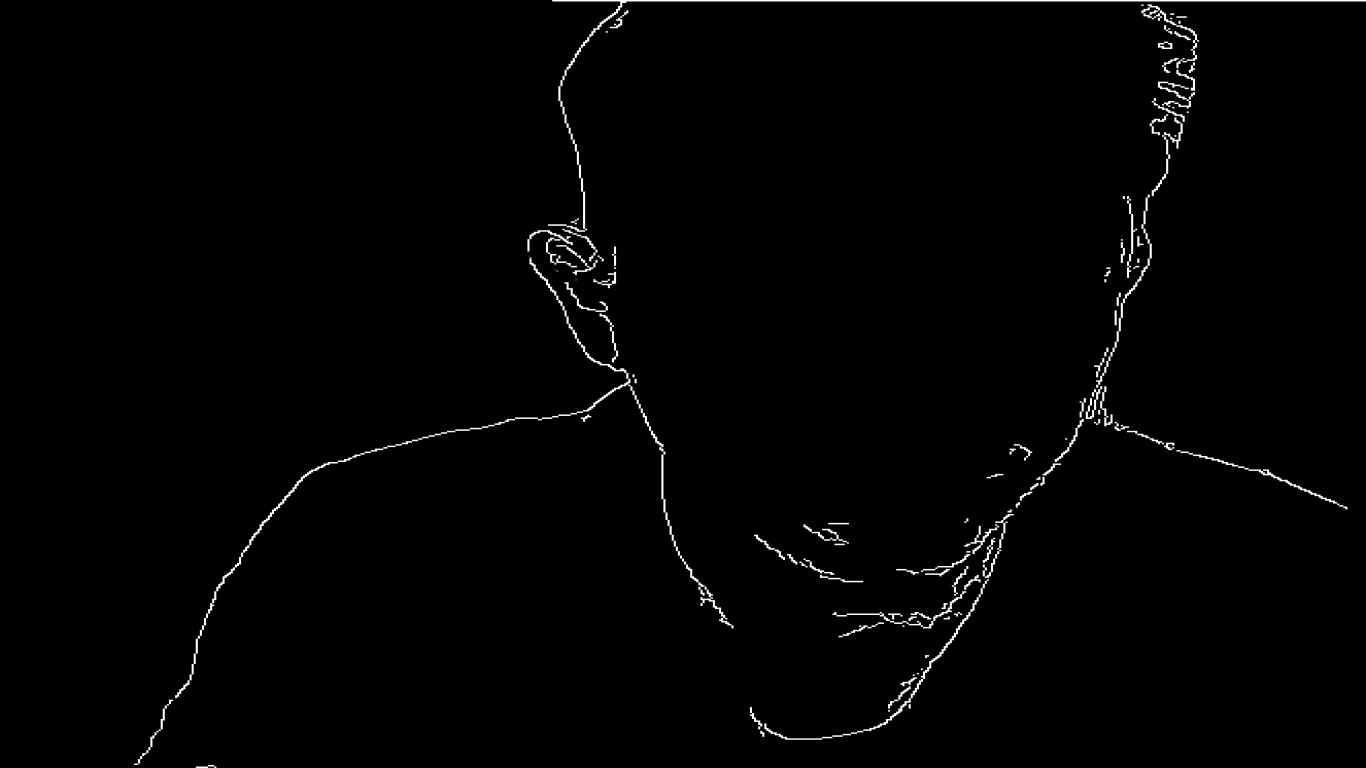}
\caption{}
\label{fig:i-position-1}
\end{subfigure}
\begin{subfigure}[b]{0.31\linewidth}
\includegraphics[width=\linewidth]{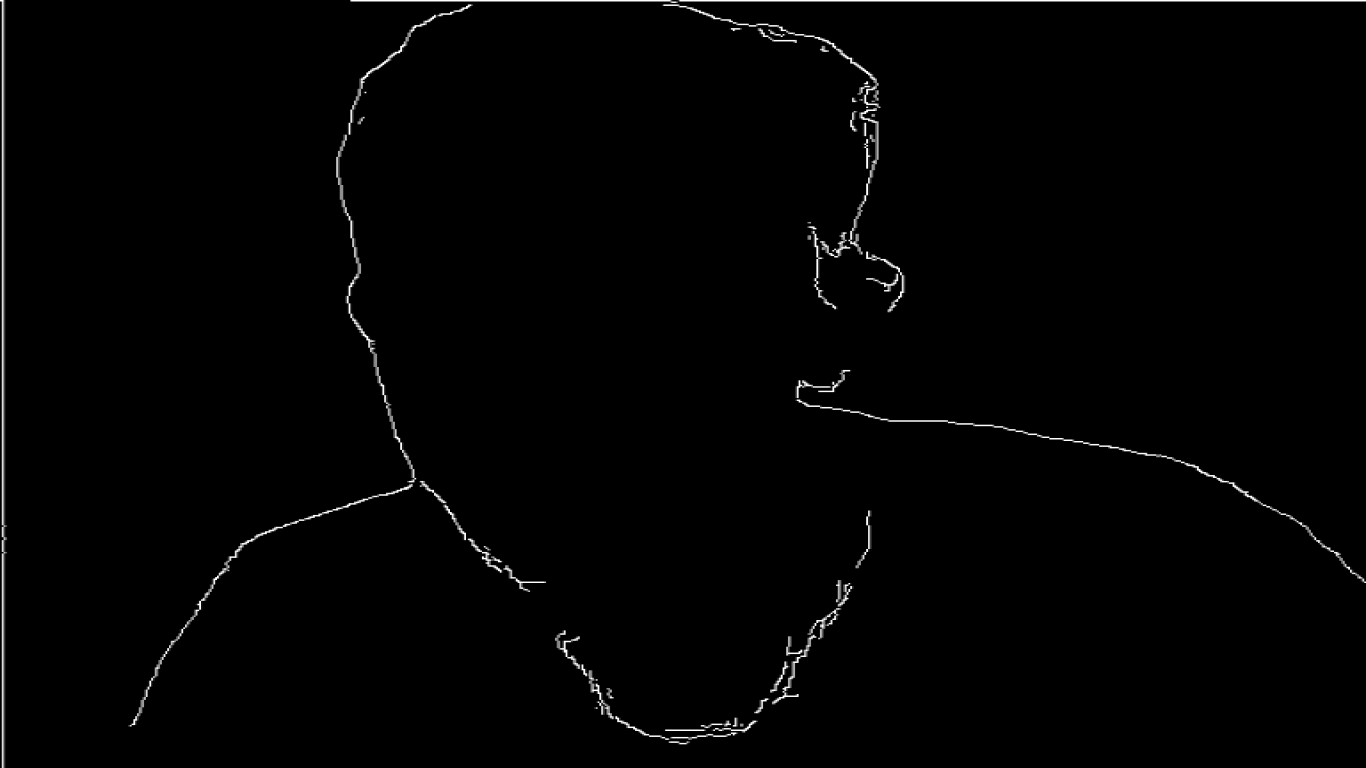}
\caption{}
\label{fig:i-position-2}
\end{subfigure}
\begin{subfigure}[b]{0.31\linewidth}
\includegraphics[width=\linewidth]{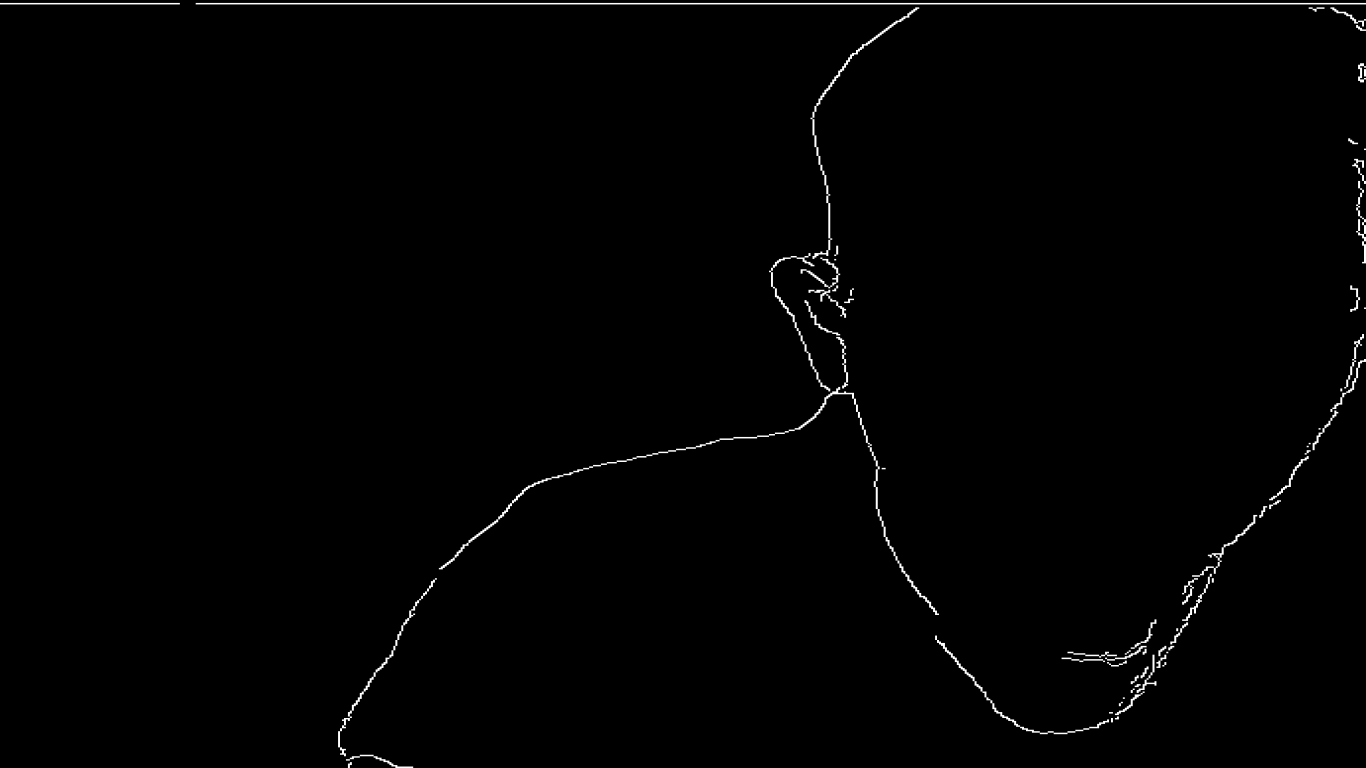}
\caption{}
\label{fig:i-position-3}
\end{subfigure}
\caption{Three frames from participant I's dataset, which shows the different postures and positions during the video call; (c) Shows a frame where one of the shoulders and arms are not visible.  }
\label{fig:typing-movement}
\end{figure}

\section{At-Home Participant Demographics and Experimental Settings}
\label{tab:demographics}

{
\begin{table*}[b]
\newcommand{\specialcell}[2][c]{%
\begin{tabular}[#1]{@{}c@{}}#2\end{tabular}}
\renewcommand*{\arraystretch}{1.33}
\resizebox{\textwidth}{!}{
\begin{tabular}{|c|c|c|c|c|c|c|c|c|c|c|}
\hline
\textbf{Participant} &
  \textbf{A} &
  \textbf{B} &
  \textbf{C} &
  \textbf{D} &
  \textbf{E} &
  \textbf{F} &
  \textbf{G} &
  \textbf{H} &
  \textbf{I} &
  \textbf{J} \\ \hline
\textbf{Gender} &
  M &
  M &
  M &
  F &
  F &
  F &
  M &
  M &
  M &
  M \\ \hline
\textbf{Age} &
  21 &
  24 &
  24 &
  28 &
  29 &
  23 &
  28 &
  21 &
  22 &
  25 \\ \hline
\textbf{Dominant Hand} &
  Right &
  Right &
  Right &
  Right &
  Ambidextrous &
  Right &
  Right &
  Right &
  Right &
  Right \\ \hline
\textbf{Typing Style} &
  Touch &
  Touch &
  Hunt-and-Peck &
  Hunt-and-Peck &
  Hybrid &
  Touch &
  Touch &
  Hybrid &
  Hunt-and-Peck &
  Touch \\ \hline
\textbf{Webcam} &
  \begin{tabular}[c]{@{}c@{}}Dell XPS 15 \\ 7590\end{tabular} &
  \begin{tabular}[c]{@{}c@{}}Mimoday\\ S2\end{tabular} &
  \begin{tabular}[c]{@{}c@{}}Mac Book Pro\\ A1990\end{tabular} &
  \begin{tabular}[c]{@{}c@{}}Lenovo Ideapad 5\\ 15IIL05\end{tabular} &
  \begin{tabular}[c]{@{}c@{}}Mac Book Pro\\ A1398\end{tabular} &
  Microsoft LifeCam Cinema  &
  \begin{tabular}[c]{@{}c@{}}Logitech C922x pro\\  stream webcam\end{tabular} &
  \begin{tabular}[c]{@{}c@{}} Acer\\  N19C3 \end{tabular}&
  \begin{tabular}[c]{@{}c@{}} Lenovo Ideapad \\ 320-15abr \end{tabular}&
  \begin{tabular}[c]{@{}c@{}} Lenovo Ideapad \\ 320-15abr \end{tabular} \\ \hline
\textbf{Keyboard} &
  \begin{tabular}[c]{@{}c@{}}Dell XPS 15 \\ 7590\end{tabular} &
  \begin{tabular}[c]{@{}c@{}}Blackwidow Chroma \\ RZ03-0122\end{tabular} &
  \begin{tabular}[c]{@{}c@{}}Mac Book Pro\\ A1990\end{tabular} &
  \begin{tabular}[c]{@{}c@{}}Lenovo Ideapad 5\\ 15IIL05\end{tabular} &
  \begin{tabular}[c]{@{}c@{}}Mac Book Pro\\ A1398\end{tabular} &
  \begin{tabular}[c]{@{}c@{}}HP \\ W2M75UA \end{tabular} &
  \begin{tabular}[c]{@{}c@{}}Dell Wireless \\ WK636p\end{tabular} &
  \begin{tabular}[c]{@{}c@{}}Acer \\ N19C3 \end{tabular}&
  \begin{tabular}[c]{@{}c@{}}Lenovo Ideapad \\ 320-15abr \end{tabular}&
  \begin{tabular}[c]{@{}c@{}}Lenovo Ideapad \\ 320-15abr \end{tabular} \\ \hline
\textbf{Sleeves} &
  Short &
  Short &
  Short &
  Short &
  Short &
  Short &
  Long &
  Short &
  Short &
  Short \\ \hline
\textbf{Laptop/Desktop} &
  Laptop &
  Desktop &
  Laptop &
  Laptop &
  Laptop &
  Laptop &
  Desktop &
  Laptop &
  Laptop &
  Laptop \\ \hline
\textbf{Height ($cm$)} &
  180 &
  168 &
  170 &
  168 &
  152 &
  168 &
  168 &
  173 &
  178 &
  168 \\ \hline
\textbf{\begin{tabular}[c]{@{}c@{}}Typing Speed\\ (Keystrokes/Second)\end{tabular}} &
  8.83 &
  3.67 &
  2.75 &
  3 &
  3 &
  3.67 &
  5.25 &
  2.83 &
  1.41 &
  2.58 \\ \hline
\textbf{\begin{tabular}[c]{@{}c@{}}Typing Accuracy\\ (\%)\end{tabular}} &
  96.9 &
  93.3 &
  73.1 &
  90.5 &
  98.5 &
  96.1 &
  93.8 &
  91.2 &
  49 &
  82 \\ \hline
\textbf{\begin{tabular}[c]{@{}c@{}} Sample Typing \\ Posture \end{tabular}} &
  \raisebox{-0.4\height}{\includegraphics[width=0.13\textwidth]{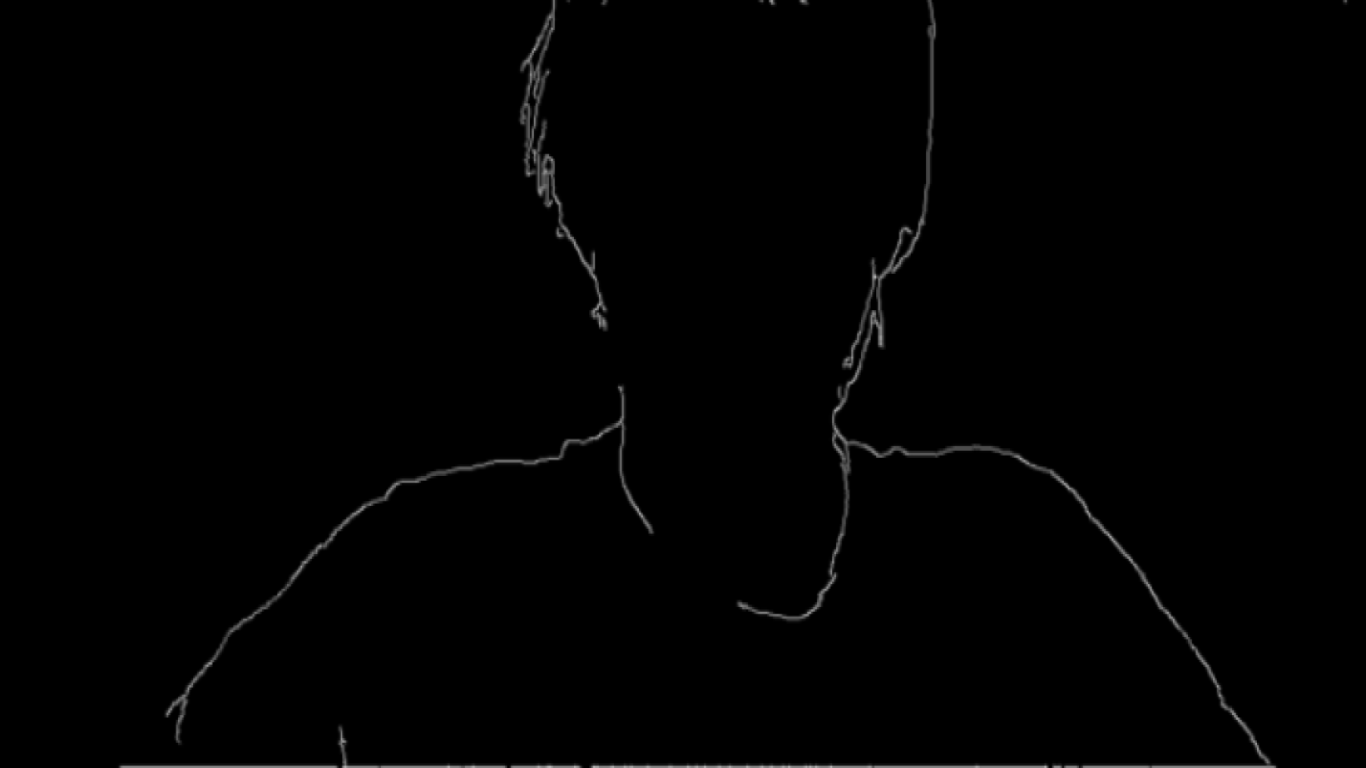}} &
  \raisebox{-0.4\height}{\includegraphics[width=0.13\textwidth]{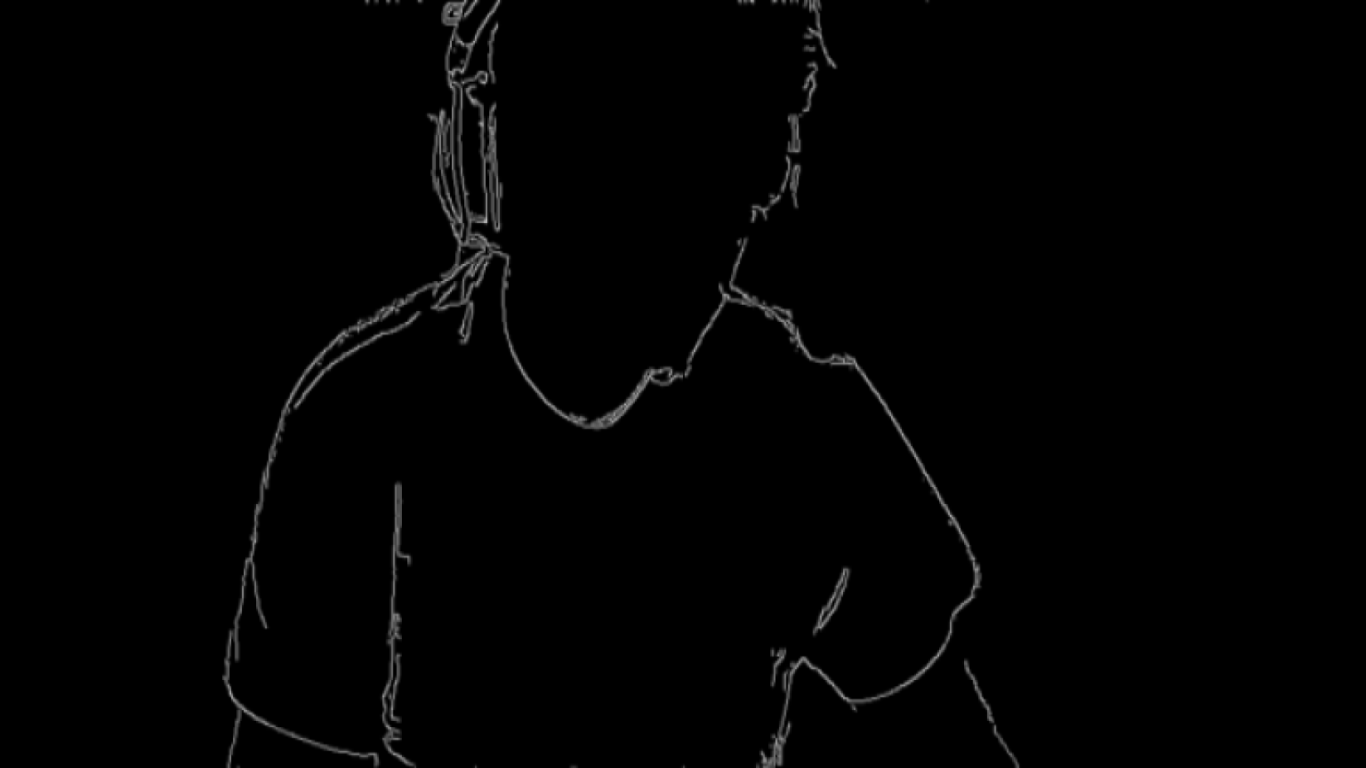}} &
  \raisebox{-0.4\height}{\includegraphics[width=0.13\textwidth]{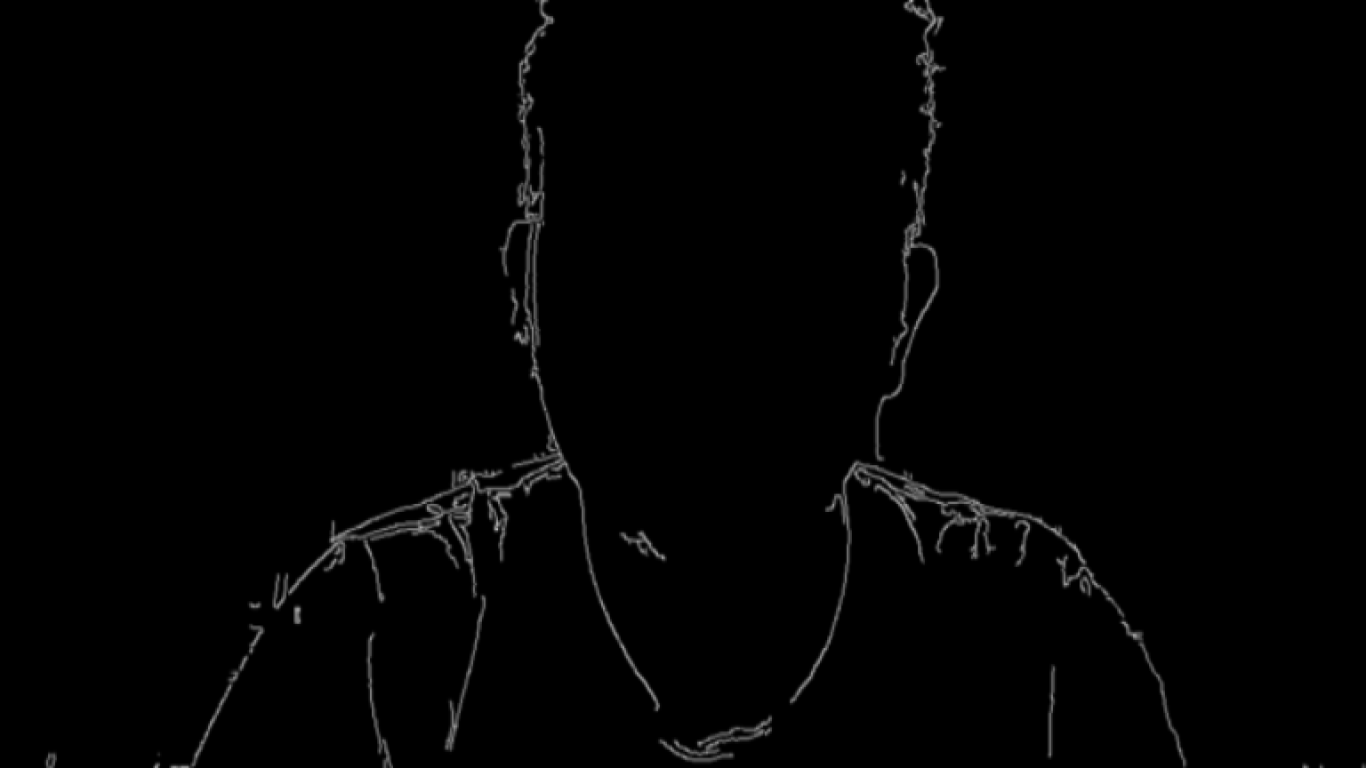}} &
  \raisebox{-0.4\height}{\includegraphics[width=0.13\textwidth]{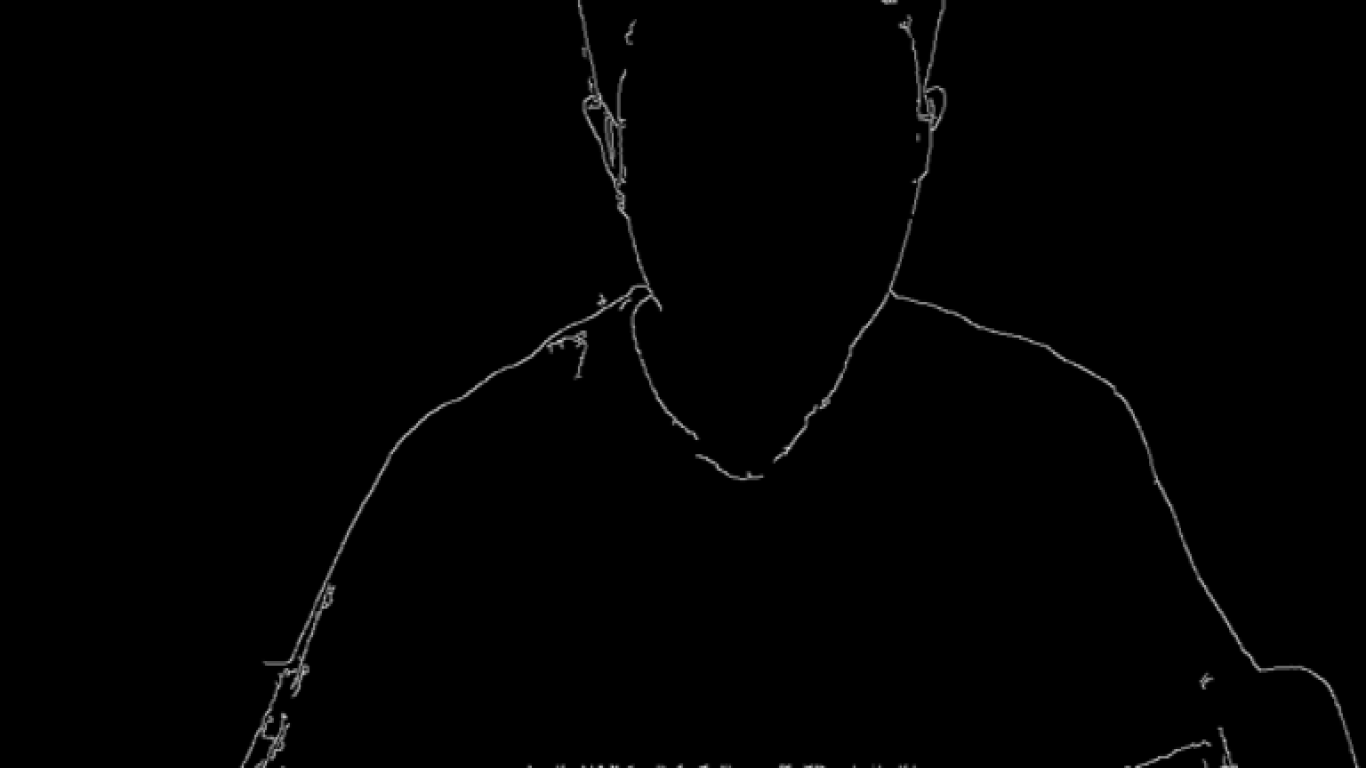}} &
  \raisebox{-0.4\height}{\includegraphics[width=0.13\textwidth]{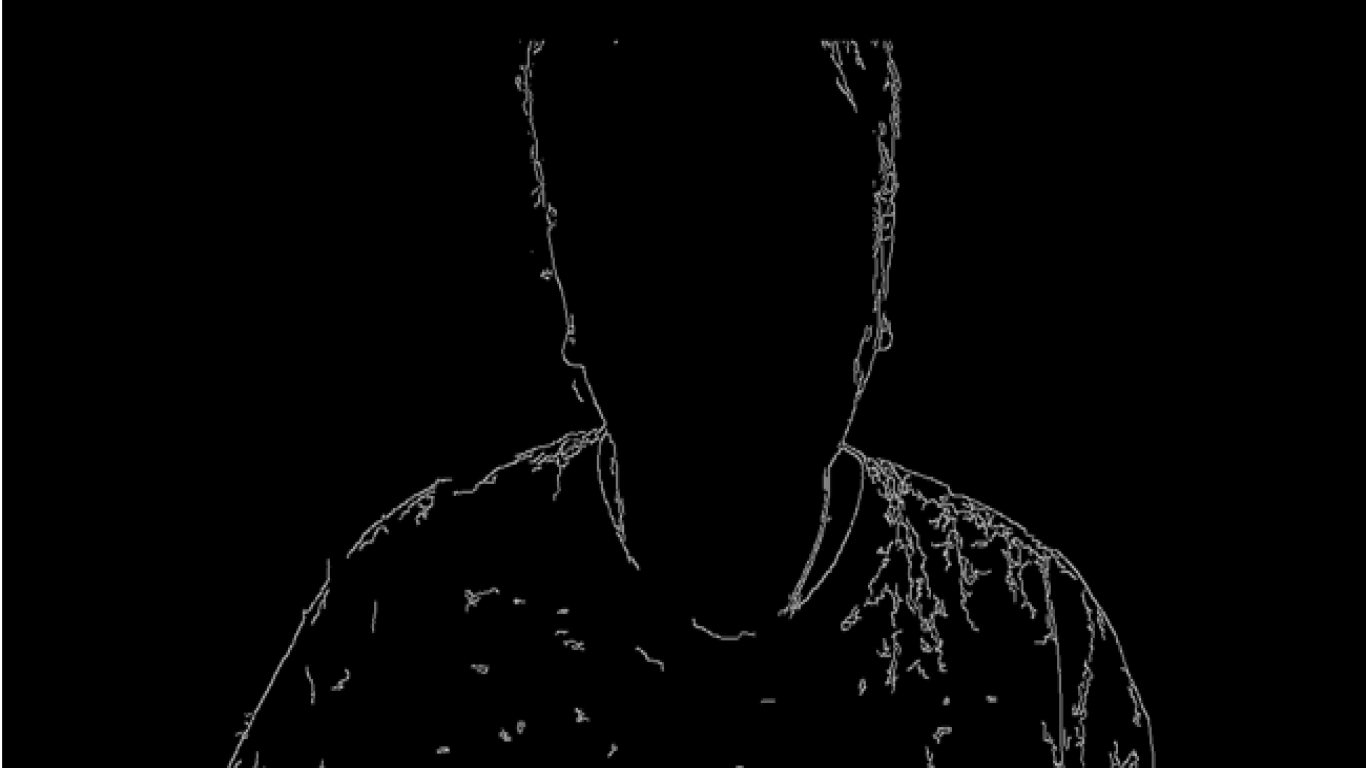}} &
  \raisebox{-0.4\height}{\includegraphics[width=0.13\textwidth]{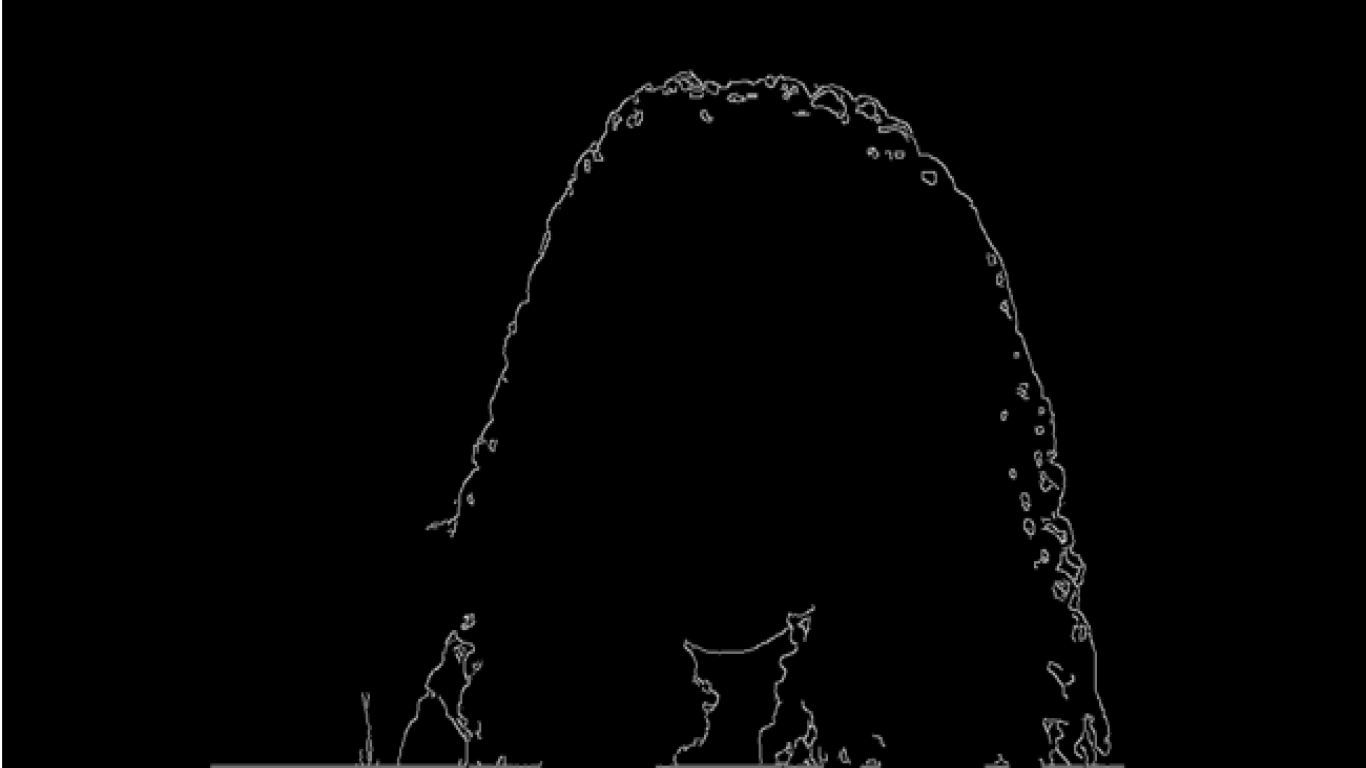}} &
  \raisebox{-0.4\height}{\includegraphics[width=0.13\textwidth]{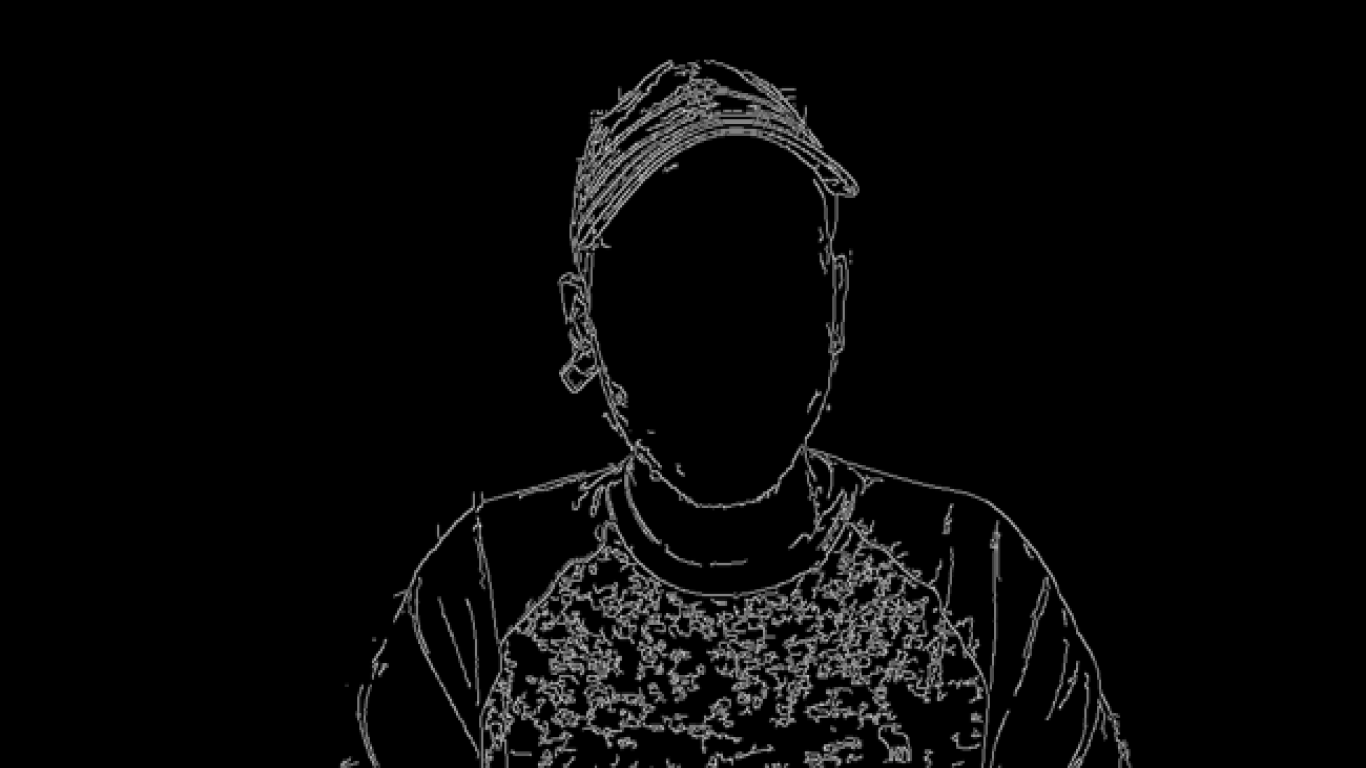}} &
  \raisebox{-0.4\height}{\includegraphics[width=0.13\textwidth]{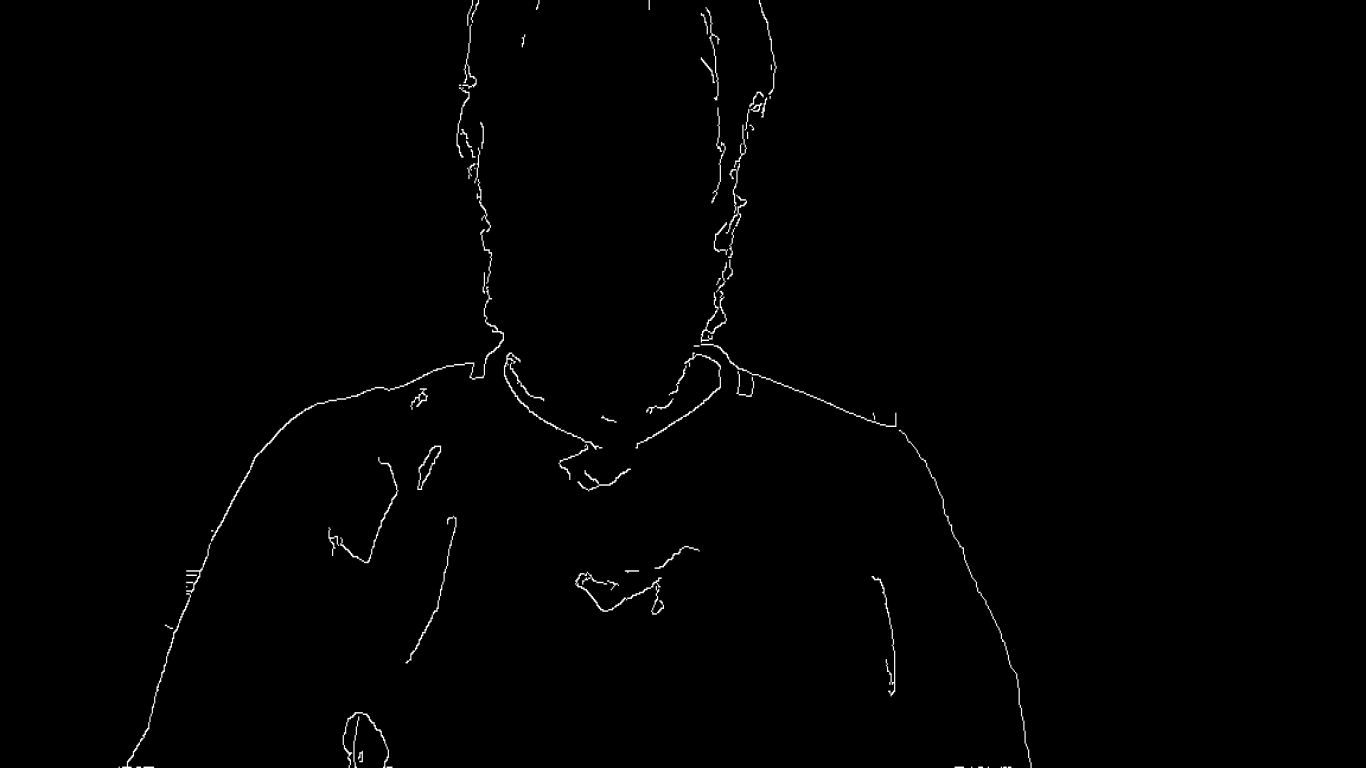}} &
  \raisebox{-0.4\height}{\includegraphics[width=0.13\textwidth]{fig-i1.png}} &
  \raisebox{-0.4\height}{\includegraphics[width=0.13\textwidth]{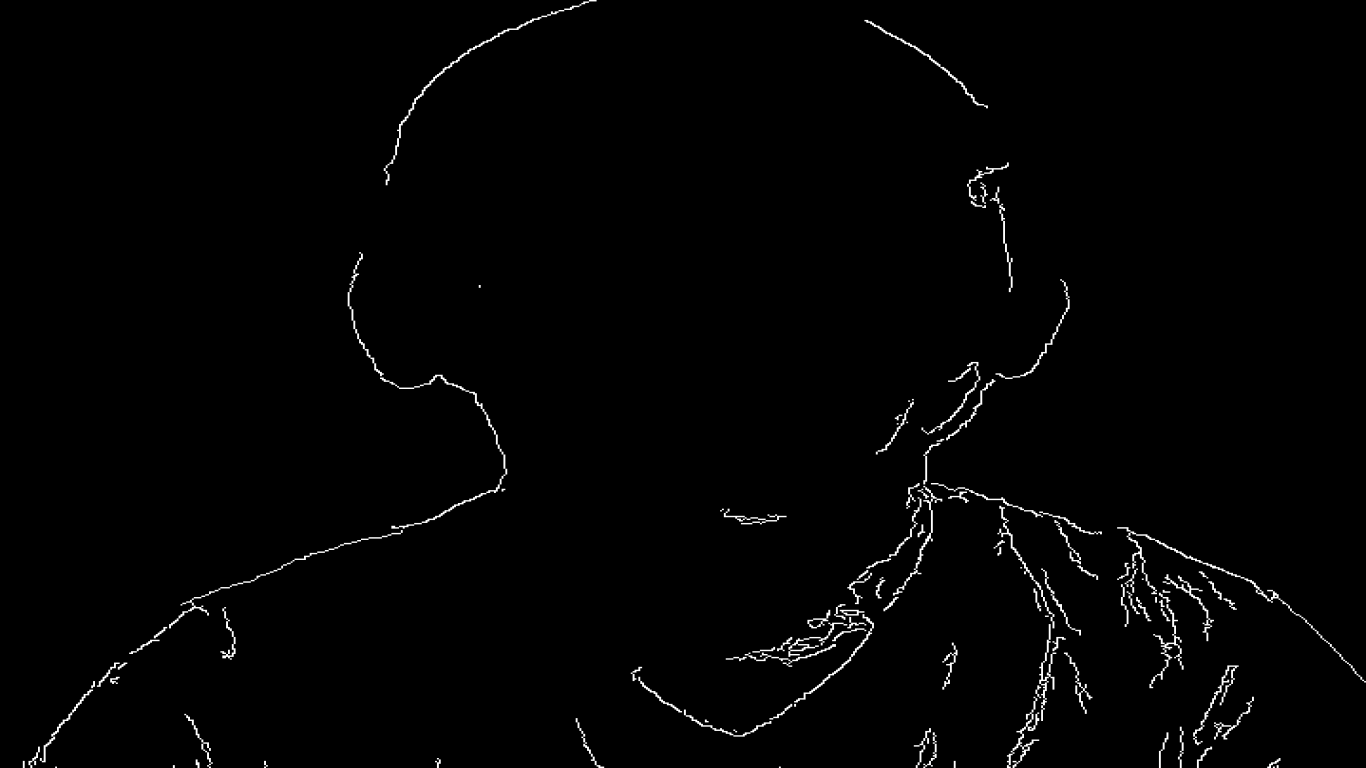}}  \\ \hline    
\end{tabular}
}
\caption{Participant demographics and experimental settings for At-Home setup.}
\vspace{4in}
\end{table*}
}

\end{appendices}

\end{document}